\documentclass{elsarticle}
\usepackage{a4wide,amsmath,amssymb,graphics,epsfig}
 \allowdisplaybreaks[2]
\newcommand{\hatA}[0]{\hat{A}}

\newcommand{\tr}[0]{\textnormal{tr}}

\usepackage{grffile}
\begin{document}

\begin{frontmatter}
\title{Complete Monopole Dominance of the Yang-Mills Confining Potential}

\author[a]{Nigel Cundy}
\author[b,c]{Y. M. Cho}
\author[a]{Weonjong Lee}
\address[a]{    Lattice Gauge Theory Research Center, FPRD, and CTP, Department of Physics \&
    Astronomy,\\ Seoul National University, Seoul, 151-747, South Korea}
\address[b]{Administration Building 310-4, Konkuk University,
Seoul 143-701, Korea}
\address[c]{ Department of Physics \&
    Astronomy,\\ Seoul National University, Seoul, 151-747, South Korea}


\begin{abstract}
We continue our investigation of quark confinement using a particular variant of the Cho-Duan-Ge gauge independent Abelian decomposition. The decomposition splits the gauge field into a restricted Abelian part and a coloured part in a way that preserves gauge covariance. Furthermore the restricted part of the gauge field can be divided into a Maxwell term and a topological term. Previously, we showed that by a particular choice of this decomposition we could fully describe the confining potential using only the restricted gauge field. We proposed that various topological objects (a form of magnetic monopole) could arise in the restricted field which would drive confinement. Our mechanism does not explicitly refer to a dual Meissner effect, nor does it use centre vortices. We did not need to gauge fix or introduce any new dynamical fields.

In this work, we show that if we do gauge fix in addition to performing the Abelian decomposition then it is possible to ensure that the topological part of the restricted field fully accounts for the confining potential. Our relationship is exact: there is no approximation or model involved. This isolates the objects responsible for confinement from non-confining contributions to the gauge field and allows us to directly search for our proposed topological objects. Using numerical studies in SU(2), we confirm that our proposed monopoles are present in the field, and the winding number associated with these monopoles is a key factor driving quark confinement.

In SU(2), our monopoles are described by two parameters, which we label as $\cos 2a$ and $c$. We show that it is possible to re-parametrise the Yang Mills action and the functional integration measure in terms of these variables (plus the necessary additional parameters). We can thus treat the monopoles as dynamical variables in the functional integral. This might be the first step in a future analytical computation to complement our numerical results.  
\end{abstract}
\end{frontmatter}
\section{Introduction}
An enduring problem in QCD is to find the mechanism which causes quark confinement, which is known to be non-perturbative in its origin. Although several models have been proposed -- for example, center vortices~\cite{Vortices}, and a dual Meissner effect due to magnetic monopoles~\cite{Monopoles1,Monopoles2,Mandelstam:1976,thooft:1976} -- there has not yet been a convincing demonstration that any of them are correct.
Our work~\cite{Cundy:2015caa,Cundy:2013xsa} investigates the Cho-Duan-Ge (CDG) Abelian decomposition (sometimes referred to as the Cho-Faddeev-Niemi decomposition)~\cite{Cho:1980,Cho:1981,F-N:98,Shabanov:1999,Duan:1979}. Unlike Dirac and 't Hooft (Maximum Abelian Gauge) monopoles, the CDG decomposition allows for monopole solutions while respecting the gauge symmetry and does not require a singular gauge field or an additional Higgs field. The decomposition is constructed from a colour field, $n$, which may be built from a SU($N_C$) matrix $\theta$, where SU($N_C$) is the gauge group of QCD. Each $n$ defines a different decomposition, and an important question is the best way of choosing this field.
Recent work~\cite{Kondo:2008su,Shibata:2009af,Kondo:2010pt,Kondo:2005eq} has demonstrated that the magnetic part of the field strength dominates the confining string, using a decomposition constructed from one particular choice of $\theta \in
SU(N_C)/U(N_C-1)$; however in this case only one
of the possible $N_C-1$ types of monopole is visible.

 We consider a different choice of $\theta \in SU(N_C)/(U(1))^{N_C-1}$. The initial goal of our study was to investigate whether monopoles apparent in this construction may also lead to confinement\footnote{A different choice of an $SU(N_C)/(U(1))^{N_C-1}$ Abelian decomposition was described in~\cite{Shibata:2007pi},  but without a discussion of the relationship to quark confinement.}. Concentrating on the Wilson Loop,
an observable used to measure the string tension, we showed that the path ordering may be removed by diagonalising the gauge links along the Wilson Loop by an $SU(N_C)/(U(1))^{N_C-1}$ field $\theta$.
This, in principle, allowed us to use Stokes' theorem to express the Wilson Loop in terms of a surface integral over an Abelian restricted gauge field strength tensor.
 We used the CDG decomposition, constructed from a colour field $n^j = \theta\lambda^j\theta^\dagger$ for a diagonal Gell-Mann matrix $\lambda^j$, to find a consistent construction of this restricted field across space-time, and not just for those gauge links along the Wilson Loop where it was originally defined.
Our relationship for the string tension in terms of this restricted field is exact: we do not require any approximations or additional path integrals. We also searched for topological structures in the field strength. Among the parameters that defined the $\theta$ field, one class of them, which we labelled as $c_i$ 
-- one parameter in SU(2), or three parameters in SU(3) -- could be split into various homotopy classes characterised by an integer winding number. This winding number survives smooth gauge transformations, and most importantly the trace of the Wilson Loop could be shown to depend explicitly on this winding number. With the number of topological objects contributing to the winding proportional to the area of the Wilson Loop, this provided a possible mechanism for quark confinement. We deduced the field strengths associated with these topological objects, and showed that the Yang-Mills field strength tensor in restricted lattice QCD showed the expected structures. However, in our previous numerical work we restricted ourselves to gauge invariant observables. We showed that the Abelian restricted field could exactly account for the confining potential, but we did not show that the topological term can account for it, in part because this depends on the gauge.

The problem we faced was that the parameters $a_i$ and $c_i$ which described the $\theta$ field are not gauge invariant -- only the winding number survives continuous gauge transformations, and can be destroyed by a discontinuous transformation. This means that it is necessary to fix the gauge before measuring these quantities. However it was not clear to us which gauge to fix to. The issue was that our theoretical formulation is only valid if the $\theta$ field is smooth and continuous. In the continuum, this is certainly true in some gauges (one can choose a gauge where $\theta$ is a constant matrix), but not every gauge, and there was no obvious reason why it should be satisfied using the standard Landau or Coloumb gauges. This is even before we encounter the difficulty of defining `smooth' or `continuous' on the lattice. We therefore decided to leave the question of gauge dependent quantities to one side, and concentrate on gauge invariant quantities such as the string tension and Field Strength.

However, to demonstrate that our mechanism for confinement is correct, we need to look at the gauge dependent quantities, and show that in some gauge the winding of $c_i$ exists in practice and can account for the string tension. In this work, we suggest that there is at least one particular gauge where a) the $\theta$ term wholly dominates confinement; b) the $\theta$ field is smooth; and c) we can observe the winding and how it can lead to confinement. The question is then whether this picture survives gauge transformations: we see that the effect of the gauge transformation is just to move the same topological effects from the topological part of the restricted field into the Maxwell part of the restricted field. In other words, the same topology drives confinement in other gauges, but they are hidden within the gauge field $A_\mu$. Another way of expressing what we are proposing is that we can extract the relevant degrees of freedom that drive confinement through a particular choice of gauge fixing and Abelian decomposition, and place them in the $\theta$ field, where they can be isolated from the rest of the gauge field and studied in closer detail.

One other issue that we left undone in our earlier work was the question of the quantum theory. In our previous work, we took each gauge configuration individually and examined them independently. From a lattice field theorist used to Monte-Carlo simulations, this is perhaps a natural way of doing things, but maybe not so much for those used to trying to perform path integrals analytically. We therefore also give the full quantum formulation of the Yang Mill fields in terms of our Abelian decomposition. Our idea is to re-write the path integral in terms of new fields, which include those which parametrise the Abelian decomposition and the restricted gauge field ($\theta$ and $\hat{A}$). Thus the parameters we are studying become dynamical quantum fields. We find the measure of the path integral by demanding that it is gauge invariant. The corresponding Yang-Mills action, re-written in terms of the new fields, shows some interesting features. Most importantly, we see hints that the same topological solutions that drive the area law for the Wilson Loop might also lead to a dynamically generated gluon mass. We do not, however, complete this analysis by constructing gauge invariant states (the glueballs) and measuring their mass spectrum; this still looks challenging. 

In section \ref{sec:2} we briefly review the main theoretical points of our previous work,~\cite{Cundy:2015caa}. In section \ref{sec:3}, we prepare for our discussion of the new formulation of Yang-Mills theory by discussing the gauge transformations of the $\theta$ and restricted field, while in section \ref{sec:4} we re-parametrise the Yang-Mills action to show how our decomposition can be easily visualised within the path integral. In section \ref{sec:5}, we discuss our gauge fixing which allows us to isolate the topological objects which dominate either the Wilson Loop or Polyakov line. In section \ref{sec:6} we show some numerical results in SU(2), while we conclude in section \ref{sec:7}.

Because of the expense of our numerical simulations, in this work we restrict ourselves to the quicker SU(2) theory. Some of our theoretical results are only given in SU(2), while others we have derived in both SU(2) and SU(3). The SU(3) theory proceeds in much the same way as SU(2) -- it involves the same concepts, in almost the same way -- but the algebra is considerably more cumbersome. For these reasons we intend to delay our full investigation of SU(3).

\section{Abelian Dominace of the Confining Potential}\label{sec:2}

The Wilson Loop in an SU($N_C$) gauge theory is defined as
\begin{align}
W_L[C_s,A] = & \frac{1}{N_C} \tr \left(W[C_s,A]\right) & W[C_s,A] = \mathcal{P}[e^{-ig\oint_{C_s} dx_\mu A_\mu(x)}]\label{eq:1}
\end{align}
for a closed curve $C_s$ of length $L$ which starts and finishes at a position $s$, where $\mathcal{P}$ represents path ordering and the gauge field, $A_\mu$, can be written in terms of the Gell-Mann matrices, $\lambda^a$, as $\frac{1}{2}A_\mu^a \lambda^a$. We will use the summation convention that the superscripts $a,b,\ldots$ on a Gell-Mann matrix implies that it should be summed over all values of $a$, $\lambda^a A^a \equiv \sum_{a = 1}^{N_C^2-1} \lambda^aA^a$, while the indices $j,k,\ldots$ are restricted only to the diagonal Gell-Mann matrices, so, in the standard representation, $A^j\lambda^j \equiv \sum_{j = 3,8,\ldots, N_C^2 - 1}\lambda^jA^j$. We shall often leave the gauge field dependence of $W$ and $W_L$ implicit.

The Wilson Loop when $C_s$ is an $R\times T$ rectangle, with spatial extent $R$ and temporal extent $T$, can be used to measure the confining static quark potential, $V(R)$,~\cite{Wilson:1974}
\begin{gather}
V(R)=-\lim_{T\rightarrow\infty} \log(\langle W_L[C_s]\rangle)/T,
\end{gather}
where $\langle \ldots \rangle$ denotes the vacuum expectation value. It is expected, and observed in lattice simulations, that for intermediate distances the confining potential is linear in $R$, so $V(R) \sim \rho R + k$, where $\rho$ is the string tension, and $k$ is a constant. At very small distances, the potential is expected to be Coulomb, while in the presence of fermion loops at very large distances the string is broken and the potential becomes independent of $R$~\cite{Bali:2005fu}. The main focus of this work is on the intermediate regime, so we expect that the expectation value of the Wilson Loop will scale with the spatial and temporal extents of the Wilson Loop as
\begin{gather}
\langle W_L[C_s]\rangle \sim e^{-\rho R T}.\label{eq:al1}
\end{gather}
This is known as the area law scaling of the Wilson Loop. As discussed in~\cite{Cundy:2015caa}, this is satisfied, if on each individual configuration, the Wilson Loop scales as
\begin{gather}
W_L[C_s] \sim e^{i F}, \label{eq:al2}
\end{gather}
where $F$ is randomly distributed from configuration to configuration (according to one of a certain set of distributions which includes those relevant for this work) with a mean value proportional to the area contained within the curve $C_s$. The eventual goal (and this work is intended as a step towards that goal) is to demonstrate from first principles that equation (\ref{eq:al2}) is satisfied in pure gauge QCD (and later full QCD), and thus that the quarks are linearly confined.

A difficulty with evaluating equation (\ref{eq:1}) is the path ordering. If the fields $A_\mu(x)$ at different $x$ and $\mu$ commuted with each other, then we could ignore the path ordering, and use Stokes' theorem to convert the line integral to a surface integral. The problem would then reduce to showing that there was some flux flowing through the surface so that the surface integral was proportional to the area (perhaps by counting lines of flux). However, $A_\mu$ are non-Abelian fields: we cannot immediately do this. Our approach is then first to construct an Abelian field $\hat{A}_\mu(x)$ so that $W_L[C_s,\hat{A}] = W_L[C_s,A]$, and the calculated string tension does not depend on which of the two fields we use. We may then remove the path ordering, apply Stokes' theorem to replace the line integral with a surface integral over some field strength, and then show that the surface integral is proportional to the area enclosed within the loop.

First we shall define what is meant by path ordering. We split $C_s$ into infinitesimal segments of length $\delta \sigma$, and define the gauge link as $U_\sigma \in SU(N_C) = \mathcal{P}[e^{-ig\int_{\sigma}^{\sigma + \delta \sigma} A_\sigma d\sigma}] \sim e^{-ig \delta \sigma A_\sigma}$. $0\le\sigma\le L$ represents the position along the curve and we write $A_\sigma \equiv A_{\mu(\sigma)}(x(\sigma))$. We have assumed and will require throughout this work that the gauge field, $A$, is differentiable. This limits us to only a certain subset of gauges, and once we have found a suitable gauge we are restricted to continuous gauge transformations, i.e. only those gauge transformations which can be built up from repeatedly applying infinitesimal gauge transformations, $A_\mu \rightarrow A_\mu + g^{-1}\partial_\mu \alpha + i [\alpha,A_\mu]$, where $\alpha \equiv \alpha^a \lambda^a$ and $\partial_\mu \alpha$ are both infinitesimal. We also neglect the effects of the corners of the Wilson Loop; the discontinuity in $A_\sigma$ at the corner can, for example, be avoided by using a rounded corner or particular choices of gauge. The path ordered integral over gauge fields is defined as the ordered product of the gauge links around the curve in the limit $\delta\sigma \rightarrow 0$.

$W[C_s]$ can be written in this lattice representation as
\begin{gather}
W[C_s] = \lim_{\delta \sigma \rightarrow 0} \prod_{\sigma = 0,\delta \sigma,2\delta\sigma,\ldots}^{L-\delta\sigma} U_\sigma.
\end{gather}
We wish to now replace $U_\sigma$ by an equivalent Abelian field.

We introduce a field $\theta_\sigma \equiv \theta(x(\sigma))$, which, for the moment, we shall take to be an element of U($N_C$), at each location along $C_s$ and insert the identity operator $\theta_\sigma \theta_\sigma^\dagger$ between each of the gauge links. $\theta$ is chosen so that $\theta^\dagger_\sigma U_\sigma \theta_{\sigma + \delta\sigma}$ is diagonal. There are $L/\delta \sigma$ gauge links along the path, and we introduce $L/\delta \sigma$ $\theta$ fields, so there is no obvious reason why the system cannot be solved. In fact, it is easy to construct a solution: it is easy to show (the proof is given in~\cite{Cundy:2015caa}) that $\theta_s$ contains the eigenvectors of $W[C_s]$: $W[C_s]\theta_s = \theta_s e^{i\sum_{\lambda^j \text{ diagonal}}\rho^j \lambda^j}$, for some real $\rho^j$.

 As the phases of the eigenvectors are arbitrary, this definition only determines $\theta$ up to a $(U(1))^{N_C}$ transformation $\theta \rightarrow \theta \chi$. $\chi$ makes no difference to any physical observable, but for practical purposes it is useful to select the phases and ordering of the eigenvectors by some arbitrary \textit{fixing condition} to give a unique choice of $\theta \in SU(N_C)/(U(1))^{N_C-1}$. We define $SU(N_C)/(U(1))^{N_C-1}$  by considering the following parametrisation of a $U(N)$ matrix:
 \begin{multline}
  \left(\begin{array}{cccc}
                  \cos a_1&i\sin a_1 e^{ic_1}&0&\hdots\\
                  i \sin a_1 e^{-ic_1}&\cos a_1&0&\hdots\\
                  0&0&1&\hdots\\
                  \vdots&\vdots&\vdots&\ddots
                 \end{array}\right)  \left(\begin{array}{cccc}
                  \cos a_2&0&i\sin a_2 e^{ic_2}&\hdots\\
                  0&1&0&\hdots\\
                  i \sin a_2 e^{-ic_2}&0&\cos a_2&\hdots\\
                  \vdots&\vdots&\vdots&\ddots
                 \end{array}\right)\ldots\\
                  e^{i\left(d_0+ \sum_{\lambda^j \text{diagonal}} d_j \lambda^j\right)}\label{eq:deftheta}
 \end{multline}
 $a_i$, $c_i$, $d_0$ and $d_j$ are real parameters, and there are $N_C(N_C-1)/2$ Givens matrices (i.e. one for each of the possible ways of embedding a $2\times 2$ matrix into a $N_C \times N_C$ matrix) parametrised by one particular $a$ and $c$. An $SU(N_C)/(U(1))^{N_C-1}$ matrix is parametrised in the same way, but without the final $(U(1))^{N_C}$ term (i.e. by setting $d_j$ and $d_0$ to some arbitrary fixed value, most conveniently $d_j = d_0= 0$).

Under a gauge transformation, $U_\sigma \rightarrow \Lambda_\sigma U_{\sigma}\Lambda_{\sigma + \delta\sigma}^\dagger$ for $\Lambda = e^{il^a \lambda^a}\in SU(N_C)$, $\theta \rightarrow \Lambda \theta \chi$,
where the $(U(1))^{N_C-1}$  factor $\chi$ depends on the fixing condition (in this case $\Lambda\theta \in SU(N)$ so there is no contribution to $\chi$ from $d_0$). This follows from the definition of $\theta$ as containing the eigenvectors of the Wilson Loop. The Wilson Loop transforms under a gauge transformation as $W[C_s,U] \rightarrow \Lambda_s W[C_s,U] \Lambda_s^\dagger$, so the operator which diagonalises it transforms according to $\theta_s \rightarrow \Lambda_s \theta_s$; although we also need to reselect the $U(1)^{N_C-1}$ factor so that the fixing condition remains satisfied. With $\theta^\dagger_\sigma U_\sigma \theta_{\sigma + \delta\sigma} = e^{i\sum_{\lambda^j \text{ diagonal}} \delta \sigma u^j \lambda^j}$ for real $u$,
\begin{gather}
\theta^\dagger_s W[C_s]\theta_s = e^{i \sum_{\lambda^j \text{ diagonal}} \lambda^j \oint_{C_s} d\sigma u^j_\sigma},\label{eq:evth}
\end{gather}
removing the non-Abelian structure and the path ordering without introducing an additional path integral.

Our goal is to apply Stokes' theorem to convert this line integral into a surface integral, and this requires extending the definition of $\theta$ and ${u}^j$ across the surface bounded by $C_s$. In practice, we construct these fields across all of space time. To generalise $\theta$, we construct nested curves, $C^i$, in the same plane as $C_s$ and then stack these rectangles on top of each other in the other dimensions, so that every location in Euclidean space-time is contained within one and only one curve. We then define $\theta$ so it diagonalises the gauge links (and only these gauge links) which contribute to $W[C^i,U]$.

We cannot naively extend $u^j$ across all of space-time, because its definition requires that all the gauge links $U$ are diagonalised by $\theta$, not just those that contribute to the Wilson Loop, and in general this cannot be satisfied. Instead, we replace the gauge links $U$ with a field $\hat{U}$, defined in a consistent way so that it is both diagonalised by $\theta$ across all of space time, and equal to $U$ along the path $C_s$.  The first of these conditions means that
\begin{align}
[\lambda^j,\theta_x^\dagger \hat{U}_{\mu,x}\theta_{x + \hat{\mu}\delta \sigma}] = &0, \\
\intertext{for each diagonal $\lambda^j$, which can be re-written in the form}
\hat{U}_{\mu,x} n^j_{x+\delta\sigma\hat{\mu} }\hat{U}^\dagger_{\mu,x} - n^j_x= &0&n_x^j \equiv &\theta_x \lambda^j \theta^\dagger_x.\label{eq:defeq1}
\end{align}
This condition is satisfied across all of space-time and for all directions $\mu$. Note that $n^j$ is independent of the choice of $\chi$. As we shall see later, the $\theta$-dependence of the restricted field strength $F_{\mu\nu}[\hatA]$ only appears within $n^j$, and objects contributing to the restricted field strength drive confinement. This is the justification of our earlier statement that the choice of $\chi$ does not affect the physical observable, which is the restricted field strength. To give this $\hat{U}$ field a physical meaning we need to relate it to the gauge field $U$, and we do so via a second field $\hat{X}$ defined according to
\begin{gather}
 \hat{X}_\mu(x) = U_\mu(x) \hat{U}^\dagger_\mu(x).\label{eq:9star}
\end{gather}
 For later convenience (equation (\ref{eq:9})), we restrict $\hat{X}_\mu$ by imposing the condition
\begin{align}
\tr [n^j_x(\hat{X}^\dagger_{\mu,x} - \hat{X}_{\mu,x})] = &0
\label{eq:defeq2}.
\end{align}
 Under a gauge transformation $n$ transforms as $n_x \rightarrow \Lambda_x n_x \Lambda^\dagger_x$ (which follows from the transformation rule for $\theta$) and the requirement that equations (\ref{eq:defeq1}) and (\ref{eq:defeq2}) are satisfied in every gauge leads to the transformation rules $\hat{U}_\mu(x) \rightarrow \Lambda_x \hat{U}_{\mu,x} \Lambda^\dagger_{x+\hat{\mu}\delta\sigma}$ and $\hat{X}_{\mu,x} \rightarrow \Lambda_x \hat{X}_{\mu,x} \Lambda^\dagger_x$.
Equations (\ref{eq:defeq1}) and (\ref{eq:defeq2}) are the lattice versions of the defining equations of the gauge independent CDG decomposition~\cite{Cho:1980,Cho:1981,F-N:98,Shabanov:1999,Duan:1979}, which in the continuum is described by\footnote{The original discoverers of this decomposition prefer to call it a gauge independent decomposition rather than gauge invariant decomposition. In these earlier models, the choice of $\theta$ was left arbitrary. The procedure was to fix to some arbitrary gauge, and then select some $\theta$. Since the decomposition only depends on the choice of $\theta$, and proceeds regardless of which gauge was originally chosen, the decomposition was referred to as gauge independent to distinguish it from other approaches which required fixing to particular gauges, such as the decomposition based on the Maximum Abelian Gauge. Our work differs in philosophy from the original approach because we do not need to perform any gauge fixing to extract our observables, since our key quantities (such as the field strength and Wilson Loop) are gauge covariant and thus the corresponding observables are gauge invariant. However, our particular choice of $\theta$ is gauge dependent, and thus the quantities which we use to parametrise it can only be examined after gauge fixing to some arbitrary gauge. The authors of ~\cite{Cho:1980,Cho:1981,F-N:98,Shabanov:1999,Duan:1979} gauge fixed to an arbitrary gauge and then performed an Abelian decomposition; we decompose and then (if required, which isn't the case for any physical observable) gauge fix.}
\begin{align}
A_\mu = &\hat{A}_\mu + X_\mu\label{eq:11}\\
D_\mu[\hat{A}] n^j = &0\label{eq:contde1} \\
\tr(n^j X) =&0\label{eq:contde2}\\
D_\mu[\hat{A}] \alpha \equiv& \partial_\mu \alpha - i g [\hat{A}_\mu,\alpha]\label{eq:14}\\
\hat{A}_\mu =& \sum_j\left[\frac{1}{2}n^j\tr(n^j A_\mu) + \frac{i}{4g}  [n^j,\partial_\mu n^j]\right],\label{eq:hatA1}
\end{align}
with
\begin{align}
\hat{U} \sim & e^{-i \delta\sigma g \hat{A}}& \hat{X} \sim & e^{i \delta \sigma X}.
\end{align}
Equation (\ref{eq:11}) is the naive continuum limit of equation (\ref{eq:9star}); equations (\ref{eq:contde1}) and (\ref{eq:14}) are the naive continuum limit of equation (\ref{eq:defeq1}), and equation (\ref{eq:contde2}) is the continuum limit of equation (\ref{eq:defeq2}). Proof that equation (\ref{eq:hatA1}) solves the continuum decomposition has previously been given in (for example)~\cite{Cundy:2015caa,Cho:1980,Cho:1981,Kondo:2008su}.

The condition (\ref{eq:defeq2}) may then be interpreted as requiring that $\tr(n^j \hatA) = \tr (n^jA)$, the component of the gauge field $A$ parallel to $n$ is fully contained within $\hatA$. In the continuum, the solution for $\hat{A}$ and $X$ is unique, but on the lattice we found that there were sometimes several distinct solutions to equations (\ref{eq:defeq1}) and (\ref{eq:defeq2}). In this case, we choose the solution which had the largest value of $\tr( \hat{X})$, a condition which is both gauge invariant and satisfied along $C_s$ where $\hat{U} = U$ and thus $\hat{X}=1$.

The continuum defining equations ensure that the field strength $\hat{F}_{\mu\nu}$ associated with $\hat{A}$, defined by
\begin{gather}
[D_\mu[\hat{A}],D_\nu[\hatA]] \alpha = -ig [\hat{F}_{\mu\nu}[\hat{A}],\alpha]\label{eq:f7}
\end{gather}
for any field $\alpha$ in the adjoint representation of the gauge group, satisfies $\hat{F}_{\mu\nu}[\hatA] = \beta_{\mu\nu}^j n^j$ for some real scalars $\beta^j$. The proof of this follows by substituting each of the $n^j$ fields in turn in place of $\alpha$, using equation (\ref{eq:contde1}) to show that $[\hat{F},n^j] = 0$ for all the $n^j$, and noting that the only objects which commute with each of the $n^j$ are proportional to the other $n^j$ fields.

We express the restricted field as $\hat{U}_{\mu,x} \equiv \theta_{x}e^{i \lambda^j \delta\sigma\hat{u}^j_{\mu,x}} \theta^\dagger_{x+\hat{\mu}\delta\sigma}$ for real $\hat{u}$, and since $\hat{U} = U$ along the curve $C_s$, we see that $W[C_s,U] = W[C_s,\hat{U}]=\theta_s W[C_s,\theta^\dagger \hat{U}\theta]\theta^\dagger_s = \theta_s e^{i \lambda^j \oint_{C_s} \hat{u}^j_\sigma d\sigma}\theta^\dagger_s$.
 Applying Stokes' theorem to the Abelian field $\theta^\dagger_{x} \hat{U}_{\mu,x} \theta_{x+\hat{\mu}\delta\sigma}$ gives, if $\hat{u}$ is differentiable,
\begin{align}
\theta^\dagger_s W[C_s]\theta_s =& e^{i \lambda^j \int_{x \in \Sigma} d\Sigma_{\mu\nu} \hat{F}^j_{\mu\nu}},\label{eq:7b}\\
\hat{F}^j_{\mu\nu} =& \partial_\mu \hat{u}^j_\nu - \partial_\nu \hat{u}^j_\mu,\label{eq:7}
\end{align}
where $\hat{F}^j$ is gauge invariant ($\hat{u}$ transforms as an Abelian field), $\Sigma$ the (planar) surface bound by the curve $C_s$, and $d\Sigma$ an element of area on that surface. Note that $\hat{u}$ does depend on the fixing condition, although $\hat{F}$ is independent of it. $\hat{u}^j$ can be given explicitly as~\cite{Cundy:2015caa}
\begin{gather}
\hat{u}^j_\mu = -\frac{1}{2}\tr( n^j g A_\mu  + {i} \lambda^j \theta^\dagger \partial_\mu \theta). 
\end{gather}
This expression contains two terms. The first, which we call the Maxwell term, is proportional to $\tr (n^j  A_\mu)$ and depends on both $\theta$ and the gauge field directly. The second we call the topological or $\theta$ term, is proportional to $\tr \lambda^j \theta^\dagger \partial_\mu \theta$, and is only a function of the $\theta$ field, and the gauge field indirectly through $\theta$.

We can now consider the Wilson Line around an infinitesimal plaquette $p$, which for a smooth $\hat{u}$ field gives
\begin{gather}
W[p,\hat{U}]=\hat{U}_{x,\mu} \hat{U}_{x+\delta\sigma \hat{\mu},\nu} \hat{U}^\dagger_{x+\delta\sigma\hat{\nu},\mu} \hat{U}^\dagger_{x,\nu} = e^{i \hat{F}_{\mu\nu}},
\end{gather}
which leads to
\begin{gather}
W[p,\theta^\dagger \hat{U}\theta] = e^{i \lambda^j (\partial_\mu\hat{u}^j_\nu -\partial_\nu \hat{u}^j_\mu) },
\end{gather}
and building the integral over the surface bounded by $C_s$ from the product of integrals over these small plaquettes, using that the exponent is Abelian, gives equations (\ref{eq:7b}) and (\ref{eq:7}).

Finally,  as alluded to earlier, we note that $\hat{F}_{\mu\nu}[\hatA] $ can be written in the forms
\begin{align}
\hat{F}_{\mu\nu}[\hatA] =& \frac{1}{2}n^j (\partial_\mu \tr n^j A_\nu - \partial_\nu \tr n^j A_\mu) - \frac{i}{2g} n^j\tr(n^j[\theta\partial_\mu \theta^\dagger,\theta\partial_\nu \theta^\dagger])\\
=&\frac{1}{2}n^j (\partial_\mu \tr n^j A_\nu - \partial_\nu \tr n^j A_\mu) + \frac{i}{8g} n^j\tr(n^j[\partial_\mu n^k,\partial_\nu n^k]),\label{eq:26}
\end{align}
as is proved in ~\cite{Cundy:2015caa}. These functions solely depend on $n$, and $\theta$ only indirectly through $n$, and since $n$ is independent of $\chi$ and $\hat{F}$ is the physical observable we want to study, we conclude that the choice of $\chi$ will not affect any of the physical observables we need.

Equation (\ref{eq:7b}) is only valid if $\hat{u}$ is differentiable. Equation (\ref{eq:7b}) is also similar to what we see in QED, which is, of course, not confining. In analogy to QED, we may expect the contribution of those portions of space time where $\hat{u}$ is continuous to have little contribution to the string tension.  However, we must also add to this equation the effects of discontinuities in $\hat{u}$. We do so by only extending the area integral over those areas where $\hat{u}$ is continuous, and add additional line integrals around the areas where it is discontinuous. The linear string tension will, at least in part, arise from these discontinuities. Since $\hat{u}^j$ is built from the gauge field and $\theta$, and we are working on a gauge where the gauge field is assumed to be differentiable,  we must therefore consider whether $\theta$ is differentiable.

In practice, we found that non-analyticities in $\theta$ occurred when the gauge field was not differentiable, when the Wilson Loop had non degenerate eigenvalues, and when $a_i = 0$ or $\pi/2$. It was this last part which interested us, because at these points the parameter $c$ is undefined. This means that in principle, $c_i$ can wind around these points. Since $e^{ic_i}$ must be a continuous function of position if the $\theta$ field is smooth, this means that as we transverse a closed path $c_i$ can only change by an integer multiple of $2\pi$. These factors of $2\pi$ can only emerge at the $a_i = 0$ or $\pi/2$ discontinuities. We expect the number of these discontinuities to be proportional to the area contained within the curve, since they might appear at any point in space time. When we convert from a line integral to a surface integral bound by the closed loop via Stokes' theorem, each of these objects contributes to the Abelian restricted  field strength, and thus we can also consider the Wilson Loop as the sum of these contributions. The field strength associated with these discontinuities (in four dimensional space time) is not point-like, but extends in lines of high electric or magnetic field, meaning that it is impossible to evade these objects by distorting the surface bounded by the curve. 

Another way of seeing how the winding number contributes to the path integral is to expand $\theta^\dagger \partial_\mu \theta$ in terms of the parameters $a$ and $c$. In SU(2) we find,
\begin{gather}
\theta^\dagger\partial_\sigma \theta =  i \partial_\sigma a \phi  + i \sin a \cos a \bar{\phi}\partial_\sigma c - i \sin^2 a \partial_\sigma c \lambda^3 \label{eq:dtheta},
\end{gather}
with
\begin{align}
 \phi =& \left(\begin{array}{cc}
               0&e^{ic}\\
               e^{-ic}&0
              \end{array}\right),&
\bar{\phi} =& \left(\begin{array}{cc}
               0&ie^{ic}\\
               -ie^{-ic}&0
              \end{array}\right),& \text{ and }
\lambda^3 = & \left(\begin{array}{cc}1&0\\0&-1\end{array}\right) .             
\end{align}

This $\sin^2 a\partial_\sigma c$ term is what we believe generates quark confinement. When we integrate this around the curve, it will contain a term proportional to the winding number $\nu_c$. For example, writing $\sin^2a = \frac{1}{2}(1-\cos 2a)$ we find that this term contributes $\oint dx_\mu \frac{1}{2}(1-\cos2a)\partial_\mu c = \pi \nu_c - \frac{1}{2}\oint dx_\mu \cos2a\;\partial_\mu c$ to the Wilson Loop. The first term is proportional to the winding number; the second is harder to analyse: if the distributions of $a$ and $c$ around the loop were independent of each other distributed around the loop, then on average this second term would give $\pi \nu_c \langle \cos2a \rangle$, also proportional to the winding number; but in practice this analysis is likely to be too naive, since the winding is generated at points where $\cos 2a = \pm 1$. 

Suppose that $\hat{u}^j$ contains a non-analyticity. We integrate the field around a loop $\tilde{C}$ parametrised by $\tilde{\sigma}$ surrounding the discontinuity, bounding the surface integral by an additional line integral
$
\oint_{\tilde{C}} d\tilde{\sigma} \hat{u}^j_{\tilde{\sigma}},
$
far enough away from the discontinuity that $\hat{u}^j_{\tilde{\sigma}}$ is analytic along $\tilde{C}$.
We define $\{\tilde{C}_n\}$ as the set of curves surrounding all these discontinuities, and $\tilde{\Sigma}$ the area bound within these curves.
We can write
\begin{gather}
e^{i \lambda^j\delta\tilde{\sigma} \hat{u}^j_{\mu,x}} = \theta^\dagger_x \hat{X}^\dagger_{\mu,x} \theta_x \theta^\dagger_x U_{\mu,x} \theta_{x + \delta\tilde{\sigma}},\label{eq:str1}
\end{gather}
and since $\hat{u}$ is continuous on $\tilde{C}$, after fixing the gauge we can expand $U = 1 -i\frac{1}{2}g\delta\tilde{\sigma} A^a \lambda^a$ and $\theta^\dagger_x \theta_{x+\delta\tilde{\sigma}} = 1 + \delta\tilde{\sigma} \theta^\dagger \partial_{\tilde\sigma} \theta$. We define $X_0 \equiv \frac{1}{2}\theta^\dagger(X + X^\dagger)\theta$.  For smooth fields, we expect $X_0 = I + O(\delta \sigma^2)$, where $I$ is the identity operator.

This gives
\begin{multline}
i\delta\tilde{\sigma} \hat{u}^j_{\mu,x} = \frac{1}{\tr (\lambda^j)^2}\text{Im}\left(\;\tr \left[\lambda^j\theta^\dagger_x \hat{X}^\dagger_{\mu,x} \theta_x \theta_x^\dagger U_{\mu,x}\theta_{x+\delta\tilde{\sigma} \hat{\mu}}\right]\right)=\\
\frac{1}{2\tr (\lambda^j)^2}\tr[\lambda^j \theta^\dagger_x (\hat{X}^\dagger_{\mu,x} - \hat{X}_{\mu,x}) \theta_x - \frac{1}{2}i \lambda^j \delta \tilde{\sigma}X_{0\mu,x}\theta^\dagger_x  gA^a_{\mu,x} \lambda^a\theta_x + \lambda^j X_{0\mu,x} \delta \tilde{\sigma}\theta_x^\dagger\partial_{\tilde{\sigma}} \theta]\label{eq:9}.
\end{multline}

Using (\ref{eq:defeq2}) the first term in equation (\ref{eq:9}) gives zero, while the second term will not contribute to an integral around $\tilde{C}$ if $U$ and $X_0$ are continuous and the area of the loop is small enough. We therefore concentrate on the contribution from the final term. Equation (\ref{eq:7}) is then replaced by
\begin{gather}
\theta^\dagger_s W[C_s]\theta_s = e^{i  \lambda^j\left[ \int_{(x \in \Sigma) \cap (x\not{\in} \tilde{\Sigma})} d\Sigma_{\mu\nu} \hat{F}^j_{\mu\nu} + \sum_n\oint_{\tilde{C}_n} d\tilde{\sigma}\frac{1}{\tr (\lambda^j)^2} \tr [\lambda^j X_0  \theta^\dagger \partial_{\tilde{\sigma}}\theta]\right]},
\end{gather}
where $d\Sigma$ is an element of area.

\section{Gauge transformations}\label{sec:3}

A gauge transformation can be parametrised as
\begin{gather}
U_{\mu}(x) \rightarrow \Lambda_x U_\mu(x) \Lambda^\dagger_{x + \epsilon \hat{\mu}},
\end{gather}
where $\Lambda$ is an element of SU($N$). For the infinitesimal transformation, $\Lambda \sim e^{i l_a \lambda^a}$ this corresponds to the usual gauge transformation rule $gA_\mu \rightarrow gA_\mu + \partial_\mu l - i[l,gA_\mu]$ with $l = l_a\lambda^a$ and $U_\mu(x) \sim e^{-igA_\mu(x)\delta x}$. We can also write (if the lattice spacing is small enough and the fields sufficiently smooth) $gA_\mu \rightarrow g\Lambda_x A_\mu \Lambda^\dagger_x +i \Lambda\partial_\mu \Lambda^\dagger$.

The definition of $\theta$ as the eigenvectors of the Wilson Loop projected into $SU(N)/U(1)^{(N-1)}$ means that under a gauge transformation $\theta$ transforms as
\begin{gather}
\theta \rightarrow \Lambda \theta e^{i \delta_j \lambda_j},
\end{gather}
where $\delta_j$ are chosen to re-project $\theta$ into the required format. This means that $\hat{u}$ transforms as
\begin{align}
-2\hat{u}^j_\mu = &\tr( n^j g A_\mu  + {i} \lambda^j \theta^\dagger \partial_\mu \theta)
\nonumber\\
\rightarrow&\tr( \Lambda n^j \Lambda^\dagger (g  \Lambda A_\mu \Lambda^\dagger  +i \Lambda\partial_\mu \Lambda^\dagger) + {i} \lambda^j ( e^{-i\delta_k\lambda^k} \theta^\dagger \Lambda^\dagger \partial_\mu (\Lambda \theta e^{i\delta_k\lambda^k})))\nonumber\\
= & \tr( n^j  (g   A_\mu   +i \partial_\mu (\Lambda^\dagger) \Lambda) + {i} \lambda^j (\theta^\dagger \partial_\mu \theta + i\lambda^k \partial_\mu \delta_k+ \theta^\dagger \Lambda^\dagger \partial_\mu (\Lambda) \theta))
\nonumber\\
=& \tr (n^j  (g   A_\mu) + i \lambda^j \theta^\dagger \partial_\mu \theta) - 2 \partial_\mu \delta_j
\end{align}
Thus we have $\hat{u}^j_\mu \rightarrow \hat{u}^j_\mu + \partial_\mu \delta_j$, as we would expect for an Abelian field.

Note, however, that neither the Maxwell term contribution nor the topological contribution to $\hat{u}$ (and thus $\hat{A}$) are by themselves gauge invariant. Indeed, the Maxwell term transforms to
\begin{gather}
\tr( n^j  (g   A_\mu)) \rightarrow \tr( n^j  (g   A_\mu   +i \partial_\mu (\Lambda^\dagger) \Lambda))
\end{gather}
while the topological term transforms to
\begin{gather}
\tr ({i} \lambda^j (\theta^\dagger \partial_\mu \theta))\rightarrow \tr ({i} \lambda^j (\theta^\dagger \partial_\mu \theta + \theta^\dagger \Lambda^\dagger \partial_\mu (\Lambda) \theta)).
\end{gather}

\section{Re-parametrisation of the Yang-Mills action.}\label{sec:4}
We seek to explicitly provide a path integral formulation of this decomposition. The first step is to re-write the path integral in terms of the variables $a$, $c$ and $d$ used to parametrise the Abelian decomposition, plus the remaining variables required to describe the other gauge fields. We begin by writing the lattice gauge links as
\begin{gather}
U_{\mu,x} = \tilde{\theta}_{\mu,x} e^{-i \epsilon \partial_\mu \tilde{d}^j_\mu\lambda_j} \tilde{\theta}^\dagger_{\mu,x + \epsilon \hat{\mu}},
\end{gather}
where (in SU(2)),
\begin{gather}
\tilde{\theta}_{\mu,x} = \left(\begin{array}{cc} 
\cos \tilde{a}_\mu& i \sin \tilde{a}_\mu e^{i\tilde{c}_\mu}\\
i \sin \tilde{a}_\mu e^{-i\tilde{c}_\mu}&\cos \tilde{a}_\mu
\end{array}\right)
\end{gather}
The corresponding object in higher gauge groups can be constructed in analogy to the prescription in equation (\ref{eq:deftheta}). There are $(N^2 - N)/2$ $\tilde{a}_\mu$ and $\tilde{c}_\mu$ parameters per $\tilde{\theta}$ field, and $N-1$ $\tilde{d}^j_\mu$ parameters, so the total number of variables described by this theory is the expected $N^2 - 1$ per lattice site per direction (we have not yet gauge fixed). Note that, despite the notation, $\tilde{c}_\mu$, $\tilde{a}_\mu$ and $\tilde{d}^j_\mu$ are not four vectors since they do not transform canonically under Lorentz transformations.

We can find the measure for the path integral through the requirement that it should be gauge invariant. This is the same requirement that leads to the standard Haar measure in lattice gauge theory.

We can easily express the variables $a$ and $c$ which parametrise the Abelian decomposition in terms of these parameters. The easiest example is when we are investigating the Polyakov Loop. Here we just identify $\tilde{\theta}_{\hat{t},x}\equiv \theta_x$, where $\theta_x$ is the Abelian decomposition $\theta$ matrix, and the first index in $\tilde{\theta}$ indicates the direction and the second the location. This implies that  $\hat{X}_{\hat{t}} = 1$, and the defining equations of the Abelian decomposition are then automatically satisfied in that direction. For the Wilson Loop, we need to switch from a Cartesian coordinate system. Instead, when considering Wilson Loops in the $xt$ plane, we use the direction index $\mu = 0$ to indicate those links that lie along the various nested Wilson Loops, and $\mu = 1$ to indicate the other gauge links in the $xt$ plane. $\mu = 2$ and $\mu = 3$ then represent the Cartesian $y$ and $z$ directions respectively as usual. In this case we can identify $\theta_{0,x} \equiv \theta_x$, and once again this allows us to write the parameters in the Abelian decomposition as dynamical variables.

For most of the following, we will use the Polyakov Loop, i.e. straight Cartesian, representation of the gauge links, as this is easier and we do not have to worry about effects from the corners of the Wilson Loop. This choice does not affect the computation of the measure, but does influence our expression for the field strength tensor.

\subsection{SU(2)}
We write $\hat{\theta}_{\mu,x}$ as
\begin{align}
\hat{\theta}_{\mu,x} = &\left(\begin{array}{cc}
\cos \tilde{a}_{\mu,x}& i \sin \tilde{a}_{\mu,x} e^{i\tilde{c}_{\mu,x}}\\
i \sin \tilde{a}_{\mu,x} e^{-i\tilde{c}_{\mu,x}}&\cos \tilde{a}_{\mu,x}
\end{array}\right) \left(\begin{array}{cc} e^{i \tilde{d}_{\mu,x}}&0\\0&e^{-i\tilde{d}_{\mu,x}}\end{array}\right)\nonumber\\
=&\tilde{\theta}_{\mu,x} e^{i \tilde{d}_{\mu,x} \lambda_3}
\end{align}
and
\begin{align}
 \tilde{\phi} = &\left(\begin{array}{cc}0& e^{i\tilde{c}_{\mu,x}}\\ e^{-i\tilde{c}_{\mu,x}}&0\end{array}\right) &\tilde{\bar{\phi}} = & \left(\begin{array}{cc}0& ie^{i\tilde{c}_{\mu,x}}\\ -ie^{-i\tilde{c}_{\mu,x}}&0\end{array}\right). 
\end{align}
We then parametrise an infinitesimal gauge transformation as
\begin{gather}
 \Lambda = e^{il_3\lambda_3} e^{il_2\lambda_2}e^{il_1 \lambda_1} = e^{i\Lambda_d \lambda_3}e^{i\Lambda_b \tilde{\bar{\phi}}} e^{i\Lambda_a \tilde{\phi}},
\end{gather}
with $l_3$, $l_1$, $l_2$ and $\Lambda_a$, $\Lambda_b$ and $\Lambda_d$ infinitesimal. The parametrisation in terms of $\Lambda_x$ is easier to work with, though obviously these $\Lambda_x$ are going to be functions of $a,c,d$ so we will have to convert back to the $l_x$ formulation of the gauge transformation at the end of the calculation. In fact, we find (up to terms of $O(l^2)$ or $O(\Lambda^2)$)
\begin{align}
 l_3 =& \Lambda_d\nonumber\\
 l_1 = &\Lambda_a \cos \tilde{c}_{\mu,x} - \Lambda_b \sin \tilde{c}_{\mu,x}& \Lambda_a = & l_1 \cos \tilde{c}_{\mu,x} + l_2 \sin \tilde{c}_{\mu,x}\nonumber\\
 l_2 = &\Lambda_a \sin \tilde{c}_{\mu,x} + \Lambda_b \cos \tilde{c}_{\mu,x}& \Lambda_b = & -l_1 \sin \tilde{c}_{\mu,x} + l_2\cos \tilde{c}_{\mu,x}.
\end{align}
Applying this transformation to $\hat{\theta}$ gives (neglecting terms of $O(l^2)$ throughout this calculation)
\begin{align}
 \Lambda \hat{\theta} = & e^{i \lambda_3 l_3} \left(\begin{array}{cc}1&-\Lambda_b e^{i\tilde{c}_{\mu,x}}\\\Lambda_b e^{-i\tilde{c}_{\mu,x}}&1\end{array}\right)\left(\begin{array}{cc}\cos(\tilde{a}_{\mu,x}+\Lambda_a)&i\sin(\tilde{a}_{\mu,x}+\Lambda_a) e^{i\tilde{c}_{\mu,x}}\\i\sin(\tilde{a}_{\mu,x}+\Lambda_a) e^{-i\tilde{c}_{\mu,x}}&\cos(\tilde{a}_{\mu,x}+\Lambda_a)\end{array}\right)e^{i \tilde{d}_{\mu,x}\lambda_3}\nonumber\\
 =& e^{i \lambda_3 l_3}\left(\begin{array}{cc}\cos(\tilde{a}_{\mu,x}+\Lambda_a) - i \sin \Lambda_b \sin(\tilde{a}_{\mu,x}+\Lambda_a)&i(\sin(\tilde{a}_{\mu,x}+\Lambda_a)+i\Lambda_b\cos(\tilde{a}_{\mu,x}+\Lambda_a)) e^{i\tilde{c}_{\mu,x}}\\i(\sin(\tilde{a}_{\mu,x}+\Lambda_a)-i\Lambda_b\cos(\tilde{a}_{\mu,x}+\Lambda_a)) e^{-i\tilde{c}_{\mu,x}}&\cos(\tilde{a}_{\mu,x}+\Lambda_a)+i \sin \Lambda_b \sin(\tilde{a}_{\mu,x}+\Lambda_a)\end{array}\right)e^{i \tilde{d}_{\mu,x}\lambda_3}.
\end{align}
We can read off,
\begin{align}
 \tilde{a}'_{\mu,x} =& \tilde{a}_{\mu,x} + \Lambda_a = \tilde{a}_{\mu,x} + l_1 \cos \tilde{c}_{\mu,x} + l_2 \sin \tilde{c}_{\mu,x}\nonumber\\
 \tilde{c}'_{\mu,x} =& \tilde{c}_{\mu,x} + \tilde{d}'_{\mu,x} -\tilde{d}_{\mu,x}+ \Lambda_b \cot \tilde{a}_{\mu,x} = \tilde{c}_{\mu,x}+l_3 + (l_1 \sin \tilde{c}_{\mu,x} - l_2 \cos \tilde{c}_{\mu,x}) (\tan \tilde{a}_{\mu,x} - \cot \tilde{a}_{\mu,x}).\nonumber\\
  \tilde{d}'_{\mu,x} =& d_{\mu,x}+l_3 - \Lambda_b \tan \tilde{a}_{\mu,x} = \tilde{d}_{\mu,x}+l_3 + (l_1 \sin \tilde{c}_{\mu,x} - l_2 \cos \tilde{c}_{\mu,x})\tan \tilde{a}_{\mu,x}\nonumber\\
  \epsilon\hat{u}_{\mu,x}' = & \epsilon\hat{u}_{\mu,x} + \tilde{d}'_{\mu,x} - \tilde{d}'_{\mu,x+\epsilon \hat{\mu}}
\end{align}
The Jacobian from the primed to original coordinates thus reads  (again neglecting terms of O($l^2$)) 
\begin{align}
J = &
\left|\begin{array}{ccc}
        \frac{\partial \tilde{a}'_{\mu,x}}{\partial \tilde{a}_{\mu,x}}& \frac{\partial \tilde{a}'_{\mu,x}}{\partial \tilde{d}_{\mu,x}}&\frac{\partial \tilde{a}'_{\mu,x}}{\partial \tilde{c}_{\mu,x}}\\
       \frac{\partial \tilde{d}'_{\mu,x}}{\partial \tilde{a}_{\mu,x}}& \frac{\partial \tilde{d}'_{\mu,x}}{\partial \tilde{d}_{\mu,x}}&\frac{\partial \tilde{d}'_{\mu,x}}{\partial \tilde{c}_{\mu,x}}\\
       \frac{\partial \tilde{c}'_{\mu,x}}{\partial \tilde{a}_{\mu,x}}& \frac{\partial \tilde{c}'_{\mu,x}}{\partial \tilde{d}_{\mu,x}}&\frac{\partial \tilde{c}'_{\mu,x}}{\partial \tilde{c}_{\mu,x}}
          \end{array}
 \right|
\nonumber\\
=&\left|\begin{array}{ccc}
           1&0&-l_1 \sin \tilde{c}_{\mu,x} + l_2 \cos \tilde{c}_{\mu,x}\\
           \frac{ l_1 \sin \tilde{c}_{\mu,x}-l_2 \cos \tilde{c}_{\mu,x}}{\cos^2 \tilde{a}_{\mu,x}}&1&\tan \tilde{a}_{\mu,x} (l_1 \cos \tilde{c}_{\mu,x} + l_2 \sin \tilde{c}_{\mu,x})\\
           4\left(\frac{l_1 \sin \tilde{c}_{\mu,x}-l_2 \cos \tilde{c}_{\mu,x} }{\sin^22 \tilde{a}_{\mu,x}} \right)&0&1+(\tan \tilde{a}_{\mu,x}- \cot \tilde{a}_{\mu,x}) (l_1 \cos \tilde{c}_{\mu,x} + l_2 \sin \tilde{c}_{\mu,x})
          \end{array}
 \right|\nonumber\\
 =& 1 - 2\frac{\cos 2\tilde{a}_{\mu,x}}{\sin 2\tilde{a}_{\mu,x}} \delta \tilde{a}_{\mu,x}\nonumber\\
 =& 1 - \frac{\partial \log (\sin 2\tilde{a}_{\mu,x})}{\partial \tilde{a}_{\mu,x}} \delta \tilde{a}_{\mu,x},
\end{align}
where $\delta \tilde{a}_{\mu,x} = \tilde{a}'_{\mu,x}-\tilde{a}_{\mu,x}$.

The Haar measure $\mu(\tilde{a}_{\mu,x},\tilde{c}_{\mu,x},\tilde{d}_{\mu,x})$ is (up to a multiplicative constant) that function which satisfies $\mu(\tilde{a}_{\mu,x},\tilde{c}_{\mu,x},\tilde{d}_{\mu,x})d\tilde{a}_{\mu,x}d\tilde{c}_{\mu,x}d\tilde{d}_{\mu,x} = \mu(\tilde{a}_{\mu,x}',\tilde{c}_{\mu,x}',\tilde{d}_{\mu,x}') d\tilde{a}_{\mu,x}' d\tilde{c}_{\mu,x}' d\tilde{d}_{\mu,x}'$, so that the measure is gauge invariant. From this, we see that
\begin{multline}
 \mu(\tilde{a}_{\mu,x},\tilde{c}_{\mu,x},\tilde{d}_{\mu,x}) = \mu(\tilde{a}_{\mu,x},\tilde{c}_{\mu,x},\tilde{d}_{\mu,x}) (1+ \frac{\partial\log\mu}{\partial \tilde{a}_{\mu,x}} \delta \tilde{a}_{\mu,x} + \frac{\partial\log\mu}{\partial \tilde{c}_{\mu,x}} \delta \tilde{c}_{\mu,x} + \frac{\partial\log\mu}{\partial \tilde{d}_{\mu,x}} \delta \tilde{d}_{\mu,x} ) \\(1-\frac{\partial \log (\sin 2\tilde{a}_{\mu,x})}{\partial \tilde{a}_{\mu,x}} \delta \tilde{a}_{\mu,x}),
\end{multline}
from which we can see that $\mu \propto \sin 2\tilde{a}_{\mu,x}$. This means that the fields at $\tilde{a}_{\mu,x} = 0$ and $\tilde{a}_{\mu,x} = \pi/2$, which drive the topological solutions, are suppressed by the measure. This makes sense, because it allows for a phase transition: if the action (a function of temperature and in full QCD chemical potential, magnetic field, etc.) encourages these solutions, and the measure suppresses them, then there might well be some temperature (and chemical potential and magnetic field etc.) which changes from the suppression being more significant to the encouragement being most important, leading to a confinement or de-confinement transition. We can construct a lattice theory by taking the continuum theory and super-imposing a lattice on top of it. Thus the value of $\tilde{a}_{\mu,x}$ on a particular lattice sites could be the continuum value at that particular location. On such a lattice theory, we will not get exactly $\tilde{a}_{\mu,x} = 0$ or $\tilde{a}_{\mu,x} = \pi/2$, but merely something close to it the lattice site surrounding the point in the continuum theory where $\tilde{a}_{\mu} = 0$. Suppose that we have a lattice spacing $\epsilon$, and a point around which there is winding ($\tilde{a}_\mu$ = 0) somewhere on the continuum: this will fall inside a $1\times 1$ square of gauge links on the lattice. Thus to have winding, we need in practice to have four neighbouring gauge links of the order of $\partial^2_\nu \tilde{a}_\mu \epsilon^2$, since we can expect $\partial_\nu \tilde{a}_\mu = 0$ near the minimum or maximum value of $a_\mu$ if we choose a continuous $\theta$. Ignoring the gauge action for simplicity, the measure term tells us that the number of lattice sites with this or smaller $\tilde{a}_\nu$ is $1-\cos 2\tilde{a}_\mu \sim (\partial^2_\nu \tilde{a}_\mu \epsilon^2)^2$ multiplied by the number of lattice sites. The number of lattice sites in a four dimensional box of fixed physical volume surrounding the singularity scales with $1/\epsilon^4$, meaning that the number of unit hyper-cubes which contain the point at $\tilde{a}_\mu = 0$ will remain roughly constant as $\epsilon \rightarrow 0$; the measure suppresses them to the degree that is necessary to ensure that their number remains stable as the lattice spacing decreases.    

Thus the path integral can be written as
\begin{gather}
\int d[\cos 2a_{\mu}] d[c_{\mu}] d[d_{\mu}] e^{-\beta \sum_x F_{\mu\nu}^2 [\tilde{a}_{\mu}(x),\tilde{c}_{\mu}(x),\tilde{d}_{\mu}(x)] },
\end{gather}
where $\beta$ is inversely proportional to the square of the gauge coupling, $d[f]$ represents the functional integral over $f$ and $F_{\mu\nu}^2$ represents the Yang-Mills action. $\cos 2\tilde{a}_\mu$ is bound between $-1$ and $1$, $\tilde{c}_\mu$ and $\tilde{d}_\mu$ can extend over the entire real axis.

\subsection{SU(3)}
In SU(3), the measure is given by
\begin{gather}
 \mu = \sin 2 \tilde{a}_{1\mu,x} \sin 2\tilde{a}_{2\mu,x} \sin 2 \tilde{a}_{3\mu,x} \cos^2 \tilde{a}_{2\mu,x}.
\end{gather}
The proof is given in appendix \ref{app:su3measure}

\subsection{Numerical Check of the Measure}
We can check that these measures are correct by comparing Monte-Carlo simulations of the standard parametrisation of lattice QCD and the new one using $\tilde{a}$, $\tilde{c}$ and $\tilde{d}$ as dynamical variables. This new formulation of QCD turned out to be considerably slower to compute, because of the need to continually convert from the gauge links (which we need for the standard Luscher-Weisz action) and the new parameters. We used a HMC algorithm~\cite{HMC} to generate the gauge fields, directly adding the measure term (in SU(2) $\tr \log \sin(2\tilde{a}_{\mu,x})$) to the action. Hybrid Monte-Carlo is a combination of a molecular dynamics evolution of the fields, introducing a conjugate momentum and then integrating over a classical trajectory (with the QCD action playing the role of the classical potential energy), which is used to generate a new trail configuration. A Metropolis accept/reject step is then applied, which ensures that the configurations will be distributed according to the correct distribution. $\tilde{a}_{\mu,x}$ is bound between $0 \le \tilde{a}_{\mu,x} \le \pi/2$ so $\sin(2\tilde{a}_{\mu,x})$ cannot be negative, and will only be zero at the points which should be forbidden by the action. However, when  $\tilde{a} \sim 0$ or $\tilde{a} \sim \pi/2$, $\log \sin 2\tilde{a}$  can become large, which caused instabilities in the HMC molecular dynamics evolution. To resolve this, we modified the action for the molecular dynamics slightly to  $\tr \log (\sin(2\tilde{a}_{\mu,x}) + \zeta)$ where $\zeta$ is some small tunable parameter. The Metropolis step, however, used the correct action (which ensures that the whole algorithm generates the desired distribution). Too small a choice of $\zeta$ means that the molecular dynamics becomes unstable, and the whole Monte Carlo breaks down. Too large a choice of $\zeta$ would mean that the molecular dynamics is no longer close to shadowing the true distribution, and the metropolis algorithm would reject most of the trail configurations. The question is then if there is a value of $\zeta$ which lies between these two bounds. We did not encounter any difficulties with stability simulating on lattice sizes of up to $16^3\times 32$ in both SU(2) and SU(3); it is possible that difficulties might arise on larger lattice volumes. However, our implementation of this algorithm required considerably more computer time than the standard lattice Monte-Carlo methods, so for the configurations used in section \ref{sec:6} we used the standard lattice QCD Monte-Carlo to generate the configurations, with the modified algorithm using the new dynamical variables only used as a consistency check. 

\begin{figure}
\begin{center}
\includegraphics[width=10cm]{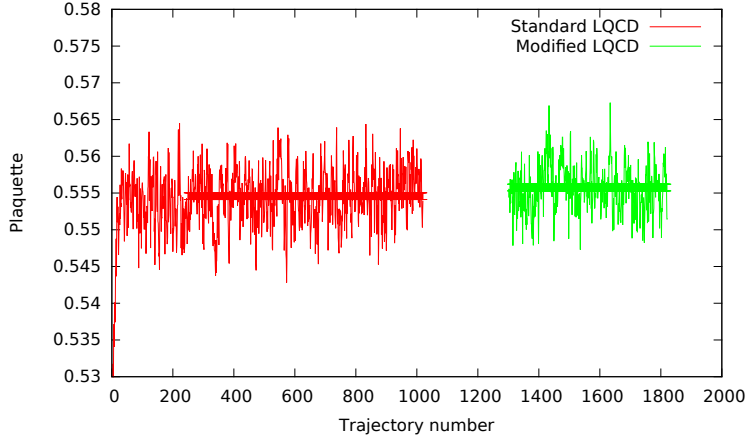}
\end{center}
\caption{The average plaquette plotted against trajectory number for a standard Lattice QCD HMC simulation of pure Yang-Mills field theory, and with the modified parametrisation. Trajectory numbers 1-1100 were computed with the standard formulation of lattice QCD. Trajectory numbers above 1300 were computed with the new formulation of lattice QCD. The thick lines represent the mean value of the plaquette and the $1\sigma$ error bars. The first 250 trajectories were excluded from the average because the ensemble was still thermalising. The mean value of the plaquette for the standard HMC run is $0.5547(4)$, for the modified HMC $0.5560(5)$, a discrepancy of less than $2\sigma$.}\label{fig:1}
\end{figure}

Having generated ensembles with both the standard and this new formulation of lattice QCD, we can test various observables. We found a good agreement on all those we tested, although we were restricted to small lattices for these tests so that we could generate enough configurations with the new parametrisation. Figure \ref{fig:1} shows the average plaquette $U_{\mu\nu}$ (the path ordered product of gauge links around a $1\times1$ rectangle; the trace of the plaqutte is a representation of the Yang-Mills action $F_{\mu\nu}^2$ plus a constant) across our molecular dynamics trajectories for both a standard lattice simulation and one built from these new variables. We show data from an SU(2) ensemble, but the situation is similar in SU(3). We observe that there is no noticeable difference either in the average plaquette or the variation from the average. The ensemble used a Tadpole Improved Luscher Wesiz gauge action~\cite{TILW} with $\beta = 3.0$ and a lattice size of $6^3\times 12$.

\subsection{Abelian Decomposition}

As discussed earlier, we identify $\tilde{\theta}_{\mu,x}$ with $\theta_x$ in one direction, and in this direction the Abelian decomposition is straight forward. In the remaining directions 
\begin{align}
U_{\mu,x} =& \tilde{\theta}_{\mu,x} e^{-i \epsilon \partial_\mu\tilde{d}^j\lambda_j} \tilde{\theta}^{\dagger}_{\mu,x+\epsilon \hat{\mu}} \nonumber\\
=& \tilde{\theta}_{\mu,x} e^{i \delta_{\mu,x}^j\lambda_j} \theta_x^\dagger \theta_x e^{-i\delta_{\mu,x}^j\lambda_j} e^{i \epsilon\partial_\mu\tilde{d}^j\lambda_j} e^{i\delta_{\mu,x + \epsilon \hat{\mu}}^j\lambda_j} \theta^\dagger_{x+\epsilon \hat{\mu}} \theta_{x+\epsilon \hat{\mu}} e^{-i\delta_{\mu,x + \epsilon \hat{\mu}}^j\lambda_j}\tilde{\theta}^{\dagger}_{\mu,x+\epsilon \hat{\mu}}. 
\end{align}
$\delta$ is a parameter which we will have to determine to ensure that the defining equations of the Abelian decomposition are satisfied. We write
\begin{align}
\hat{U}_{\mu,x} = & \theta_x  e^{i \epsilon \partial_\mu(-\tilde{d}_{\mu,x}^j + \delta^j_{\mu,x})\lambda_j}  \theta^\dagger_{x+\epsilon \hat{\mu}}\nonumber\\
Y_{\mu,x} =& \tilde{\theta}_{\mu,x} e^{i \delta_{\mu,x}^j\lambda_j} \theta_x^\dagger. 
\end{align}
In this case,
\begin{align}
U_{\mu,x} = Y_{\mu,x} \hat{U}_{\mu,x} Y^\dagger_{\mu,x+\epsilon \hat{\mu}}.
\end{align}
This isn't quite the form that we need, so we write $(Y'_{\mu,x})^\dagger \hat{U}_{\mu,x} = \hat{U}_{\mu,x} (Y_{\mu + \hat{\mu}\epsilon,x})^\dagger$, which means that our $\hat{X}$ field is
\begin{gather}
\hat{X}_{\mu,x} = Y_{\mu,x} {Y'_{\mu,x}}^\dagger 
\end{gather}
In SU(2), we parametrise $\theta_x$ as,
\begin{gather}
 \theta = \left(\begin{array} {cc} \cos a& i \sin a e^{ic}\\
                 i \sin a e^{-ic}& \cos a
                \end{array}\right).
\end{gather}
 In this case, we have,
\begin{align}
 Y_{\mu,x} = & \left(\begin{array}{c c} 
                      \cos \tilde{a} \cos a e^{i\delta} + \sin \tilde{a} \sin a e^{i (\tilde{c} - c - \delta)} & 
                      i \sin \tilde{a} \cos a e^{i(\tilde{c}-\delta)} - i \cos \tilde{a} \sin a e^{i (\delta + c)}\\
                      i \sin \tilde{a} \cos a e^{-i(\tilde{c}-\delta)} - i \cos \tilde{a} \sin a e^{-i (\delta + c)}&
                      \cos \tilde{a} \cos a e^{-i\delta} + \sin \tilde{a} \sin a e^{-i (\tilde{c} - c - \delta)} 
                     \end{array}\right)
                     \nonumber\\        
            =& \cos \tilde{a} \cos a \cos \delta + \sin \tilde{a} \sin a \cos (\tilde{c}-c-\delta) + i \lambda_3(\cos \tilde{a} \cos a \sin \delta + \sin \tilde{a} \sin a \sin (\tilde{c}-c-\delta))\nonumber\\&
            + i \lambda_1(\sin \tilde{a} \cos a \cos (\tilde{c}-\delta) - \cos \tilde{a} \sin a \cos(\delta + c))\nonumber\\&
            + i \lambda_2(\sin \tilde{a} \cos a \sin (\tilde{c}-\delta) - \cos \tilde{a} \sin a \sin(\delta + c))
            \nonumber\\
            =&Y^0_{\mu,x} + i \lambda_1 Y^1_{\mu,x} + i \lambda_2 Y^2_{\mu,x} +i \lambda_3 Y^3_{\mu,x} \nonumber\\
\hat{U}_{\mu.x} = &\left(\begin{array}{c c}
                      1+i\epsilon \partial_\mu \chi \cos 2a - i \epsilon \partial_\mu c \sin^2 a&
                      -i\epsilon \partial_\mu a e^{ic} +  \sin 2a e^{i c} (\partial_\mu \chi - \frac{1}{2}\partial_\mu c)\nonumber\\
                      -i\epsilon \partial_\mu a e^{-i c} - \sin 2a e^{-i c} (\partial_\mu \chi - \frac{1}{2}\partial_\mu c) &
                      1-i\epsilon \partial_\mu \chi \cos 2a - i \epsilon \partial_\mu c \sin^2 a
                     \end{array}\right)
                     \nonumber\\
                 = & 1 + i \epsilon \lambda_3 (\partial_\mu \chi \cos 2a - \partial_\mu c \sin^2 a) + i \epsilon \lambda_1 (-\cos c \partial_\mu a + \sin c \sin 2a (\partial_\mu \chi - \frac{\partial_\mu c}{2}))\nonumber\\&
                 + i \epsilon \lambda_2 (\sin c \partial_\mu a + \cos c  \sin 2a (\partial_\mu \chi - \frac{\partial_\mu c}{2}))
                 \nonumber\\
            =&1 + i \epsilon \lambda_1 \hat{U}^1_{\mu,x} + i \epsilon \lambda_2 \hat{U}^2_{\mu,x}+  i\epsilon \lambda_3 \hat{U}^3_{\mu,x}
            \nonumber\\
Y'_{\mu,x} = & Y_{\mu,x+\hat{\mu}} + 2 i \epsilon\lambda_3 (Y^2_{\mu,x+\hat{\mu}} \hat{U}_{\mu,x}^1) + 2 i \epsilon \lambda_1(Y^3_{\mu,x+\hat{\mu}} \hat{U}_{\mu,x}^2) + 2 i \epsilon \lambda_2(Y^1_{\mu,x+\hat{\mu}} \hat{U}_{\mu,x}^3)\nonumber\\
X_{\mu,x} = &1 + i \epsilon \lambda_3(\partial_\mu Y_1 Y_2 - \partial_\mu Y_2 Y_1 + \partial_\mu Y_0 Y_3 - \partial_\mu Y_3 Y_0)\nonumber\\&
               + i \epsilon \lambda_1(\partial_\mu Y_2 Y_3 - \partial_\mu Y_3 Y_2 + \partial_\mu Y_0 Y_1 - \partial_\mu Y_1 Y_0)\nonumber\\&
               + i \epsilon \lambda_2(\partial_\mu Y_3 Y_1 - \partial_\mu Y_1 Y_3 + \partial_\mu Y_0 Y_2 - \partial_\mu Y_2 Y_0)\nonumber\\&
               + 2i \epsilon \lambda_1 (Y_0 Y_3 \hat{U}^2 + \hat{U}^1 (Y^2)^2 - \hat{U}^1 (Y^3)^2)\nonumber\\&
               + 2i \epsilon \lambda_2 (Y_0 Y_1 \hat{U}^3 + \hat{U}^2 (Y^3)^2 - \hat{U}^2 (Y^1)^2)\nonumber\\&
               + 2i \epsilon \lambda_3 (Y_0 Y_2 \hat{U}^1 + \hat{U}^3 (Y^1)^2 - \hat{U}^3 (Y^2)^2)\nonumber\\&
               -2\epsilon (Y^1 Y^3\hat{U}^2 + Y^2 Y^3\hat{U}^1 + Y^1 Y^2\hat{U}^3) 
\end{align}
We have here assumed that we are in a gauge such that $\theta$ is a smooth function, so that $a_{x+\hat{\mu}\epsilon} - a_{x} \sim \epsilon \partial_\mu a$, $c_{x + \epsilon \hat{\mu}} - c_x \sim \epsilon \partial_\mu c$, and similarly for $\tilde{a}$ and $\tilde{c}$. We have also written $\tilde{d}^j - \delta^j = \chi^j$, and since for SU(2) there is only a single $j$ index, we have suppressed it.

To find $\delta$, we need to use the condition
\begin{gather}
 0 = \tr \theta \lambda_3 \theta^\dagger (X_\mu - X_\mu^\dagger)
\end{gather}
which means that 
\begin{gather}
0=\Im \tr \lambda_3 \theta^\dagger_x \tilde{\theta}_{\mu,x} e^{-i\epsilon \partial_\mu d \lambda_3} \tilde{\theta}^\dagger_{\mu,x+\hat{\mu}} \theta_{x+\hat{\mu}} e^{i\epsilon \partial_\mu d \lambda_3}e^{-i\epsilon \partial_\mu \delta \lambda_3},
\end{gather}
where $\Im$ denotes the imaginary part,
and
\begin{gather}
 \theta^\dagger_x \tilde{\theta}_{\mu,x} = \left(\begin{array}{cc}
                                          \cos \tilde{a} \cos a + \sin \tilde{a} \sin a e^{i(c - \tilde{c})}&
                                          i(\sin \tilde{a} \cos a e^{i\tilde{c}}-\cos \tilde{a} \sin a e^{ic})\\
                                          i(\sin \tilde{a} \cos a e^{-i\tilde{c}}-\cos \tilde{a} \sin a e^{-ic})&\cos \tilde{a} \cos a + \sin \tilde{a} \sin a e^{-i(c - \tilde{c})}
                                         \end{array}\right)
\end{gather}
which means that the $(0,0)$ component of $ \theta^\dagger_x \tilde{\theta}_{\mu,x} e^{i\epsilon \partial_\mu \tilde{d} \lambda_3} \tilde{\theta}^\dagger_{\mu,x+\hat{\mu}} \theta_{x+\hat{\mu}} e^{i\epsilon \partial_\mu \tilde{d} \lambda_3}$ is
\begin{align*}
& \cos \tilde{a} \cos a \cos \tilde{a}' \cos a'  + \cos \tilde{a} \cos a \sin \tilde{a}' \sin a' e^{ i(\tilde{c}'-c')}+\nonumber\\
&\sin \tilde{a} \sin a \sin \tilde{a}' \sin a' e^{i(c - c' + \tilde{c}'-\tilde{c})} + \sin \tilde{a} \sin a e^{i(c-\tilde{c})} \cos \tilde{a}'\cos a'+\nonumber\\
&\cos a \sin \tilde{a} \cos a' \sin \tilde{a}' e^{i(\tilde{c}-\tilde{c}'+2\epsilon\partial_\mu \tilde{d})} + \sin a \cos \tilde{a} \sin a' \cos \tilde{a}' e^{i(c - c' + 2\epsilon \partial_\mu \tilde{d})}-\nonumber\\
& \cos a \sin \tilde{a} \sin a' \cos \tilde{a}' e^{i(\tilde{c}-c' + 2\epsilon \partial_\mu \tilde{d})} - \sin a \cos \tilde{a} \cos a' \sin \tilde{a}' e^{i(c-\tilde{c}'+2\epsilon\partial_\mu \tilde{d})}
\end{align*}
where the primes indicate that the position of the variables is at $x+ \epsilon \hat{\mu}$, while without the prime they are at $x$, while the direction index on the $\tilde{a},\tilde{c}$ variables has been suppressed. This implies that
\begin{align}
 \partial_\mu \delta =&  2 \partial_\mu \tilde{d} (\cos^2 a \sin^2 \tilde{a} + \sin^2 a \cos^2 \tilde{a} - \frac{\sin 2a \sin 2\tilde{a}}{2}\cos(c-\tilde{c}))\nonumber\\
 &-\partial_\mu \gamma \sin^2 a - \partial_\mu \tilde{c} \sin^2 \tilde{a} \cos2a + \partial_\mu \tilde{c} \frac{\sin 2a \sin 2\tilde{a}}{2} \cos(\tilde{c}-c) \nonumber\\
 &+  \partial_\mu \tilde{a} \sin 2a \sin (\tilde{c}-c)
\end{align}

\subsection{Yang-Mills action}
We will concentrate on SU(2). 

The gauge link is defined as
\begin{align}
U_\mu(x) =& \tilde{\theta}_{\mu,x} e^{-i \epsilon \partial_\mu \tilde{d}_\mu \lambda_3} \tilde{\theta}^\dagger_{\mu,x+\epsilon \hat{\mu}}\nonumber\\
=& \left(\begin{array}{cc} \cos \tilde{a}_\mu& i \sin \tilde{a}_\mu e^{i\tilde{c}_\mu}\\
i \sin \tilde{a}_\mu e^{-i\tilde{c}_\mu} & \cos \tilde{a}_\mu\end{array}\right)
\left(\begin{array}{cc}e^{-i \epsilon \partial_\mu \tilde{d}_\mu}&0\\
0&e^{i\epsilon \partial_\mu \tilde{d}_\mu}\end{array}\right)
\left(\begin{array}{cc} \cos \tilde{a}'_\mu& -i \sin \tilde{a}'_\mu e^{i\tilde{c}'_\mu}\\
-i \sin \tilde{a}'_\mu e^{-i\tilde{c}'_\mu} & \cos \tilde{a}'_\mu\end{array}\right)\nonumber\\
=&1 + i \epsilon \lambda_3 (\cos 2\tilde{a}_\mu(-\partial_\mu \tilde{d}_\mu + \frac{1}{2} \partial_\mu \tilde{c}_\mu) - \frac{1}{2} \partial_\mu \tilde{c}_\mu) + \nonumber\\
&i \epsilon \lambda_1 (\cos \tilde{c}_\mu \partial_\mu \tilde{a}_\mu + \sin \tilde{c}_\mu \sin 2\tilde{a}_\mu (-\partial_\mu \tilde{d}_\mu + \frac{1}{2} \partial_\mu \tilde{c})) + \nonumber\\
&i \epsilon \lambda_2 (\sin \tilde{c}_\mu \partial_\mu \tilde{a}_\mu - \cos \tilde{c}_\mu \sin 2\tilde{a}_\mu (-\partial_\mu \tilde{d}_\mu + \frac{1}{2} \partial_\mu \tilde{c}_\mu))
\end{align}
where again we have assumed that we are in a gauge where $\tilde{a}_\mu$, $\tilde{c}_\mu$ and $\tilde{d}_\mu$ are smooth functions of position, and the prime indicates that they are at a position $x + \epsilon \hat{\mu}$ while the unprimed coordinates are at a position $x$. From this, we can construct an expression for the field strength,
\begin{multline}
A_\mu(x) = \lambda_3 (\cos 2\tilde{a}_\mu(-\partial_\mu \tilde{d}_\mu + \frac{1}{2} \partial_\mu \tilde{c}_\mu) - \frac{1}{2} \partial_\mu \tilde{c}_\mu) +
\lambda_1 (\cos \tilde{c}_\mu \partial_\mu \tilde{a}_\mu + \sin \tilde{c}_\mu \sin 2\tilde{a}_\mu (-\partial_\mu \tilde{d}_\mu + \frac{1}{2} \partial_\mu \tilde{c})) \\
+\lambda_2(\sin \tilde{c}_\mu \partial_\mu \tilde{a}_\mu - \cos \tilde{c}_\mu \sin 2\tilde{a}_\mu (-\partial_\mu \tilde{d}_\mu + \frac{1}{2} \partial_\mu \tilde{c}_\mu))
\end{multline}
It is convenient to perform the change of variables $\tilde{d}_\mu\rightarrow \frac{1}{2} \tilde{c}_\mu - \tilde{d}_\mu $, which does not alter the measure of the field theory, leaving us with,
\begin{multline}
A_\mu(x) = \lambda_3 (\cos 2\tilde{a}_\mu\partial_\mu \tilde{d}_\mu - \frac{1}{2} \partial_\mu \tilde{c}_\mu) \\+ \lambda_1 (\cos \tilde{c}_\mu \partial_\mu \tilde{a}_\mu + \sin \tilde{c}_\mu \sin 2\tilde{a}_\mu \partial_\mu \tilde{d}_\mu) + \lambda_2(\sin \tilde{c}_\mu \partial_\mu \tilde{a}_\mu - \cos \tilde{c}_\mu \sin 2\tilde{a}_\mu \partial_\mu \tilde{d}_\mu).
\end{multline}
Before constructing the field strength tensor, we have one more change of variables to make. The measure contains $d\tilde{a}_\mu \sin 2\tilde{a}_\mu$; however while this measure is well defined in a lattice theory, is it difficult to take its continuum limit. We would need to absorb the $\sin 2\tilde{a}_\mu$ term into the action (which is what we did in the lattice simulations), but the products of these terms cannot easily be expressed in terms of the exponential of an integral over all space. We therefore make the substitution $g \tilde{y}_\mu = \cos 2\tilde{a}_\mu$, where $g$ is the Yang-Mills coupling. This has the effect of removing any complications in the measure. The integration range of the new variable $\tilde{y}_\mu$ is from $-1/g$ to $1/g$, which, as we approach the continuum limit at the $g =0$ fixed point, tends towards the standard integration range for a Gaussian integral from $-\infty$ to $\infty$. In these new coordinates, we have
\begin{multline}
A_\mu(x) = \lambda_3 (g \tilde{y}_\mu\partial_\mu \tilde{d}_\mu - \frac{1}{2} \partial_\mu \tilde{c}_\mu) + \lambda_1 (-\cos \tilde{c}_\mu \frac{g\partial_\mu \tilde{y}_\mu}{2\sqrt{1-g^2 \tilde{y}_\mu^2}} + \sin \tilde{c}_\mu \sqrt{1-g^2 \tilde{y}_\mu^2} \partial_\mu \tilde{d}_\mu) \\
+ \lambda_2(-\sin \tilde{c}_\mu \frac{g\partial_\mu \tilde{y}_\mu}{2\sqrt{1-g^2 \tilde{y}_\mu^2}} - \cos \tilde{c}_\mu \sqrt{1-g^2 \tilde{y}_\mu^2} \partial_\mu \tilde{d}_\mu).
\end{multline}
The field strength is then,
\begin{align}
F_{\mu\nu} = & \partial_\mu A_\nu - \partial_\nu A_\mu - \frac{i}{g} [A_\mu,A_\nu]\nonumber\\
 = & \lambda_3 \partial_\mu (g \tilde{y}_\nu \partial_\nu \tilde{d}_\nu- \frac{1}{2} \partial_\nu \tilde{c}_\nu) - \lambda_3 \partial_\nu (g \tilde{y}_\mu \partial_\mu \tilde{d}_\mu- \frac{1}{2} \partial_\mu \tilde{c}_\mu) - \nonumber\\&
 \frac{1}{g}\lambda_3(- \frac{\cos \tilde{c}_\mu g\partial_\mu \tilde{y}_\mu}{2\sqrt{1-g^2 \tilde{y}_\mu^2}} + \sin \tilde{c}_\mu \sqrt{1-g^2 \tilde{y}_\mu^2} \partial_\mu \tilde{d}_\mu)
 (- \frac{\sin \tilde{c}_\nu g\partial_\nu \tilde{y}_\nu}{2\sqrt{1-g^2 \tilde{y}_\nu^2}} - \cos \tilde{c}_\nu \sqrt{1-g^2 \tilde{y}_\nu^2} \partial_\nu \tilde{d}_\nu) +\nonumber\\
 &\frac{1}{g}\lambda_3(- \frac{\cos \tilde{c}_\nu g\partial_\nu \tilde{y}_\nu}{2\sqrt{1-g^2 \tilde{y}_\nu^2}} + \sin \tilde{c}_\nu \sqrt{1-g^2 \tilde{y}_\nu^2} \partial_\nu \tilde{d}_\nu)
 (- \frac{\sin \tilde{c}_\mu g\partial_\mu \tilde{y}_\mu}{2\sqrt{1-g^2 \tilde{y}_\mu^2}} - \cos \tilde{c}_\mu \sqrt{1-g^2 \tilde{y}_\mu^2} \partial_\mu \tilde{d}_\mu)+\nonumber\\
 & \lambda_2 \partial_\mu (-\sin \tilde{c}_\nu \frac{g\partial_\nu \tilde{y}_\nu}{2\sqrt{1-g^2 \tilde{y}_\nu^2}} - \cos \tilde{c}_\nu \sqrt{1-g^2 \tilde{y}_\nu^2} \partial_\nu \tilde{d}_\nu) -\nonumber\\
 &\lambda_2 \partial_\nu (-\sin \tilde{c}_\mu \frac{g\partial_\mu \tilde{y}_\mu}{2\sqrt{1-g^2 \tilde{y}_\mu^2}} - \cos \tilde{c}_\mu \sqrt{1-g^2 \tilde{y}_\mu^2} \partial_\mu \tilde{d}_\mu) -\nonumber\\
& \frac{1}{g}\lambda_2 (g \tilde{y}_\mu\partial_\mu \tilde{d}_\mu - \frac{1}{2} \partial_\mu \tilde{c}_\mu) (-\cos \tilde{c}_\nu \frac{g\partial_\nu \tilde{y}_\nu}{2\sqrt{1-g^2 \tilde{y}_\nu^2}} + \sin \tilde{c}_\nu \sqrt{1-g^2 \tilde{y}_\nu^2} \partial_\nu \tilde{d}_\nu) + \nonumber\\
&\frac{1}{g}\lambda_2 (g \tilde{y}_\nu\partial_\nu \tilde{d}_\nu - \frac{1}{2} \partial_\nu \tilde{c}_\nu) (-\cos \tilde{c}_\mu \frac{g\partial_\mu \tilde{y}_\mu}{2\sqrt{1-g^2 \tilde{y}_\mu^2}} + \sin \tilde{c}_\mu \sqrt{1-g^2 \tilde{y}_\mu^2} \partial_\mu \tilde{d}_\mu) +\nonumber\\
 & \lambda_1 \partial_\mu(-\cos \tilde{c}_\nu \frac{g\partial_\nu \tilde{y}_\nu}{2\sqrt{1-g^2 \tilde{y}_\nu^2}} + \sin \tilde{c}_\nu \sqrt{1-g^2 \tilde{y}_\nu^2} \partial_\nu \tilde{d}_\nu) -\nonumber\\
 &\lambda_1\partial_\nu(-\cos \tilde{c}_\mu \frac{g\partial_\mu \tilde{y}_\mu}{2\sqrt{1-g^2 \tilde{y}_\mu^2}} + \sin \tilde{c}_\mu \sqrt{1-g^2 \tilde{y}_\mu^2} \partial_\mu \tilde{d}_\mu) - \nonumber\\
 &\frac{1}{g} \lambda_1 (-\sin \tilde{c}_\mu \frac{g\partial_\mu \tilde{y}_\mu}{2\sqrt{1-g^2 \tilde{y}_\mu^2}} - \cos \tilde{c}_\mu \sqrt{1-g^2 \tilde{y}_\mu^2} \partial_\mu \tilde{d}_\mu) (g \tilde{y}_\nu\partial_\nu \tilde{d}_\nu - \frac{1}{2} \partial_\nu \tilde{c}_\nu) + \nonumber\\
 &\frac{1}{g} \lambda_1 (-\sin \tilde{c}_\nu \frac{g\partial_\nu \tilde{y}_\nu}{2\sqrt{1-g^2 \tilde{y}_\nu^2}} - \cos \tilde{c}_\nu \sqrt{1-g^2 \tilde{y}_\nu^2} \partial_\mu \tilde{d}_\nu) (g \tilde{y}_\mu\partial_\mu \tilde{d}_\mu - \frac{1}{2} \partial_\mu \tilde{c}_\mu).
\end{align}
The partition function is
\begin{gather}
Z = \int d [\tilde{y}_{\mu}] d[\tilde{c}_\mu] d[\tilde{d}_\mu] e^{-\frac{1}{4g^2} \int d^4 x F_{\mu\nu}^2},
\end{gather}
with all the fields $\tilde{y}$, $\tilde{c}$ and $\tilde{d}$ dimensionless; the integrals over $\tilde{c}_\mu$ and $\tilde{d}_\mu$ range for all real functions of these variables; for $\tilde{y}_\mu$ we are constrained to functions where $-1/g \le \tilde{y}_\mu \le 1/g$. Since all the terms in $F_{\mu\nu}$ are proportional to a derivative operator, in the absence of any winding the resultant gauge particles will be massless, as we expect in a QCD-like theory. 

The lack of any clear quadratic terms in the action $\frac{1}{4g^2} F_{\mu\nu}^2$ is discouraging from the perspective of perturbative calculations. However, in practice we would want to first gauge fix and then expand around a minima of the action when performing perturbation theory, i.e. write $\tilde{c}_\mu \rightarrow c_{0\mu} + \tilde{c}_\mu$ where $c_{0\mu}$ represents some minima of the action. This expansion may well have quadratic terms as the most dominant contribution, allowing perturbation theory. However developing a perturbative treatment of this action is going to be challenging. From the perspective of renormalisation, it is interesting that the dynamical variables are dimensionless: the propagator has the dimensions of $1/k^4$ which might mean that there are at most only logarithmic divergences (or it might not).

There is a clear minimum of the action when all the fields are constant, i.e. $\partial_\mu \tilde{c}_\mu = 0$, $\partial_\mu \tilde{d}_\mu = 0$ and $\partial_\mu \tilde{y}_\mu = 0$, or some gauge transformation of these fields. This is, however, not the only minimum. For example, if the volume is a torus with periodic boundary conditions and side lengths $L_\mu$, we can have a similar solution with some winding in at least one of the directions $\tilde{c}_\mu$, i.e. $\partial_\mu \tilde{c}_\mu = \frac{2\pi \nu_\mu}{L_\mu}$, $\partial_\mu \tilde{d}_\mu = 0$, $\partial_\mu \tilde{y}_\mu = 0$, where $\nu_\mu$ is a constant in space integer winding number (again, not a Lorentz four vector). Thus some solutions with non-zero winding will contribute to the path integral.  Aside from this winding,  fields except where $\partial_\mu \tilde{c}_\mu \sim O(g)$ and $\partial_\mu \tilde{d}_\mu = O(g)$ (or higher order) will often be heavily suppressed by factors of $O(1/g)$ or $O(1/g^2)$ and will not contribute in the $g \rightarrow 0$ continuum limit. This will ensure that the gauge fields are reasonably smooth in space. The gauge invariance of this operator is hidden by the notation (but is still present, of course, since we have done nothing except re-express the theory in terms of new variables); it would still be necessary to gauge fix as usual before computing observables. 

However, on a torus with periodic boundary conditions (or some other parametrisation of the coordinates which involves closed loops, although in these cases the form of the field strength tensor would be different at the corners of the loops), it is possible for both $\tilde{c}_\mu$ and $\tilde{d}_\mu$ to have non-trivial topology. We perform a change of variables, writing
\begin{align}
\tilde{c}_\mu \rightarrow & c_{\mu}^0 + \frac{2\pi \nu_{c\mu}x_\mu}{L} + g\tilde{c}_\mu + \sum_{\mu\neq \nu} \frac{2\pi \nu_{c\mu\nu}x_\nu}{L_\nu}\nonumber\\
\tilde{d}_\mu \rightarrow & \tilde{d}_{\mu}^0 + \frac{\pi \nu_{d\mu}x_\mu}{L} + g\tilde{d}_\mu + \sum_{\mu\neq \nu} \frac{\pi \nu_{d\mu\nu}x_\nu}{L_\nu},
\end{align}
where $\tilde{c}_{\mu}^0$ and $\tilde{d}_{\mu}^0$ are constants ($\partial_\nu \tilde{c}_{\mu}^0 = 0$), $\nu_{c\mu}$ and $\nu_{d\mu}$ are integer winding numbers, and $\tilde{c}_\mu$ and $d_\mu$ are the new dynamical fields, which are now restricted to the range $-\pi/g$ to $\pi/g$. $\tilde{d}_\mu \rightarrow \tilde{d}_\mu + \pi$ is related to the centre symmetry $\theta_\mu \rightarrow -\theta_\mu$, which does not affect the physics; thus for $\tilde{d}_\mu$ the winding numbers need only be multiplied by $\pi$ rather than $2\pi$. We have added the possibility of additional winding in the directions orthogonal to $\mu$, which is possible if we use Cartesian coordinates. This means that each $\partial_\mu \tilde{c}_\mu$ term in the action will produce something proportional to the winding number. When we replace the derivatives with the winding numbers, these can lead to mass-like terms in the action for the $y$ and $c$ fields. The relevant part of the action (excluding terms proportional to $\partial_\mu \tilde{y}_\mu$ which cannot generate a mass term in this way and also $\partial_\nu \tilde{c}_\mu$) is
\begin{align}
\mathcal{L}_M =& \frac{1}{4g^4} \sin^2 (\chi_{\mu\nu} +g(\tilde{c}_\mu - \tilde{c}_\nu)) (\partial_\mu \tilde{d}_\mu \partial_\nu \tilde{d}_\nu)^2 (1-g^2 \tilde{y}_\mu^2)(1-g^2 \tilde{y}_\nu^2) + \nonumber\\
& \frac{1}{4g^4} (g \tilde{y}_\mu \partial_\mu \tilde{d}_\mu - \frac{1}{2} \partial_\mu \tilde{c}_\mu)^2 (1-g^2 \tilde{y}_\nu^2)(\partial_\nu \tilde{d}_\nu)^2 + \nonumber\\
& \frac{1}{4g^4} (g \tilde{y}_\nu \partial_\nu \tilde{d}_\nu - \frac{1}{2} \partial_\nu \tilde{c}_\nu)^2 (1-g^2 \tilde{y}_\mu^2)(\partial_\mu \tilde{d}_\mu)^2 + \nonumber\\
& \frac{1}{2 g^4} (g\tilde{y}_\nu \partial_\nu \tilde{d}_\nu - \frac{1}{2} \partial_\nu \tilde{c}_\nu)(g \tilde{y}_\mu \partial_\mu \tilde{d}_\mu - \frac{1}{2} \partial_\mu \tilde{c}_\mu)\sqrt{(1-g^2 \tilde{y}_\nu^2)}\sqrt{(1-g^2 \tilde{y}_\mu^2)}( \partial_\nu \tilde{d}_\nu  \partial_\mu \tilde{d}_\mu).
\end{align}
with
\begin{gather}
\chi_{\mu\nu} = c^0_\mu - c^0_\nu + \frac{2\pi}{L} (\nu_{c\mu} - \nu_{c\nu})
\end{gather}
Solutions with non-zero winding for $\tilde{d}_\nu$ correspond to minima of the action close to $g y_\nu = \pm 1$ and $\partial_\nu y_\nu$ close to zero (in such a way that $\partial_\nu y_\nu/ \sqrt{1-g^2 y_\nu^2} = \partial_\nu \tilde{a}_\nu \ll 1$).
Replacing each derivative with the winding term means that the action contains the contribution,
\begin{align}
\mathcal{L}'_M =& \frac{\pi^4}{4g^4L^4} (\sin^2 (\chi_{\mu\nu}) + g (\tilde{c}_\mu - \tilde{c}_\nu) \sin 2\chi_{\mu\nu} +  g^2 (\tilde{c}_\mu-\tilde{c}_\nu)^2 \cos 2 \chi_{\mu\nu} + \ldots) \nonumber\\
& \phantom{spacespacespacespacespacespacespacespace}(\nu_{d\mu} \nu_{d\nu})^2 (1-g^2 \tilde{y}_\mu^2)(1-g^2 \tilde{y}_\nu^2) + \nonumber\\
& \frac{\pi^4}{4g^4L^4} (g \tilde{y}_\mu \nu_{d\mu} -  \nu_{c\mu})^2 (1-g^2 \tilde{y}_\nu^2)(\nu_{d\nu})^2 + \nonumber\\
& \frac{\pi^4}{4g^4L^4} (g \tilde{y}_\nu \nu_{d\nu} -  \nu_{c\nu})^2 (1-g^2 \tilde{y}_\mu^2)(\nu_{d\mu})^2 + \nonumber\\
& \frac{\pi^4}{2 g^4L^4} (g\tilde{y}_\nu \nu_{d\nu} -  \nu_{c\nu})(g \tilde{y}_\mu \nu_{d\mu} - \nu_{c\mu})\sqrt{(1-g^2 \tilde{y}_\nu^2)}\sqrt{(1-g^2 \tilde{y}_\mu^2)}( \nu_{d\nu}  \nu_{d\mu}),
\end{align}
where $L^4$ represents the appropriate products of the various box lengths.
The partition function would resemble
\begin{gather}
Z = \sum_{\nu_{c\mu}, \nu_{d\mu}} \int_{-1/g}^{1/g} [d\tilde{y}_\mu] \int_{-\pi/g}^{\pi/g} [d\tilde{c}_\mu] \int_{-\pi/g}^{\pi/g} [d\tilde{d}_\mu] e^{-\int d^4 x (\mathcal{L}'_M + \ldots)},
\end{gather}
where the $\ldots$ contain the various kinetic and interaction terms in the Lagrangian. We need to take the $L \rightarrow \infty$ limit and $g \rightarrow 0$ limit to recover continuum (Euclidean) QCD. It is convinent to take these limits simultaneously with $L\sqrt{g}$ kept constant, which keeps the various mass terms finite as we take these two limits. We can expect that the average size of the winding numbers $\nu_c$ and $\nu_d$ will be proportional to the length $L$; this constant of proportionality will provide the dynamical scale for the mass terms in the action.

Thus what we have proposed is a possible mechanism for the dynamical generation of gluon masses via the topological winding of the $\tilde{c}$ and $\tilde{d}$ parameters in the gauge fields. This remains far short of a proof: the bound states of the theory have to be gauge covariant, and so will be some function of $\tilde{y}_\mu$, $\tilde{c}_\mu$ and $\tilde{d}_\mu$ which preserves the measure and is gauge covariant.  What we still need to do is find these bound states, re-express the path integral and action in terms of these variables, shift the variables to remove the linear terms in the action, and show that the winding of $\tilde{c}_\mu$ and $\tilde{d}_\mu$ dynamically generates a mass term in that action. This is obviously a lot of effort, and we will not attempt it here. We showed in~\cite{Cundy:2015caa} that the winding number itself is invariant under smooth gauge transformations.

\section{Topological Dominance of Wilson Loop and Polyakov Line}\label{sec:5}
As shown before, though a judicious choice of coordinate basis and Abelian Decomposition, we can express the Wilson Loop (or Polyakov Loop) in terms of an Abelian field rotated into a basis determined by the Abelian Decomposition,
\begin{gather}
W_L = \theta e^{-i \oint dx_\mu \frac{1}{2} \lambda_j \tr ( \theta \lambda_j \theta^\dagger gA_\mu(x) + i\lambda_j \theta^\dagger \partial_\mu \theta)} \theta^\dagger,
\end{gather}
where the $\lambda$ matrices are normalised so that $\tr \lambda_j^2 = 2$.
From equation (\ref{eq:dtheta}) we see that the $\theta^\dagger \partial_\mu \theta$ term contains contributions proportional to $\lambda_3 \sin^2 a \partial_\mu c$ (in SU(3) there is a similar expression~\cite{Cundy:2015caa}), where $a$ and $c$ are the two parameters which parametrise the SU(2) $\theta$ matrix. In~\cite{Cundy:2015caa} we hypothesised that winding in $c$ could lead to an area law Wilson  Loop. The question remains, of course, whether the Maxwell term proportional to $A_\mu$ could also contribute to this loop.

As shown above, neither the Maxwell term nor the $\theta$ term are gauge invariant. Thus to naively ask whether the $\theta$ term dominates is impossible: it depends on the gauge. Equally, to specify $a$, $c$ and measure the winding it is necessary to first specify the gauge. It is obviously possible to fix to a gauge such that the Maxwell term dominates: we just choose a gauge where $\theta$ is the identity operator. What is less clear is whether it is possible to fix the gauge so that the $\theta$ term dominates the Wilson or Polyakov Loop. What we want is to choose the gauge transformation $\Lambda$ so that 
\begin{gather}
\oint dx_\mu \tr( n^j  (g   A_\mu   +i \partial_\mu (\Lambda^\dagger) \Lambda)) = 0.\label{eq:MaxwellIsZero}
\end{gather}
In principle, this should not be a problem: we have one condition, and three times as many variables as we have lattice sites contributing to the loop, so the system is considerably over-determined. However, there is also a second condition: the $\theta$ field needs to be a smooth function of position for the previous derivations to be valid. Ideally, therefore, we wish to switch to a gauge where the topological part dominates the Wilson loop, and the $\theta$ field is as smooth as possible given this condition. On the lattice, we did not quite do this, but instead choose a gauge condition such that $\Delta$ is minimised, where
\begin{gather}
\Delta = \Delta_1 + \Delta_2\nonumber\\
\Delta_1 = [\tr (W_L - e^{-i\oint dx_\mu \frac{1}{2} \lambda_j \tr (\lambda_j \theta^\dagger \partial_\mu \theta)})]^2 \nonumber\\
\Delta_2 = \zeta\sum_{\mu,x} \tr(1-\theta_{x} \theta^\dagger_{x + \epsilon \hat{\mu}}),
\end{gather}
and $\zeta$ is some tunable small parameter. In practice, this is considerably easier to compute than the condition needed for perfect topological dominance, which is to have $\Delta_2$ (smoothness of $\theta$) minimised given $\Delta_1 = 0$ ($\theta$ dominance of the Wilson Loop). We found that it was quite easy to have very low values for both $\Delta_1$ and $\Delta_2$, suggesting that this gauge condition is reasonable. In our numerical simulations we could also find solutions where $\Delta_1 = 0$ quite easily, suggesting that satisfying condition (\ref{eq:MaxwellIsZero}) is indeed overdetermined, although occasionally we had to try several starting guesses to our routine minimising of $\Delta$ before we found this solution. 

Fixing to this gauge allows us to study how topology contained within the $\theta$ field drives confinement, but only in this gauge. What about the other gauges? We have chosen a gauge where the Maxwell contribution to the Wilson Loop is zero, $\oint dx_\mu \tr (n^j (g A_\mu)) =0$. When we switch to another gauge, we add to this the term from the gauge fixing matrix, $- i \oint dx_\mu \tr(n^j \partial_\mu (\Lambda^\dagger) \Lambda)$, which in general will not be zero. However, this is of the same form as the $\theta$ term. In other words, we can parametrise $\Lambda$ in the same way that we do $\theta$, and the same topological objects will be present in the restricted field, only moved to the Maxwell term.

When we perform a gauge transformation in SU(2), we convert the Maxwell term to
\begin{gather}
 \tr \theta \lambda_3 \theta^\dagger A_\mu \rightarrow 
 \tr \Lambda \theta \lambda_3 \theta^\dagger \Lambda^\dagger (\Lambda A_\mu \Lambda^\dagger + i \Lambda \partial_\mu \Lambda^\dagger).
\end{gather}
We can write the gauge transformation matrix $\Lambda$ in the form $L e^{i l_3 \lambda_3} \theta^\dagger$, where $L$ is the $SU(N)/U(1)$ that $\theta$ transforms to in the new gauge (parametrised by $N_C(N_C-1)$ variables). In this case, the gauge field transforms to
\begin{gather}
 A_\mu \rightarrow L e^{i l_3 \lambda_3} \theta^\dagger (A_\mu + i\partial_\mu \theta \theta^\dagger  + \theta \partial_\mu l_3 \lambda_3 \theta^\dagger + \theta e^{-il_3\lambda_3} i\partial_\mu L^\dagger L e^{il_3\lambda_3} \theta^\dagger ) \theta e^{-il_3\lambda_3} L^\dagger 
\end{gather}
So the gauge transformation involves adding the original $\theta$ term of the action to the gauge field, rotating the gauge field, and then subtracting the new $\theta$ term from the gauge field. This means that all we are doing by finding $\theta$, performing the Abelian decomposition, and using this choice of gauge fixing is extracting that part of the gauge field that is responsible for confinement and isolating it from the rest of the gauge field. Every topological object we find in $\theta$ in this gauge is present in the Maxwell term in any other gauge which lacks that object. The gauge fixing just helps us to identify and see these topological objects without the complications of the various background noise and any topology not involved in the confinement of these particular quarks. 

We should also comment on that the choice of $\theta$ depends on which Wilson Loop we study, and in general the topological objects will be different from one Wilson Loop to another. This means that we cannot (by our methods) identify what is confining the quarks without specifying which quarks are to be confined. We are looking at the string that forms between a specific quark-anti-quark pair. This is not a real burden to our method, because any observed instance of quark confinement would confine two particular quarks.\footnote{Of course, we are here only considering the unphysical case of static or infinitely heavy quarks; so perhaps this appeal to physical observables might seem unpalatable. The solution would, of course, be to generalise the Wilson Loop so that it can also study light quarks, perhaps by offering a weighted average over possible paths the quarks can travel.}
\section{Numerical results}\label{sec:6}
Our numerical results are taken from an SU(2) pure Yang-Mills lattice ensemble, generated at $\beta = 3.4$ with a L\"uscher-Weisz gauge action. The lattice volume is $16^3 \times 32$, and we used 15 steps of stout smearing~\cite{Morningstar:2003gk,Moran:2008ra} at parameters $\rho = 0.1$, $\epsilon = 0$ to smooth the field before computing any observables. Our ensemble contained 150 configurations. All errors were computed using the bootstrap method.

We proceed by using the prescription described earlier. On each configuration we select a series of nested and stacked $X\times T$ rectangular loops in the $xt$ plane starting and ending at a fixed point, and perform the Abelian decomposition so that the Wilson Loops on these curves computed from the Yang-Mills and original gauge fields are identical, and then gauge fix by minimising the quantity 
\begin{align}
\Delta =& \sum_{x,\mu} \zeta \tr(1-\Theta_{\mu}) + (\tr W_L[U] - \tr W_L[\Theta])^2,\nonumber\\
\end{align}
where $W_L[\Theta]$ is the topological part of the restricted Wilson Loop measured using the gauge link, and $\Theta_\mu = e^{i \frac{1}{2} \lambda_3 \tr \lambda_3 \theta^\dagger_{x,\mu} \theta_{x + \hat{\mu},\mu}}$. 
The first term in $\Delta$ is minimised when the gauge field is smooth, and the second term is minimised then the topological part of the Wilson Loop is identical to the Wilson Loop from the original Yang Mills field. Our gauge fixing procedure proceeds in steps  gradually reducing the parameter $\zeta$ to a value of $10^{-5}$ divided by the lattice volume. A large value of $\zeta$ means that the minimisation routine concentrates on finding a smooth field; a small value means that it tries to match the Wilson Loops. We found that starting off with a large value and then gradually reducing it was the best way to efficiently satisfy both conditions.

We used a molecular dynamics procedure for our minimisation algorithm, introducing a fictitious momentum, writing $\Delta$ as a potential energy, and evolving along the classical equations of motion (using a numerical integration) while resetting the momentum to zero every so often when the rate of descent of the potential energy stopped increasing with each micro-canonical step. The molecular dynamics routine conserves the kinetic energy plus $\Delta$; by starting with kinetic energy zero we ensure the kinetic energy increases during the early parts of the trajectory, and consequently $\Delta$ must decrease. The initial micro-canonical step is identical to the steepest descent method, but the descent of $\Delta$ then accelerates as the trajectory continues. One can also add in some friction, i.e. a damping force proportional to the momentum, which in exact arithmetic would guarantee convergence to a local minimum of $\Delta$ within a single trajectory; in practice, however, just restarting the trajectory with zero momentum every so often proved to be more efficient. 

To compute $\Delta$ we use  periodic boundary conditions in all directions. This introduces some finite size effects on the larger loops, because we force the two sides of the larger Wilson Loops to be similar to each other. So, for instance, if we have a Wilson Loop with $X=15$ (the largest value allowed on our lattice), the smoothness condition implies that the value of $\theta$ on one side of the loop at $x=15$ should be close to the value at $x=14$ (with the same $y$, $z$ and $t$ coordinates), which will be close to the $\theta$ at $x = 13$, and so on across the lattice to $x=0$ on the other side of the original Wilson Loop. For a large Wilson Loop, these fifteen \textit{close to}s can add up to quite a substantial discrepancy, and in an infinite volume this series of intermediate steps is the only way the smoothness condition can affect the two opposite sides of the Wilson Loop. However, if we impose periodic boundary conditions in the $x$ direction then we will also have a condition forcing the $x=0$ $\theta$ field to be close to the $\theta$ field at $x=15$ on the opposite side of the Wilson Loop. This is just a finite volume effect, but given that we are using a relatively small lattice, it might be significant. It is, however, necessary that we keep a smooth $\theta$ field to keep a connection between the lattice and continuum theories, so we cannot switch to open boundary conditions. Instead, we exclude the largest Wilson Loops from our observables.

We have not encountered a serious Gribov problem (numerous local minima of the gauge fixing condition) on our smaller test lattices, but have not examined this issue on the production lattices because of lack of computer resources. We do not expect that our results will be affected by Gribov copies, but this needs investigation.    

\begin{figure}
 \begin{center}
  \begin{tabular}{cc}
  \includegraphics[width=6.5cm]{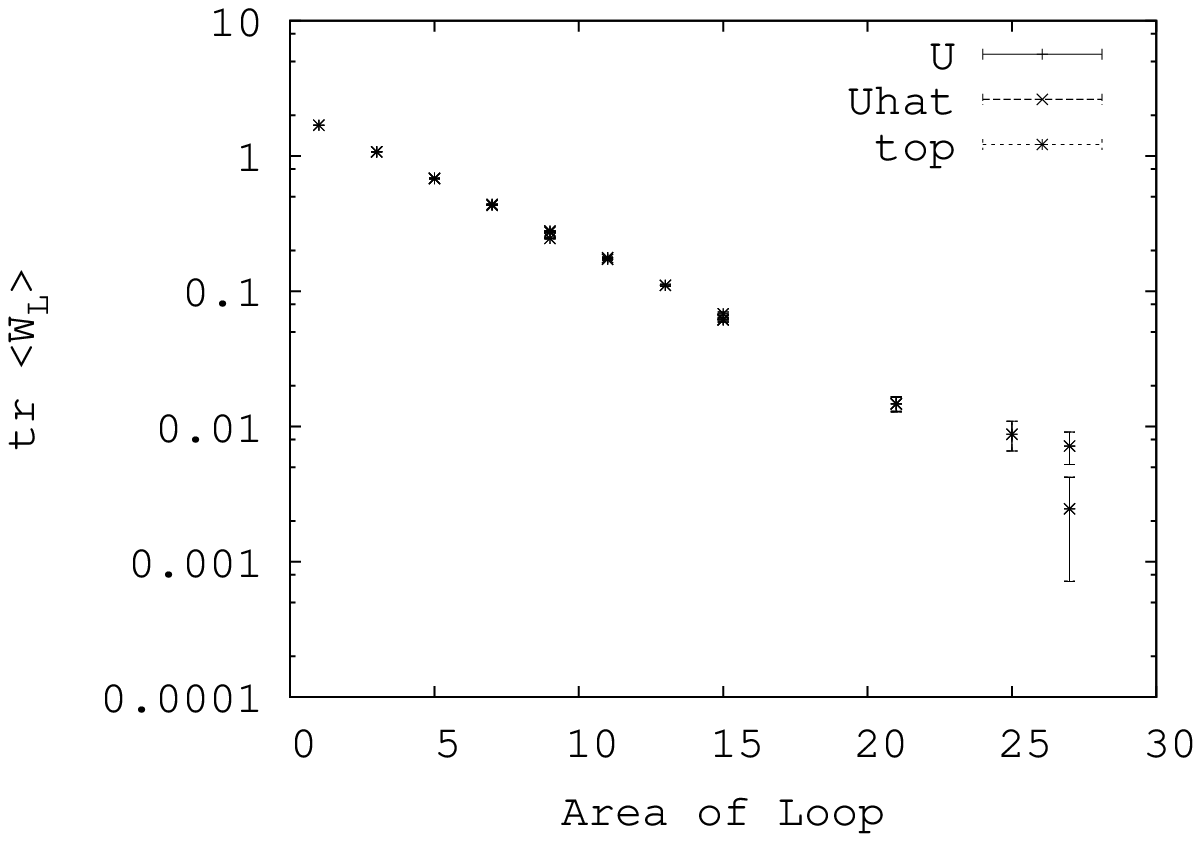}&
  \includegraphics[width=6.5cm]{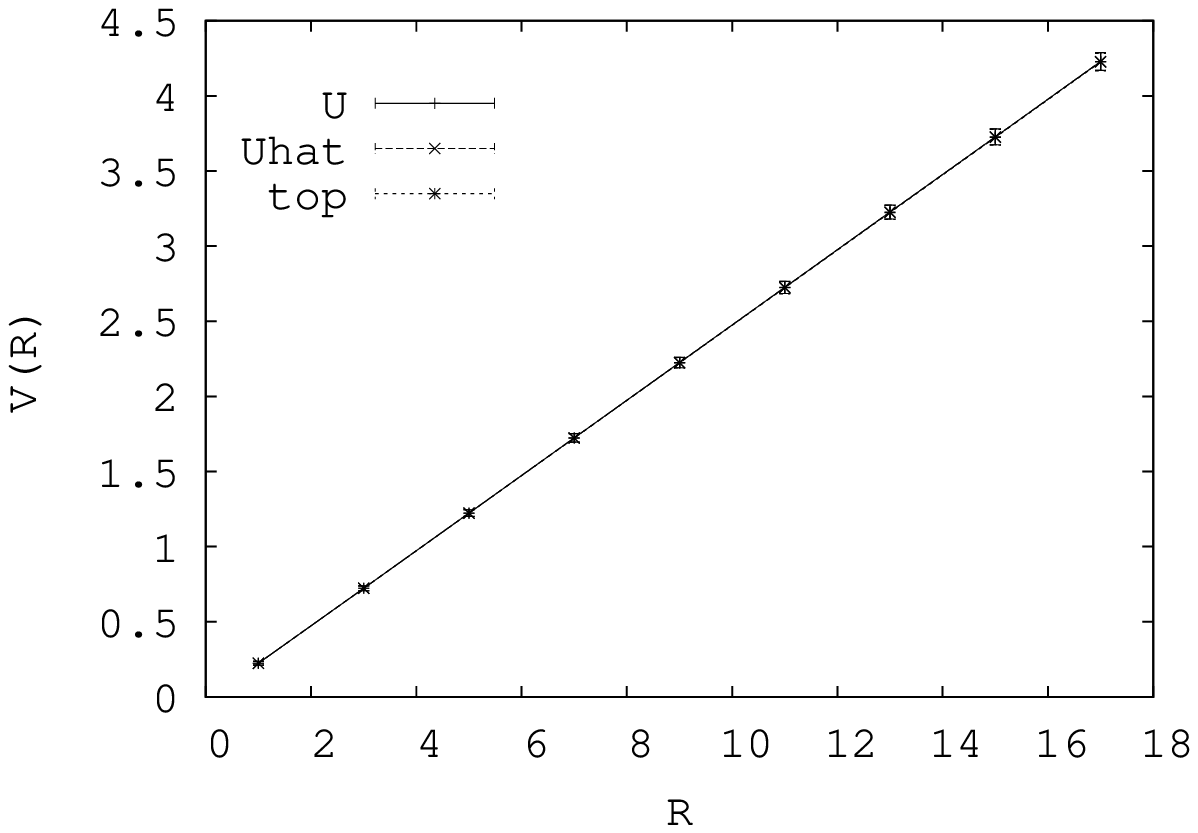}
  \end{tabular}
  \end{center}
\caption{The expected value of the Wilson Loop (left) and static quark potential (after an extrapolation to infinite time) (right), where the Wilson Loop is calculated with the Yang-Mills gauge field ($U$), the restricted gauge field ($Uhat$) and the topological part of the restricted gauge field ($top$).}\label{fig:WL}
\end{figure}

The first thing to do is to confirm that out Abelian Decomposition/gauge fixing procedure does indeed extract the confining part of the gauge field into first of all the restricted field and secondly the topological part of the restricted field. This is shown in figure \ref{fig:WL}. We provide two plots. The first one shows how expectation value of the trace of the Wilson Loop constructed using the original gauge field, the restricted gauge field, and the topological part of the restricted field scales with the area of the Wilson Loop. For large enough loop areas, we expect $\log \tr \langle W_L \rangle$ to scale linearly with the area of the loop, and we see this. This, of course, is nothing new. Our main result is that our three data sources are indistinguishable from each other. This shows two things: firstly, that it is indeed possible to gauge fix so that the topological term fully accounts for the string tension. Secondly, it shows that our gauge fixing ansatz works and actually does what it was meant to. The second plot in figure \ref{fig:WL} shows an extrapolation of the string tension to infinite time. We fitted the Wilson Loop data to the formula $\langle W_L \rangle = e^{-V(x,t) t}$, with $V(x,t) t = \rho xt + \alpha_1 x + \alpha_2 t + \alpha_3 + \alpha_4 x/t + \alpha_5 t/x + \alpha_6 /x + \alpha_7/t + \alpha_8/(xt)$, which had a $\chi^2$ per degree of freedom of $1.3$. The second figure in the plot shows the variation of $V(x,\infty) \equiv V(x)$ with $x$. Once again, the shape of this curve is standard: we see a little curvature due to the Coloumb interaction at small distances (this curvature is reduced by our smearing), and then the potential increases linearly with $x$. However, once again the key point is that the Yang-Mills, restricted and topological part string tensions are all indistinguishable from each other. This confirms that we have completely isolated the cause of quark confinement into the topological part of the restricted field.

\begin{figure}
\begin{center}
 \includegraphics{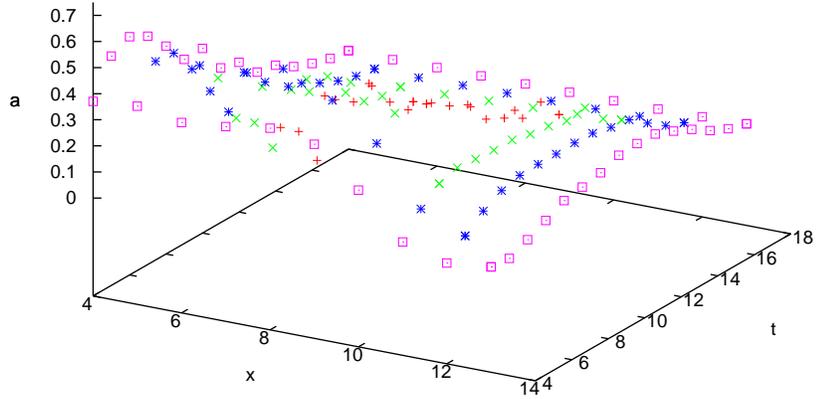}
 \end{center}
 \caption{The parameter $a$ extracted from the Abelian Decomposition $\theta$ matrix, after gauge fixing, shown on part of one $y$ $z$ slice of the lattice plotted as a function of the $x$ and $t$ coordinates of the lattice.}\label{fig:a_and_c1}
\end{figure}

\begin{figure}
\begin{center}
 \includegraphics{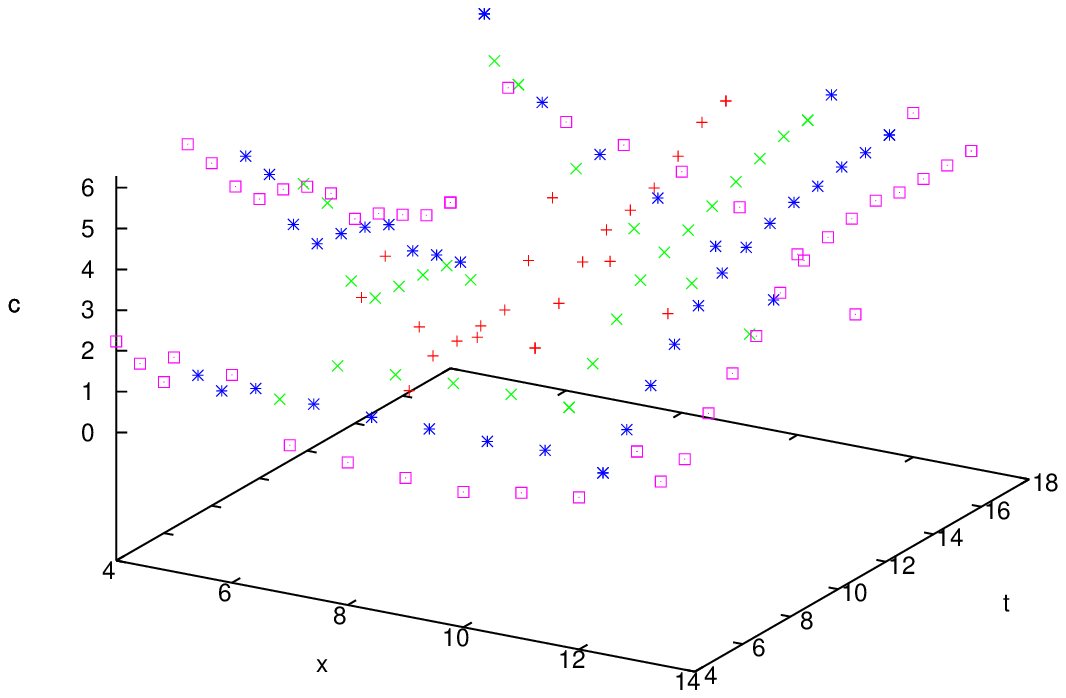}
 \end{center}
 \caption{The parameter $c$ extracted from the Abelian Decomposition $\theta$ matrix, after gauge fixing, shown on part of one $y$ $z$ slice of the lattice plotted as a function of the $x$ and $t$ coordinates of the lattice.}\label{fig:a_and_c2}
\end{figure}

\begin{figure}
 \begin{center}
  \begin{tabular}{c}
    \includegraphics{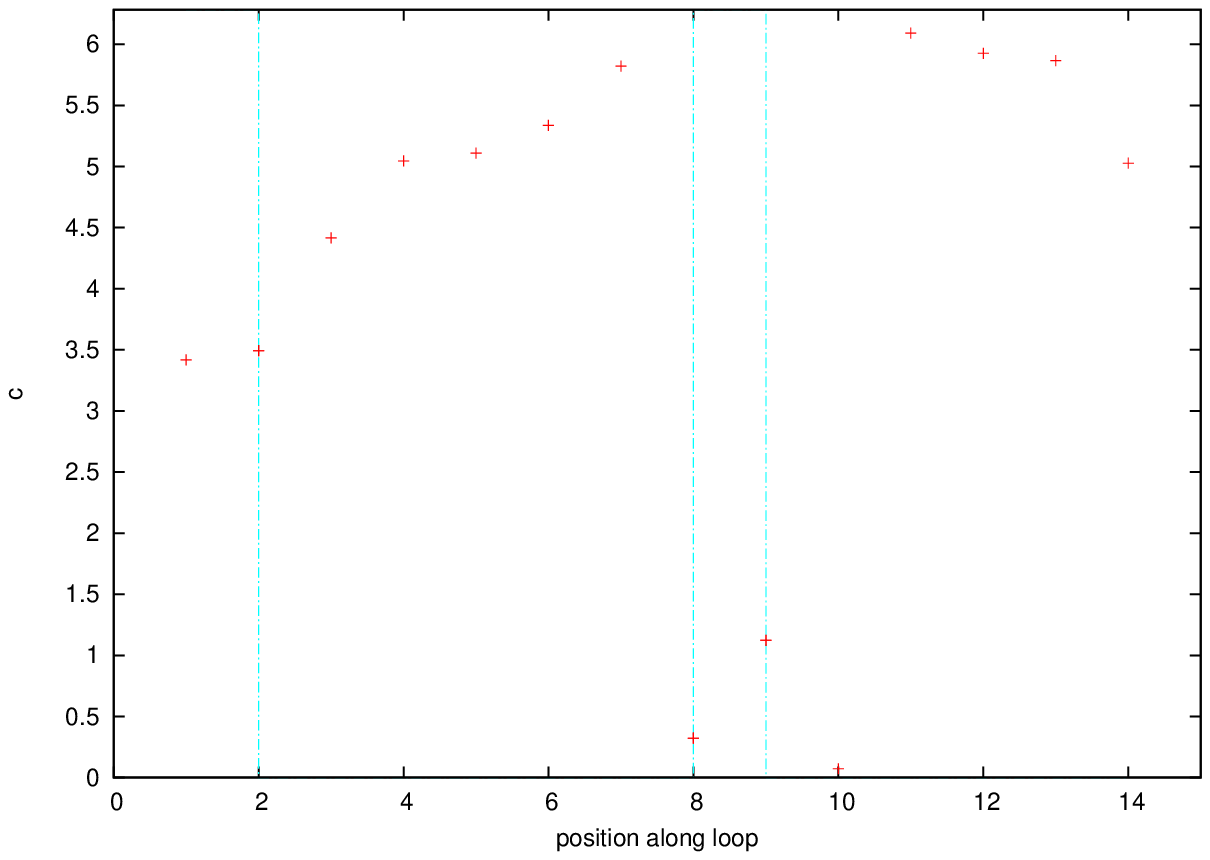}\\
    \includegraphics{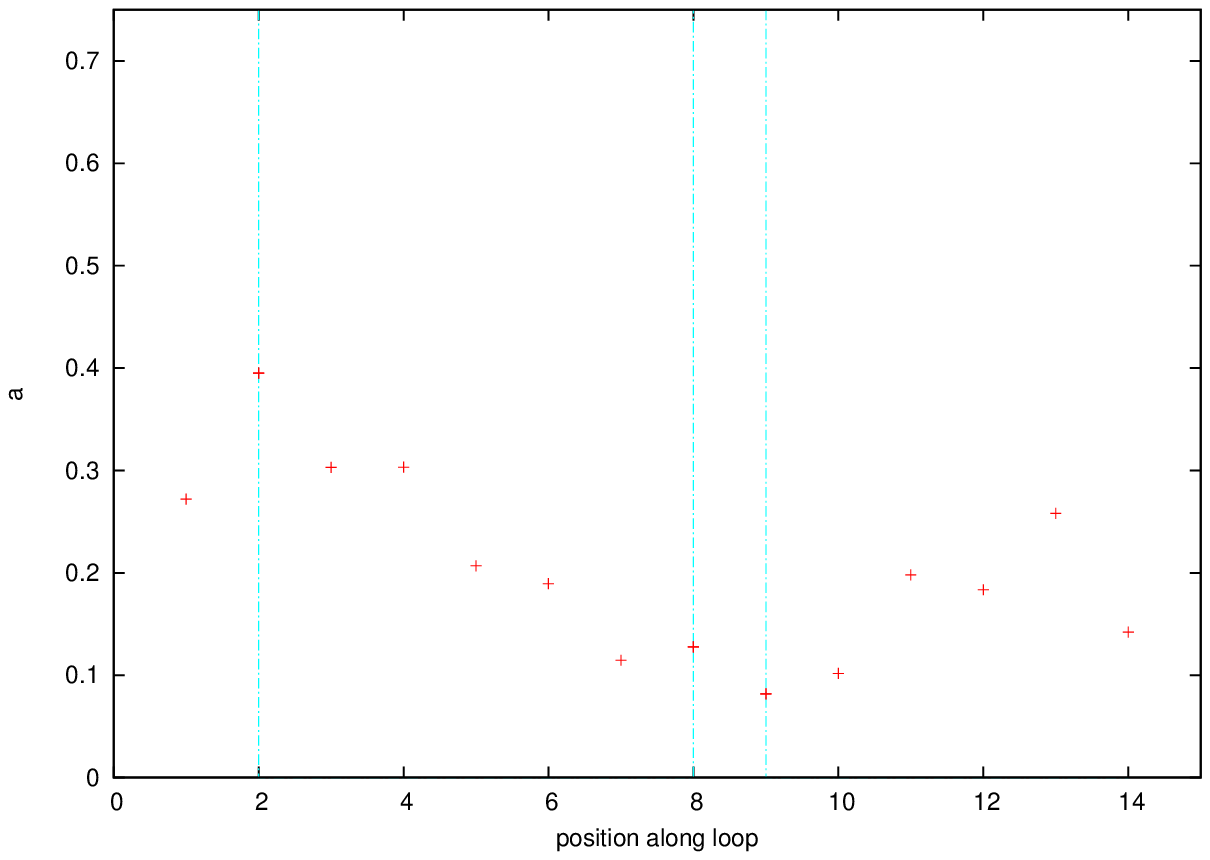}
  \end{tabular}
 \end{center}
  \caption{The parameters $a$ and $c$ extracted from the Abelian Decomposition $\theta$ matrix, after gauge fixing, shown along one Wilson Loop starting at lattice site $(x,y,z,t) = (7,6,6,7)$.}\label{fig:a_and_c3}
 \end{figure}

\begin{figure}
\begin{center}
 \begin{tabular}{c}
   \includegraphics{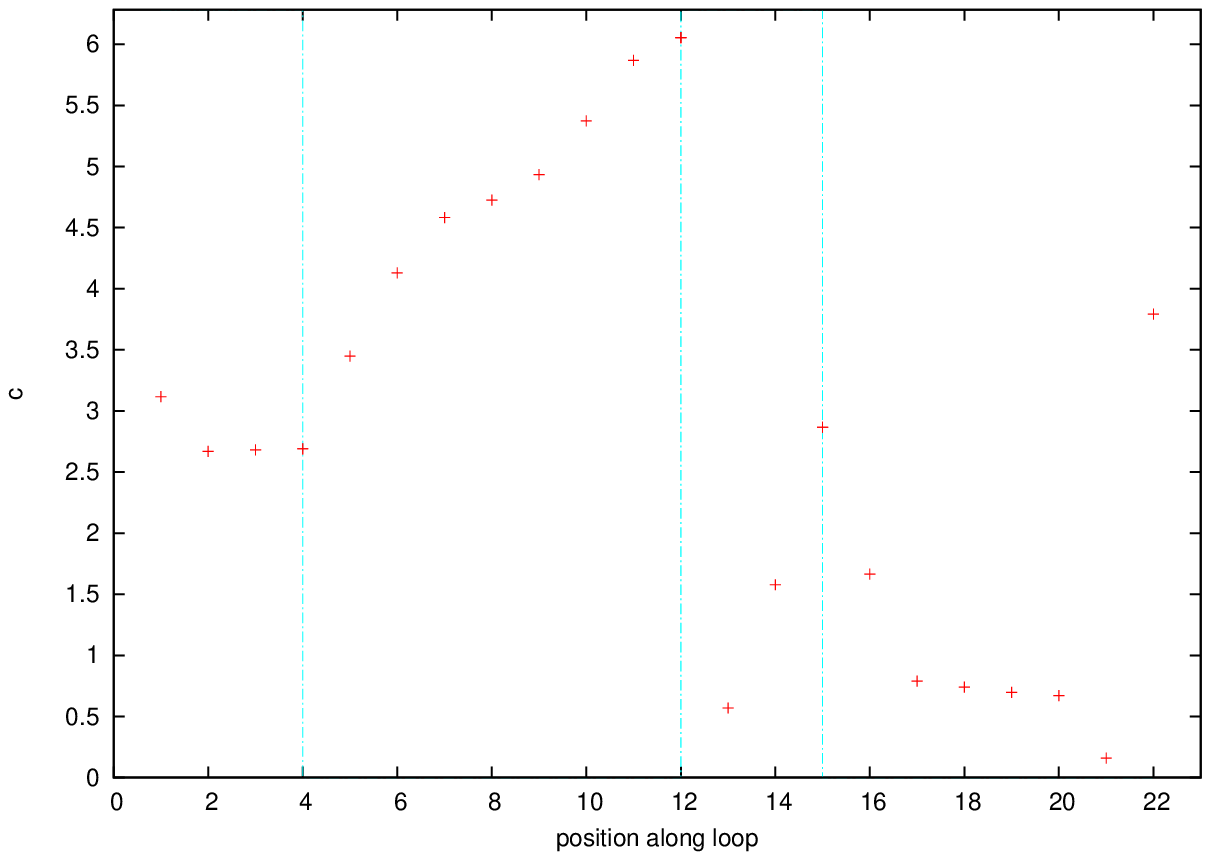}\\
   \includegraphics{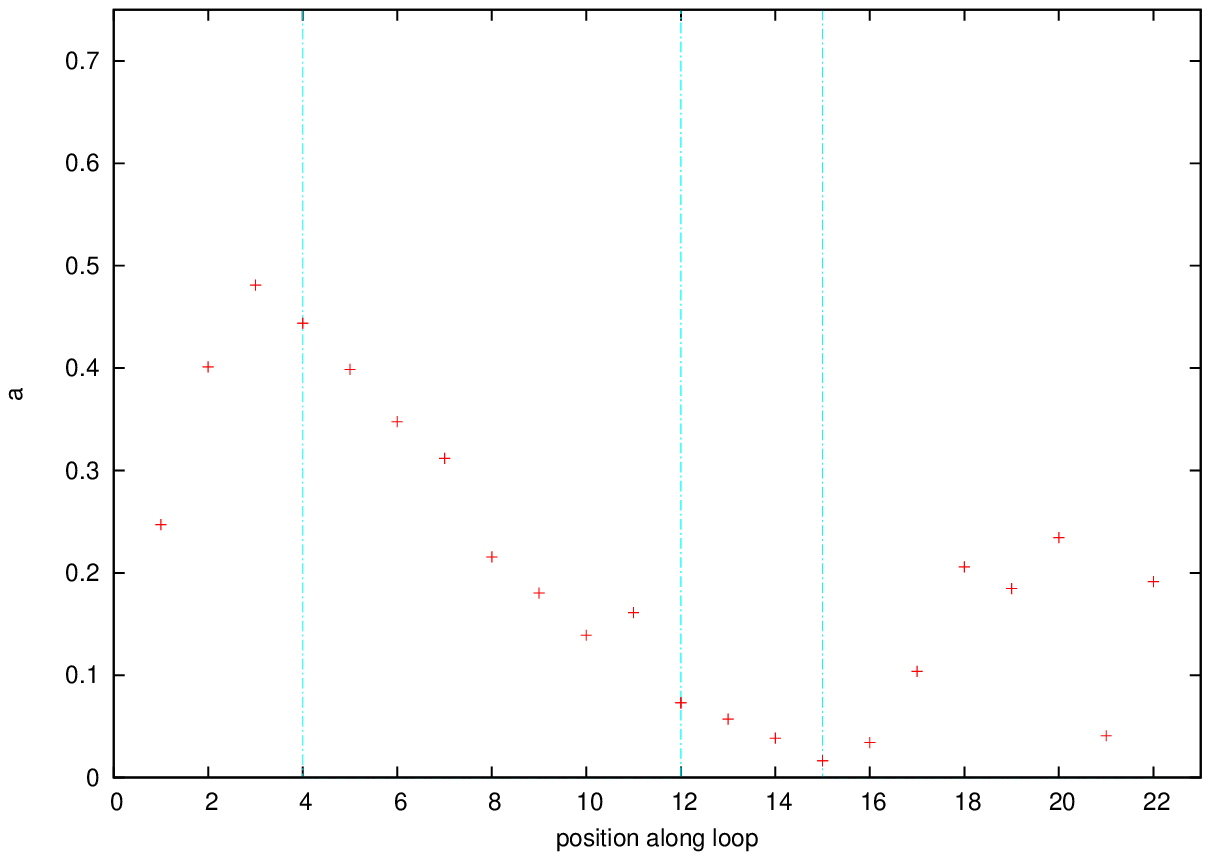}
 \end{tabular}
\end{center}
 \caption{The parameters $a$ and $c$ extracted from the Abelian Decomposition $\theta$ matrix, after gauge fixing, shown along one Wilson Loop starting at lattice site $(x,y,z,t) = (6,6,6,6)$. The vertical lines represent the corners of the Wilson Loop.}\label{fig:a_and_c4}
\end{figure}

\begin{figure}
\begin{center}
 \begin{tabular}{c}
   \includegraphics{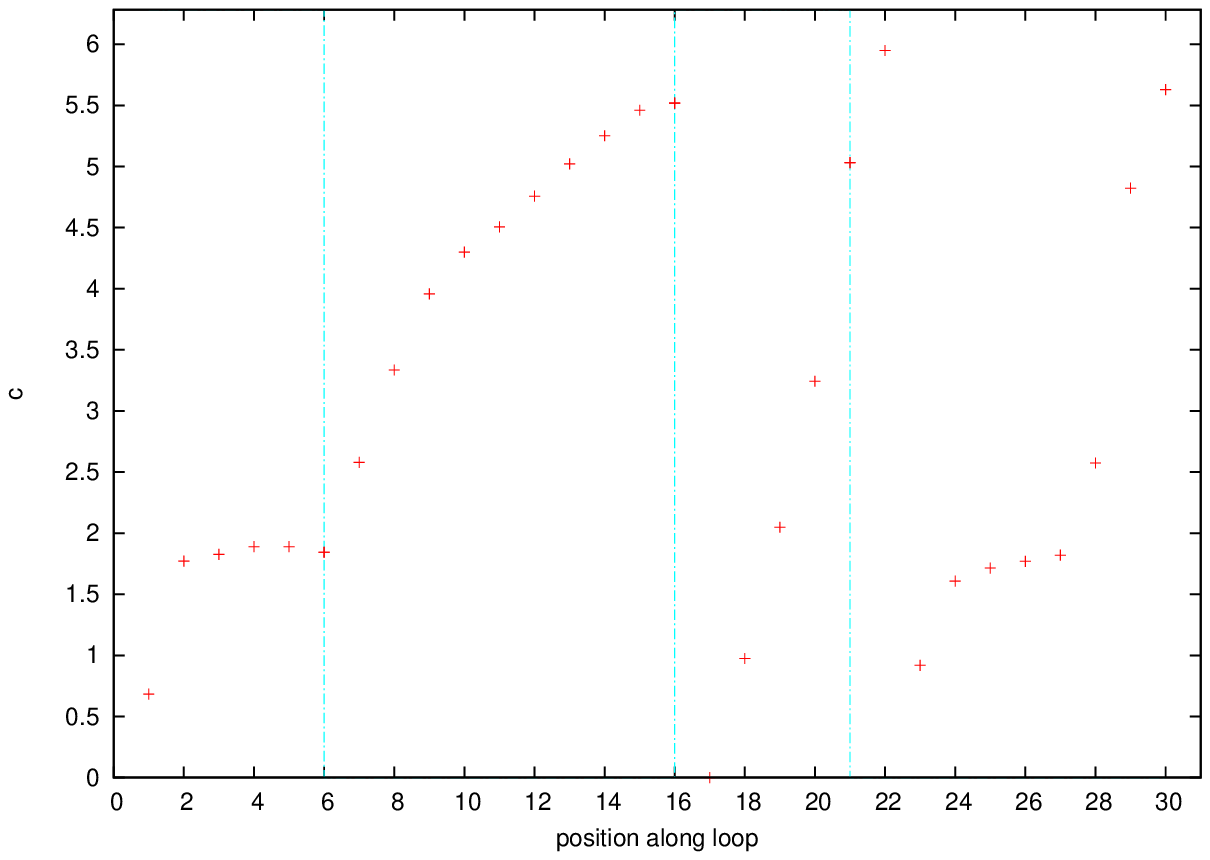}\\
   \includegraphics{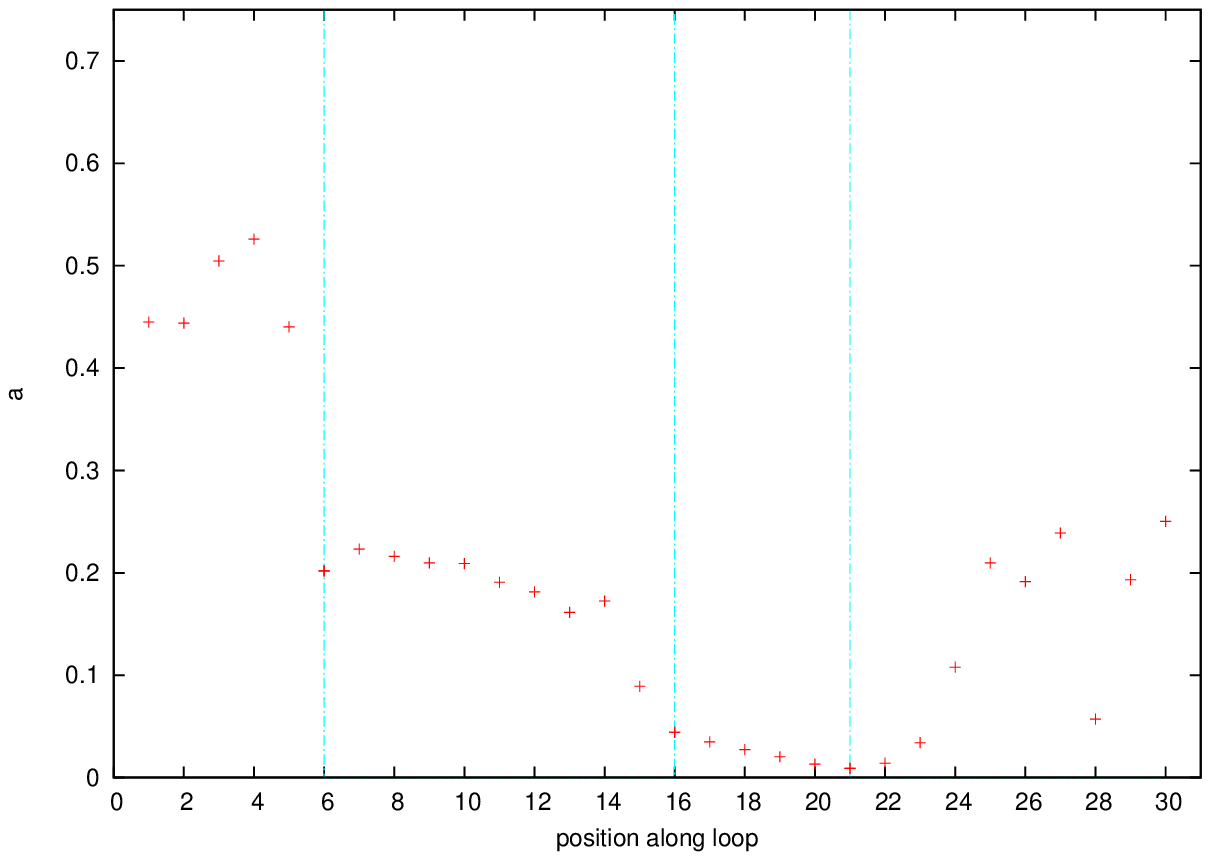}
 \end{tabular}
\end{center}
 \caption{The parameters $a$ and $c$ extracted from the Abelian Decomposition $\theta$ matrix, after gauge fixing, shown along one Wilson Loop starting at lattice site $(x,y,z,t) = (5,6,6,5)$. The vertical lines represent the corners of the Wilson Loop.}\label{fig:a_and_c5}
\end{figure}

\begin{figure}
\begin{center}
 \begin{tabular}{c}
   \includegraphics{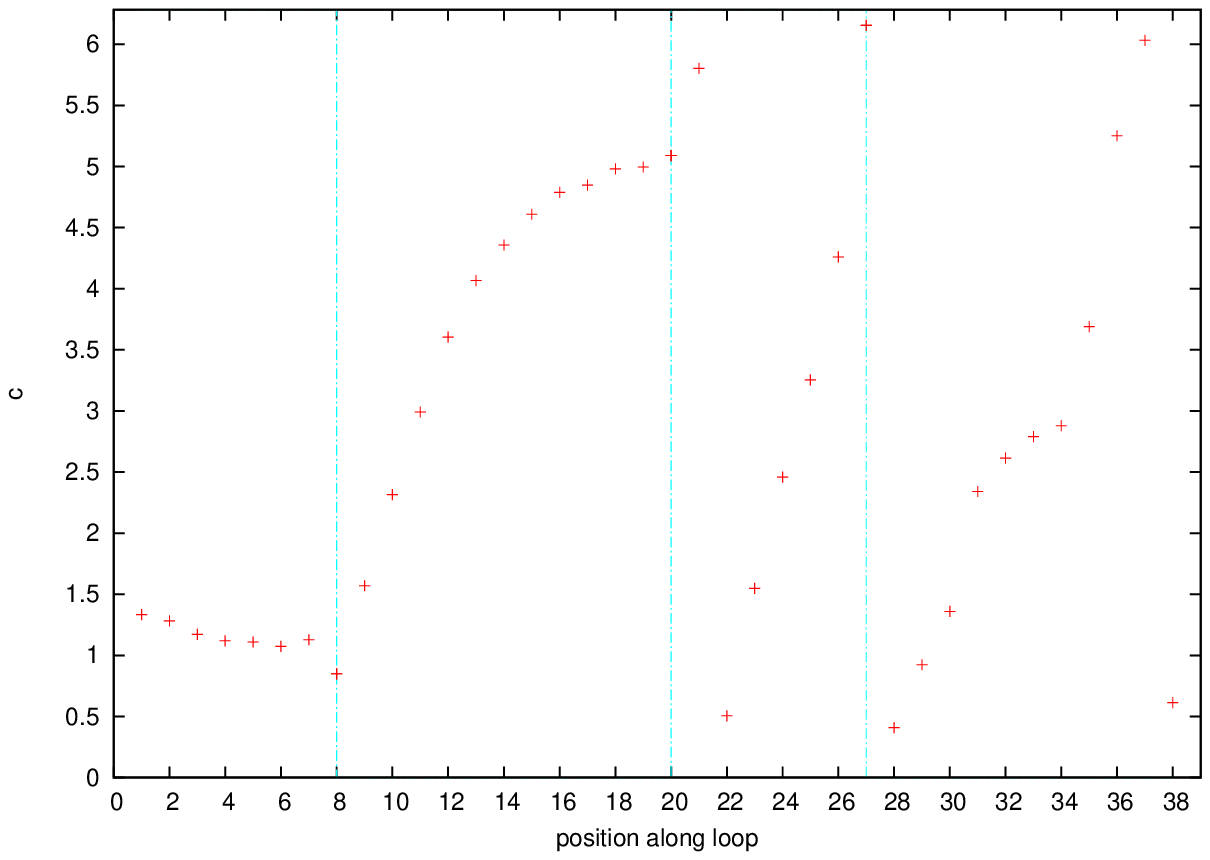}\\
   \includegraphics{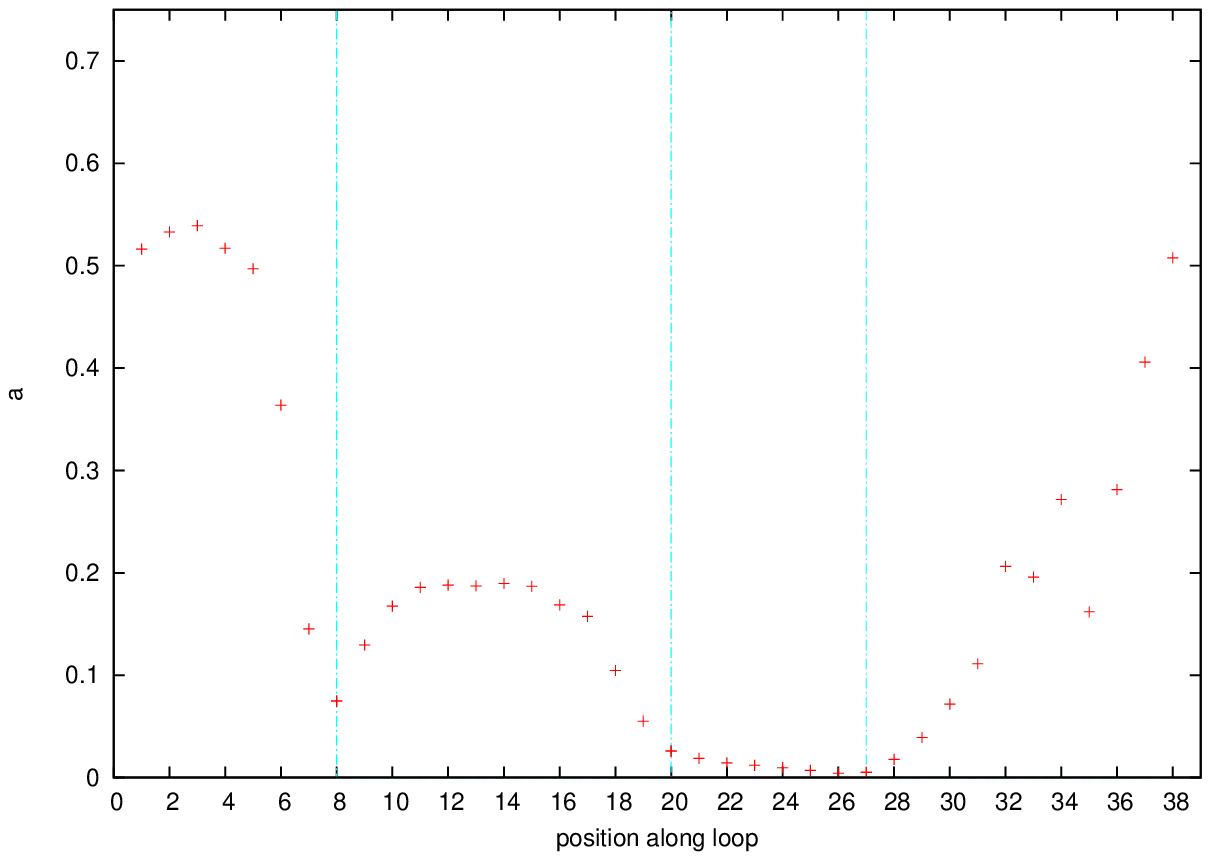}
 \end{tabular}
\end{center}
 \caption{The parameters $a$ and $c$ extracted from the Abelian Decomposition $\theta$ matrix, after gauge fixing, shown along one Wilson Loop starting at lattice site $(x,y,z,t) = (4,6,6,4)$. The vertical lines represent the corners of the Wilson Loop.}\label{fig:a_and_c6}
\end{figure}

Having gauge fixed, we can now investigate the variation of the parameters $a$ and $c$ as we move round the Wilson Loops. For figures \ref{fig:a_and_c1}-\ref{fig:a_and_c6}, we have taken one $yz$ slice of the lattice, and used this to create plots of $a$ and $c$. We looked numerous different slices of the lattice on different configurations, and saw similar features each time; the slice we show is a typical example (our only criteria in selecting the slice was to have an average amount of winding and avoid the extremes of both having too little -- where there would be no features of interest -- and too much winding -- where there would be too much going on and the picture would be confused -- on the slice we displayed. We show the first slice we looked at which satisfied this criteria). We have singled out a particular set of nested Wilson Loops to make the plots clearer. In figures \ref{fig:a_and_c1} and \ref{fig:a_and_c2}, we have presented a plot of $a$ and $c$ respectively plotted against the $x$ and $y$ coordinates of the lattice for four different Wilson Loops, which are represented by four different symbols on the plots. These are the Wilson loops starting at positions $(x,t) = $ $(4,4)$ $(5,5)$, $(6,6)$, $(7,7)$, or  with lengths in each direction $(7,12)$ (i.e. the Wilson Loop extends seven lattice sites in the $x$ direction and twelve in the $t$ direction), $(5,10)$, $(3,8)$ and $(1,6)$  respectively. To make these plots clearer, we have unwound each Wilson Loop and plotted them on a two dimensional figure, in figures \ref{fig:a_and_c3}-\ref{fig:a_and_c6}. Here the $x$ axis marks the position along the Wilson Loop. We start in the corner with smallest $x$ and $t$, then proceed along the loop in the direction of increasing $x$, then the direction of increasing $t$, then decreasing $x$ and decreasing $t$ until we return to the starting point. The  very left and very right of these plots thus indicate the same position on the Wilson Loop.
%
%
%

We can see several things from these plots. Firstly, we notice that the parameter $c$ can have a non-zero winding number as it moves along the Wilson Loop, as in figure \ref{fig:a_and_c5} where we can see it wrapping itself around the unit circle between positions 18 and 23 (and once more on the loop).  As already argued, the emergence of winding is a requirement for our explanation of how the area law of the Wilson Loop emerges. Secondly, we can see that the variation in both $a$ and $c$ is smooth most of the time as we move along the Wilson Loop.  This is not surprising, since we gauge fixed $\theta$ so that it was a reasonably smooth function of position. For the first plot, figure \ref{fig:a_and_c3}, with a $1\times 6$ Wilson Loop, we see that the evolution of $c$ is reasonably smooth (bearing in mind that $c=0$ is equivalent to $c = 2\pi$). The second plot, figure \ref{fig:a_and_c4}, with a $3\times 8$ Wilson Loop, has minima in $a$  at positions 15 and 21, and there is some discontinuity in $c$ there. In figure \ref{fig:a_and_c5}, the next of the nested Wilson Loops moving outwards, these small bumps each turn into a full  winding of the $c$ parameter around the unit circle. In the next Wilson Loop, in figure \ref{fig:a_and_c6}, the winding is maintained. The small value of $a$ at position 27  on figure \ref{fig:a_and_c6} leads to additional winding in subsequent Wilson Loops (not shown in these plots). Thirdly we notice that the values of $a$ tend to be relatively small; in principle $a$ can vary between $0$ and $\pi/2$, however we do not see any $a$ larger than about $\pi/4$. We notice that there are some points where the evolution of $c$ is not smooth, and that it jumps by a large amount. One example of this is at position $21$ on figure \ref{fig:a_and_c4}. We also observe that these jumps occur only at very small values of $a$, which again we expect since here there can be large changes in $c$ while $\theta$ evolves smoothly. 

In conclusion, we see winding emerging when $a$ has a minima close to zero. These represent the topological objects identified in our previous paper as candidate causes of confinement. The winding plays a key role in our candidate model for confinement, and that it is present is a necessary requirement for our model to be correct.

If the topological part dominates the string tension, there are three possibilities which could cause this. Recall that in SU(2) the topological part can be reduced to $\oint \sin^2 a \partial_\sigma c d \sigma$ where $\sigma$ represents the distance along the Wilson Loop. This leaves six possibilities for the area law of the Wilson Loop. 
\begin{enumerate}
 \item The average value of $\sin^2 a$ increases with the area of the Wilson Loop.
 \item There is an increasing correlation between $\sin^2 a$ and $\partial_\sigma c$, so that on larger loops $\sin^2 a$ is large where there is a large value of $\partial_\sigma c$, while on smaller loops there is no such correlation.
 \item $\oint \partial_\sigma c d \sigma$ increases with the area of the Wilson Loop, which means that the winding number scales with the area of the Wilson Loop.
\end{enumerate}
The remaining three options are various combinations of these; for example if both $\sin^2 a$ and the winding number increase with the square root of the area of the Wilson Loop.

\begin{figure}
 \begin{center}
  \begin{tabular}{cc}
   \includegraphics[width=6.5cm]{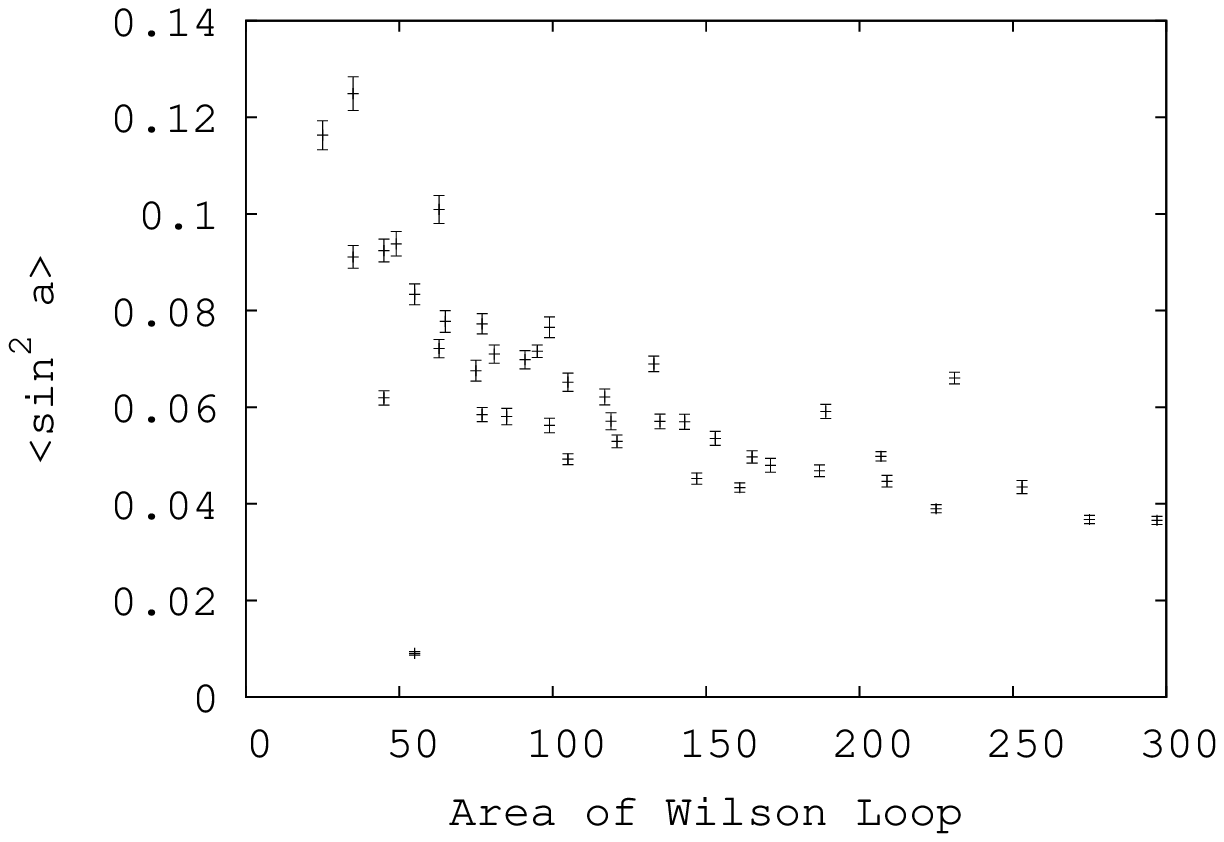} &
   \includegraphics[width=6.5cm]{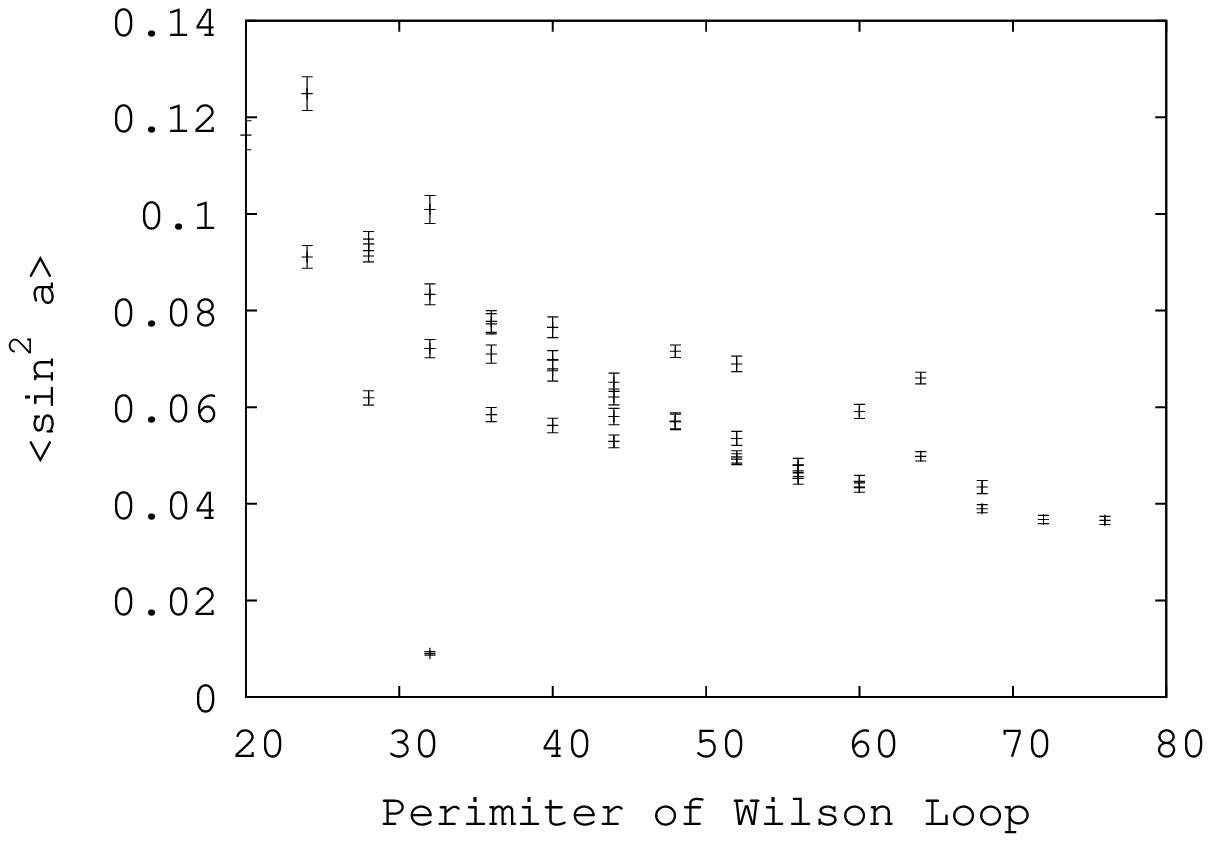} 
   \end{tabular}
 \end{center}
\caption{The average value of $\sin^2 a$ for those $\theta$ fields along a Wilson Loop plotted against the area of the Wilson Loop (left) and the perimeter of the Wilson Loop (right).}\label{fig:ava}
\end{figure}

Let us start by looking at how the average value of $\sin^2 a$ varies with the size of the Wilson Loop. This is plotted in figure \ref{fig:ava}. There are two plots here: the first shows the configuration average of $\sin^2 a$ compared to the area of the Wilson Loop, and the second shows the configuration average of $\sin^2 a$ compared to the perimeter of the Wilson Loop. We have excluded those Wilson Loops with either a very small extent in one direction (which should have some Coulomb part to the interaction), or an extent close to the lattice volume (where there might be finite volume effects). We do not see any significant change in this quantity as the size of the Wilson Loop changes. This means that, since $\sin^2 a$ does not increase with the Wilson Loop size, this cannot even partially be responsible for the area law and quark confinement.  One thing that surprises us about these results is that the average value of $\sin^2 a$ is well short of the the value of $\frac{1}{2}$ that is implied from the expected distribution of $\theta$ before gauge fixing was applied (for example as seen in the computed measure). Gauge fixing has seriously distorted the distribution of $\theta$. This is worrying because it makes our results harder to reconcile with the study of Greensite and H\"ollwieser in~\cite{Greensite:2014gra}, where it was found that circulating the Wilson Loop twice destroyed the signal for confinement. We expect the average value of the Wilson Loop to be (to a rough approximation) $2\pi \nu \langle \sin^2 a \rangle$ for each time we move around it, and if $\langle \sin^2 a \rangle \sim \frac{1}{2}$  then for a single circuit around the Wilson Loop, this will be some integer multiple of $\pi$, which can have a physical effect, while for a double circuit we would obtain (approximately) some integer multiple of $2\pi$, which is indistinguishable for $0$. This would explain why going round the Wilson Loop twice would destroy any signal for confinement. However, we find that $\langle \sin^2 a \rangle \sim 0.05$, which is considerably smaller than $\frac{1}{2}$. The issue of why the signal for confinement disappears when we circle the Wilson Loop twice is thus still uncertain. Possibly there is a finite size effect or lattice artefact at play. This will, however, need to be a topic for future research.

\begin{figure}
 \begin{center}
  \begin{tabular}{cc}
   \includegraphics[width=6.5cm]{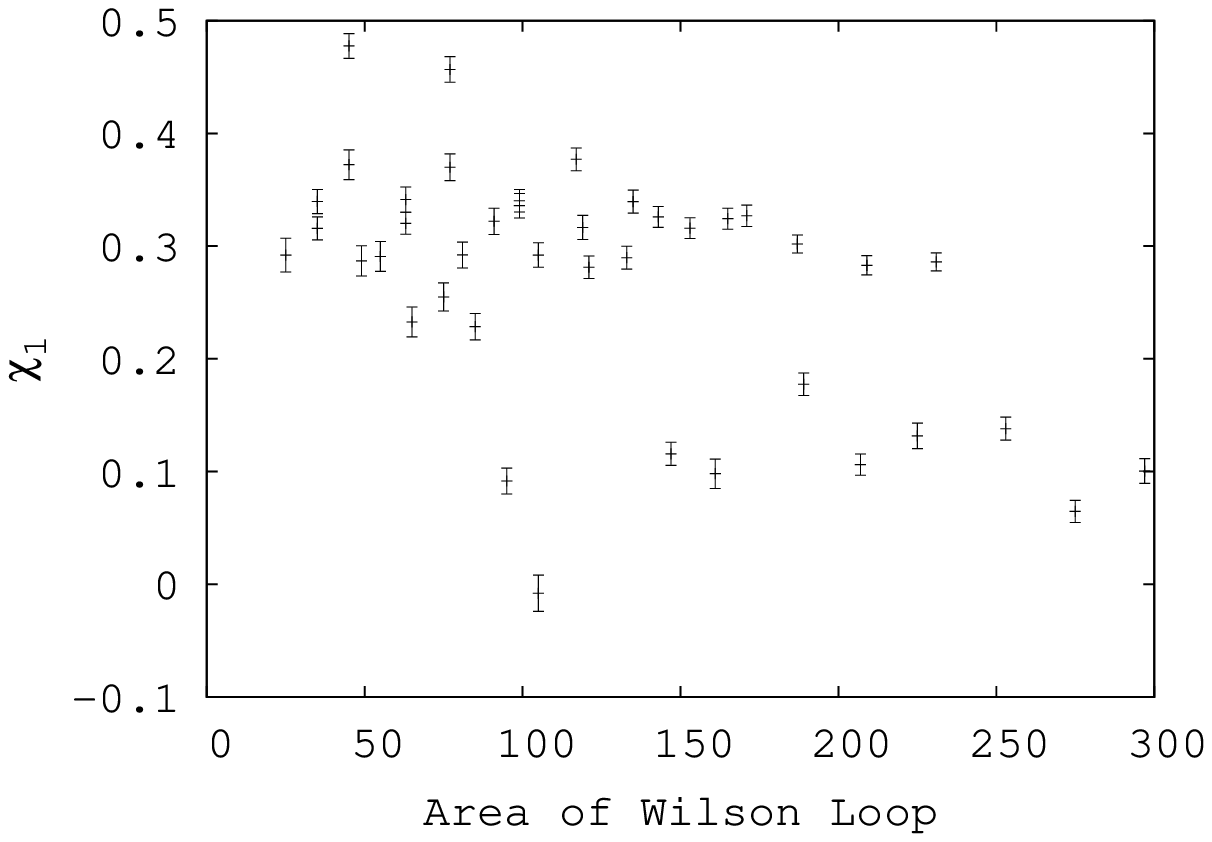} &
   \includegraphics[width=6.5cm]{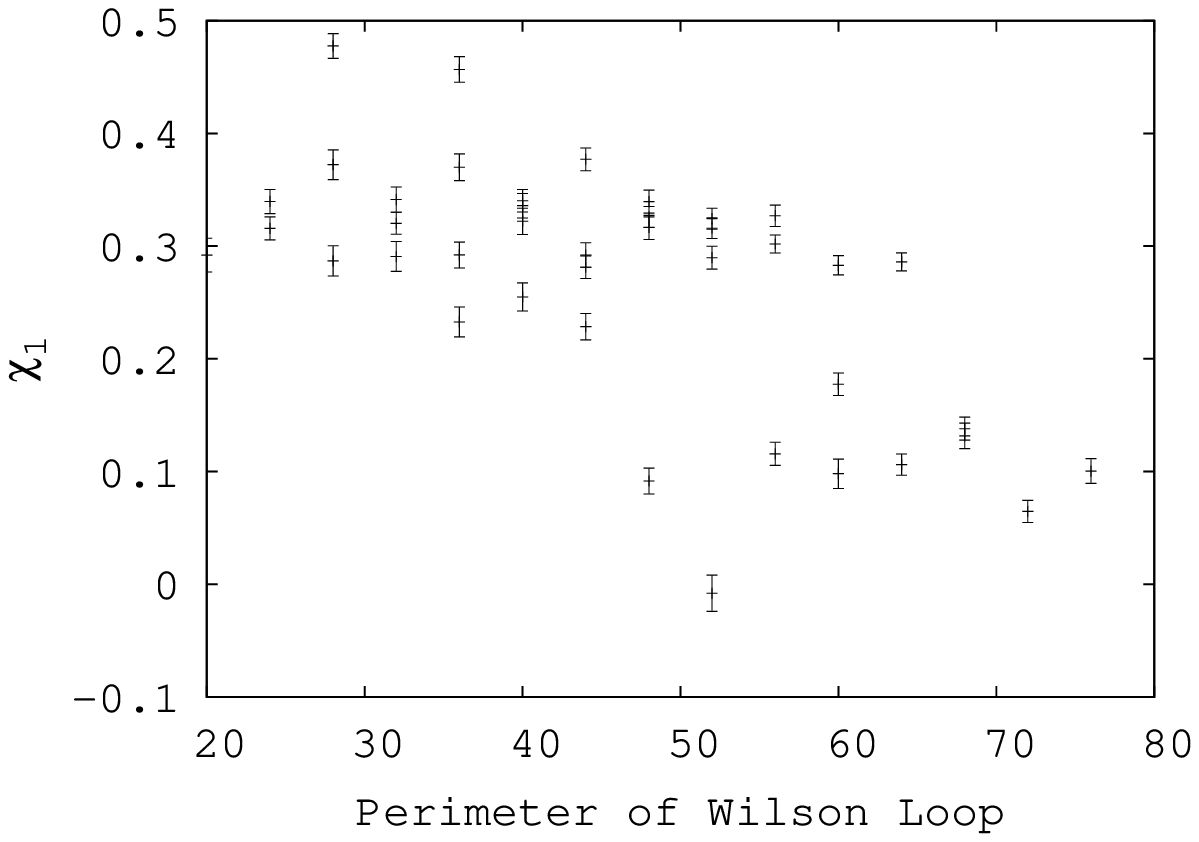} 
   \end{tabular}
 \end{center}
\caption{The average value of the correlator $\chi_1$ for those $\theta$ fields along a Wilson Loop plotted against the area of the Wilson Loop (left) and the perimeter of the Wilson Loop (right).}\label{fig:accorr1}
\end{figure}

\begin{figure}
 \begin{center}
  \begin{tabular}{cc}
   \includegraphics[width=6.5cm]{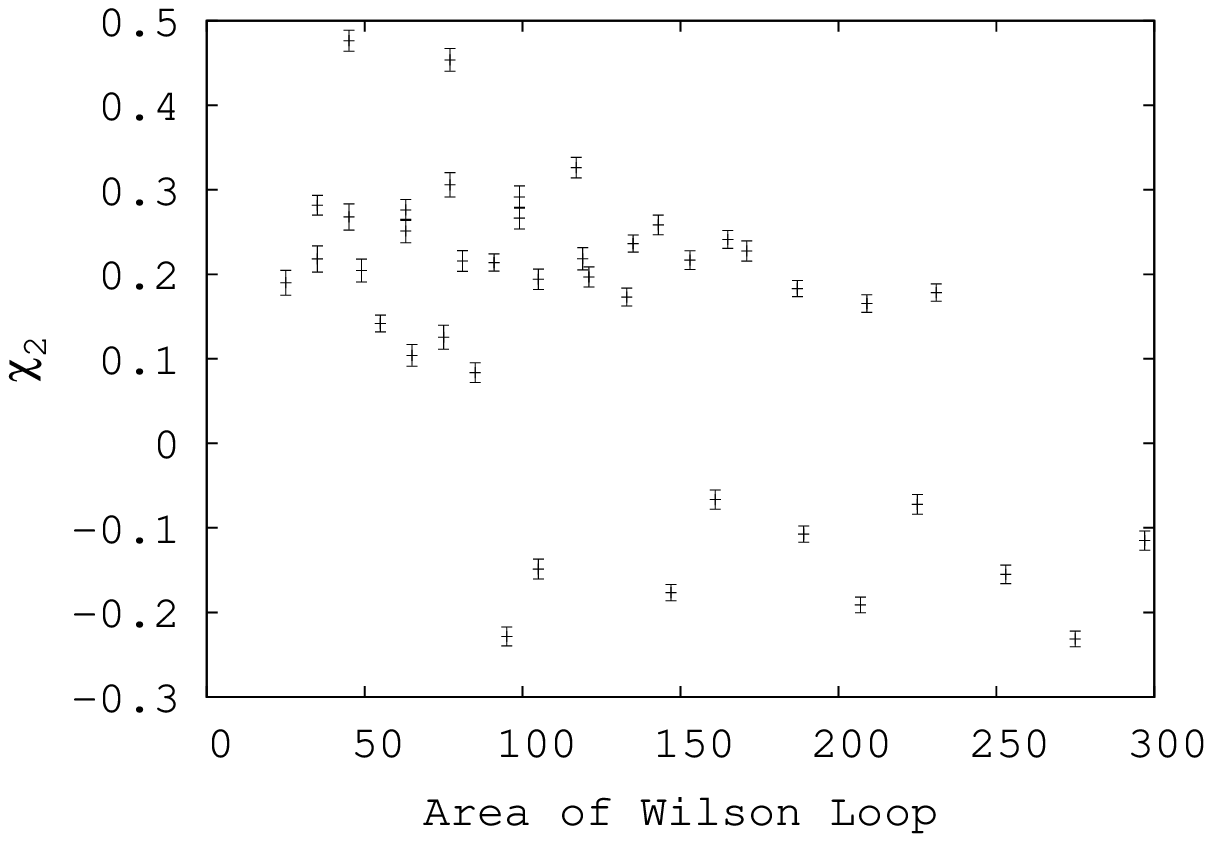} &
   \includegraphics[width=6.5cm]{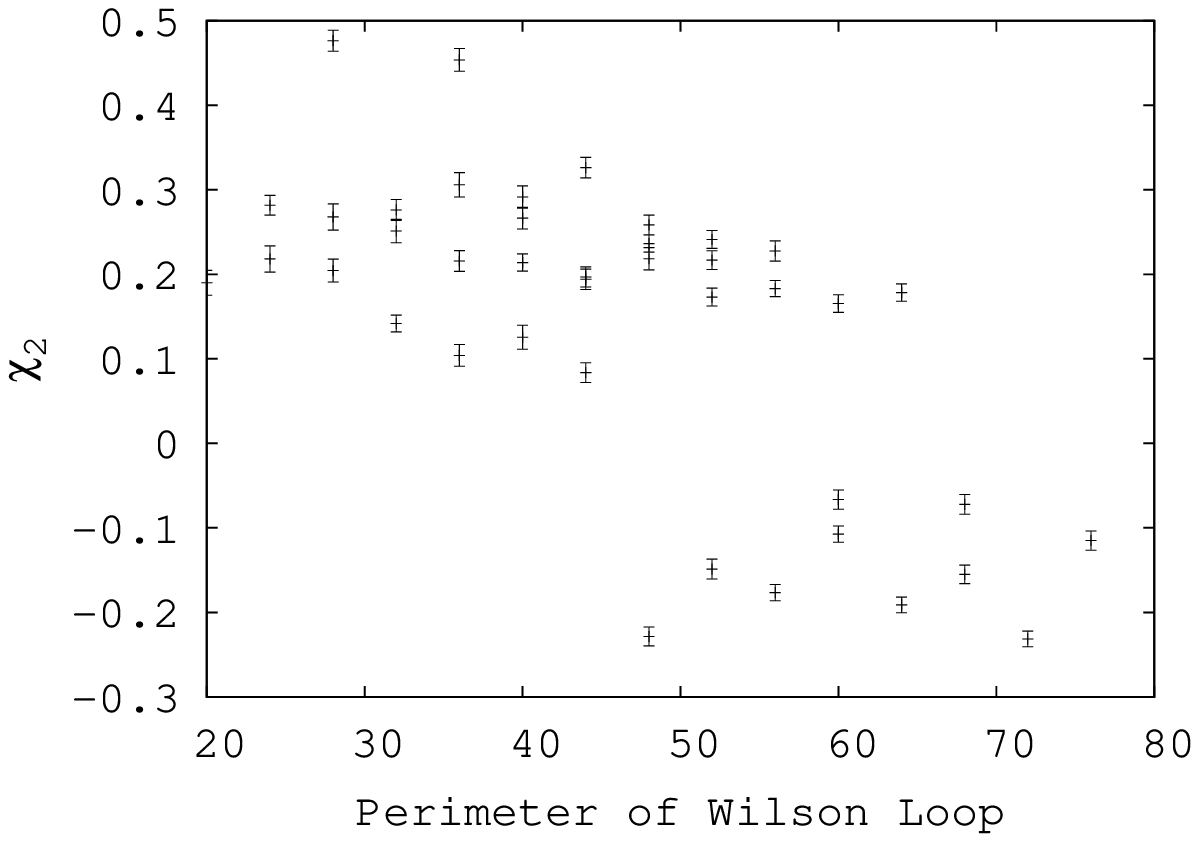} 
   \end{tabular}
 \end{center}
\caption{The average value of $\chi_2$ for those $\theta$ fields along a Wilson Loop plotted against the area of the Wilson Loop (left) and the perimeter of the Wilson Loop (right).}\label{fig:accorr2}
\end{figure}

We next check the correlation between $\sin^2 a$ and $\partial_\mu c$. We do this using two different measures,
\begin{align}
 \chi_1 = & \frac{\langle \sin^2 a \partial_\mu c\rangle - \langle \sin^2 a\rangle \langle \partial_\mu c \rangle}{\sqrt{(\langle \sin^4 a \rangle - \langle \sin^2 a\rangle^2) (\langle (\partial_\mu c)^2 \rangle - \langle \partial_\mu c \rangle^2)}}\nonumber\\
 \chi_2 = & \frac{\langle \sin^2 a |\partial_\mu c|\rangle - \langle \sin^2 a\rangle \langle |\partial_\mu c| \rangle}{\sqrt{(\langle \sin^4 a \rangle - \langle \sin^2 a\rangle^2) (\langle |\partial_\mu c|^2 \rangle - \langle |\partial_\mu c| \rangle^2)}}
\end{align}
If $\sin^2 a$ (with $a$ measured on the lattice as the average value between neighbouring lattice sites) and $\partial_\mu c$ (measured on the lattice via the difference operator) are correlated with each other, then these quantities should give 1. If there is no correlation, they should be smaller or even negative (for example if $a$ is small where $\partial_\mu c$ is large and vice versa). We plot these correlation functions against the Wilson Loop area and perimeter in figures \ref{fig:accorr1} and \ref{fig:accorr2}. Once again, we see that there is little dependence on the size of the Wilson Loop. The data is scattered with no obvious pattern at all. 
We do not see any strong evidence that this effect is driving confinement.

\begin{figure}
 \begin{center}
  \begin{tabular}{cc}
   \includegraphics[width=6.5cm]{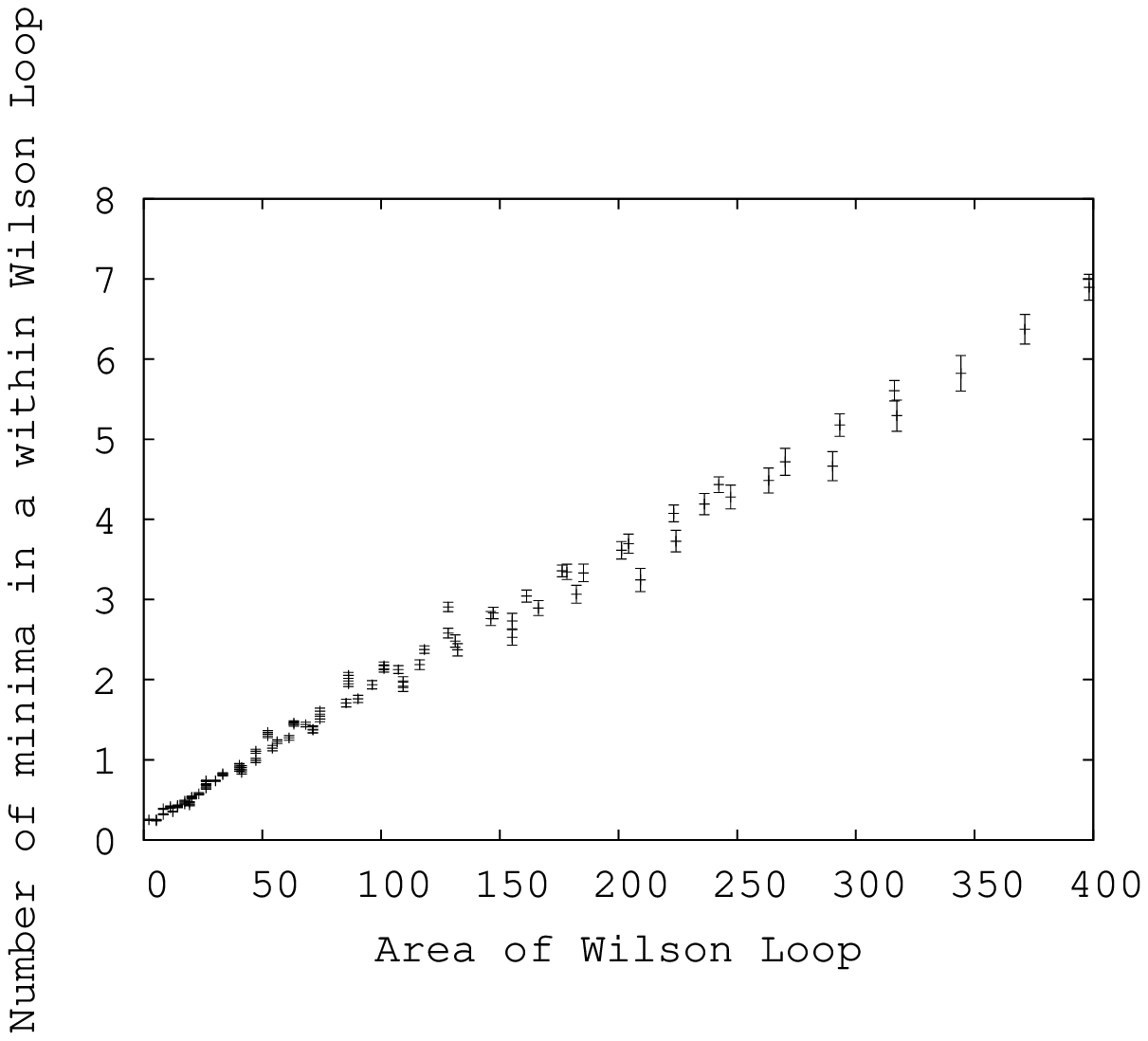} &
   \includegraphics[width=6.5cm]{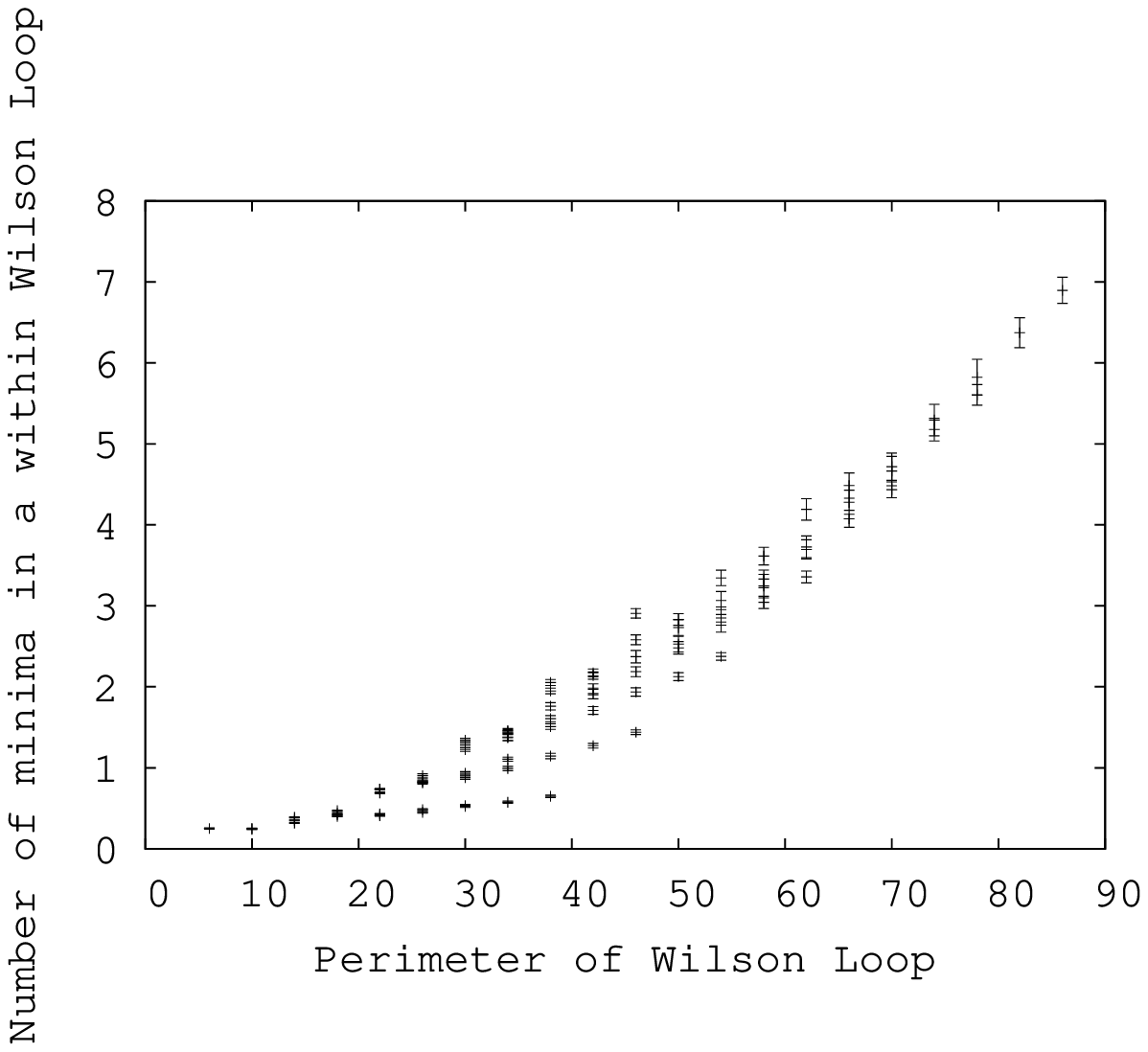} 
   \end{tabular}
 \end{center}
\caption{The average value of the number of peaks in the parameter $a$ for those $\theta$ fields on lattice sites within the area bounded by the Wilson Loop plotted against the area of the Wilson Loop (left) and the perimeter of the Wilson Loop (right).}\label{fig:peaks}
\end{figure}

So that just leaves the possibility that the winding number scales like the area of the Wilson Loop. We have already noted that the introduction of winding is related to small values of $a$, so the first thing to check is how many minima of $a$ are contained within a Wilson Loop of given area and volume. This is plotted in figure \ref{fig:peaks}, where we see, as expected, the number of minima is proportional to the area of the Wilson Loop. 

\begin{figure}
 \begin{center}
  \begin{tabular}{cc}
   \includegraphics[width=6.5cm]{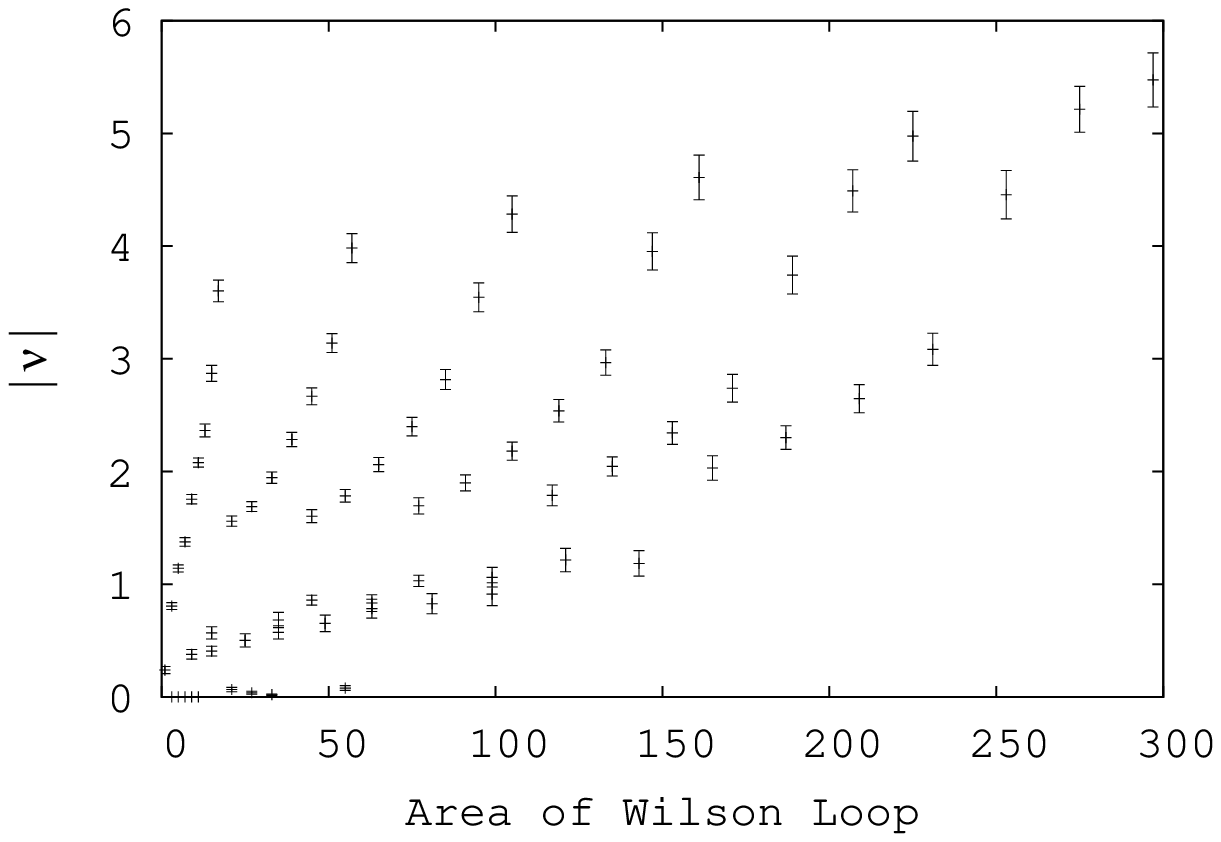} &
   \includegraphics[width=6.5cm]{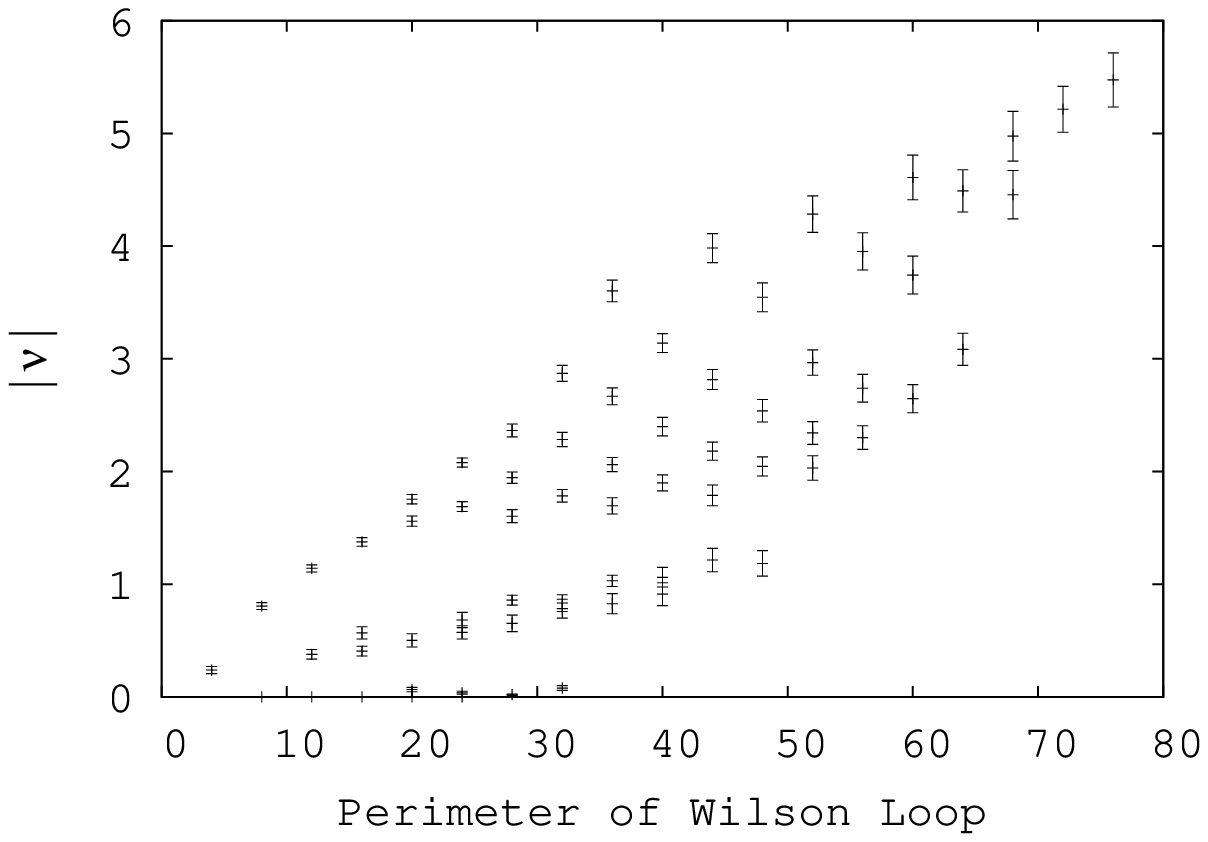} 
   \end{tabular}
 \end{center}
\caption{The average value of $|\nu|$, where $\nu$ is the winding number, for those $\theta$ fields along a Wilson Loop plotted against the area of the Wilson Loop (left) and the perimeter of the Wilson Loop (right).}\label{fig:winding}
\end{figure}

We plot the dependence of the winding number in figure \ref{fig:winding} where we show the average value of the modulus of the winding number $\nu$ (since ultimately $\pm |\nu|$ contribute equally to the final string tension) against the area of the loop and the loop's perimeter. Here we do see a clear increase in the winding number as the area of the loop increases. The data is, unfortunately, very scattered, suggesting that there might be a dependence on other factors as well as the loop area. The scatter makes it difficult to confirm a linear increase of the winding number with the area of the loop. This suggests that the statement that `the winding number alone drives the area law behaviour of the Wilson Loop' is a little too simplistic, since there is more to the question than just this. However, the increase of the winding number with the loop area and the lack of any increase in $\sin^2a$ and the correlation between $a$ and $c$ suggest that an increase in winding is the dominant effect.

\section{Conclusions}\label{sec:7}
In our previous work, we showed that the Cho-Duan-Ge Abelian decomposition was a useful tool to study quark confinement by demonstrating that the restricted field could fully account for the string tension, and by identifying the topological objects which could lead to an area law Wilson Loop and thus a confining potential. However, the restricted field is divided into two parts, the Maxwell part and the topological part, and in our previous work we did not show that the topological part dominated confinement. 

Our main result of this work is to show that by fixing to one of a particular set of gauges, the topological part of the action dominates the string tension. The topological objects we are concentrating on are present in the full Yang-Mills field in any gauge, but by performing the dual steps of the Abelian decomposition and the gauge fixing we are able to isolate them and separate them from many of the non-confining contributions to the gauge field. We can confirm via numerical simulations that in SU(2) gauge theory our proposed topological objects, which are generated when the parameter $a$ is small, and lead to a non-zero winding number for the second parameter we have labelled $c$ are indeed present. This strongly suggests that our proposed mechanism is at least partly responsible for confinement, at least in SU(2) gauge theory. The topological part alone reproduces the whole string tension, and the topological string tension is driven only by the increase of the winding number while the area of the loop increases, while the number of topological objects scales well with the area of the Wilson Loop. None of the other effects which could contribute to an area-law Wilson Loop seem to have any noticeable scaling with the area. 

We have also began the process of developing this idea into a fully dynamical theory. We have shown that it is possible to define the measure and express the action in terms of the parameters of our Abelian decomposition, with the dynamical parameters being $\cos 2{a}_\mu$, $\tilde{c}_\mu$ and $\tilde{d}_\mu$, with $\tilde{a}_\mu$ and $\tilde{c}_\mu$ describing the $\theta$ field and $\tilde{d}_\mu$ the decomposed restricted field. The Yang Mills action expressed in terms of these variables is complicated even in SU(2), and we have not yet developed the Feynman rules let alone investigated the renormalisability of this parametrisation of Yang-Mills theory. While the standard parametrisation of Yang-Mills theory is more useful since it makes Gauge and Lorentz symmetry more manifest, this new parametrisation might be helpful in trying to investigate the topological basis of confinement analytically as well as numerically. 

This work obvious still has numerous limitations, which need to be addressed in future research. Firstly, we have only studied the SU(2) theory instead of the physical SU(3). SU(2) has the advantage that it is considerably simpler and clearer, and we expect the same process to be involved in SU(3), since the SU(3) $\theta$ is built from three separate SU(2) subcomponents, each of which can contribute to the overall winding number in the same way as in SU(2), although the measure in SU(3) is only similar to the SU(2) measure for two of these sub-matrices. For the moment, SU(3) studies are too expensive for our available computer resources; however, this study needs to be reproduced in SU(3). Secondly, we have only considered one value of the lattice spacing and lattice volume. We do not expect any great difference as we change the lattice spacing or lattice volume (since we are only interested in qualitative effects rather than extracting some phenomenological result where a continuum extrapolation would be essential), but this is nonetheless something which could be improved. We also need to investigate the deconfinement transition, and the link with spontaneous chiral symmetry breaking. It would also be good to confirm our numerical results analytically, i.e. show that the topological objects do in fact arise in practice, and represent minima of the action. Having re-parametrised the Yang-Mills theory in terms of the variables which describe the topological objects, we believe that such an analytic approach will be possible, but it remains very challenging.

\section*{Acknowledgements}
Numerical calculations used servers at Seoul National University.

The research of NC was supported by Basic Science Research Program
through the National Research Foundation of Korea (NRF) funded by the
Ministry of Education (No. 2014063535).
The research of WL is supported by the Creative Research Initiatives Pro-
gram (No. 2015001776) of the NRF grant funded by the Korean government (MEST). WL would
like to acknowledge the support from KISTI supercomputing center through the strategic support
program for the supercomputing application research [No. KSC-2014-G3-002].

YMC is supporteed by NRF funded by MSFP 
(Grant 2012-002-134) and ME (Grant 2015-R1D1A1A0-1057578), 
and by Konkuk University.

\appendix
\section{Yang-Mills measure in SU(3)}\label{app:su3measure}
In SU(3), we can express the gauge transformation as
\begin{multline}
 \Lambda = e^{i\gamma_3 \lambda_3 + \gamma_8 \lambda_8} e^{i\bar{\beta}_1 \bar{\phi}_1} e^{i (\beta_1 + \tilde{a}_{1\mu,x})\phi_1} e^{i\bar{\beta}_2 \bar{\phi}_2} e^{i (\beta_2 + \tilde{a}_{2\mu,x})\phi_2} \\ e^{i\bar{\beta}_3 \bar{\phi}_3} e^{i (\beta_3 + \tilde{a}_{3\mu,x})\phi_3} e^{i (-\tilde{a}_{3\mu,x})\phi_3}e^{-i\tilde{a}_{2\mu,x} \phi_2} e^{-i\tilde{a}_{1\mu,x}\phi_1}.
\end{multline}
We will later have to express $\beta$, $\bar{\beta}$ in a form that is independent of $a$ and $c$. We know from the SU(2) case that if we apply the transformation 
\begin{gather}
\left(\begin{array}{ccc}e^{i\alpha'_1}&0&0\\0&e^{i\alpha'_2}&0\\ 0&0&e^{i\alpha'_3}\end{array}\right) e^{i\bar{\beta}_3 \bar{\phi}_3} e^{i (\beta_3)\phi_3} 
\end{gather}
to the operator
\begin{gather}
e^{i\tilde{a}_{3\mu,x} \phi_3} \left(\begin{array}{ccc}e^{i\alpha_1}&0&0\\0&e^{i\alpha_2}&0\\ 0&0&e^{i\alpha_3}\end{array}\right),
\end{gather}
then we get 
\begin{align}
 \tilde{a}_{3\mu,x}\rightarrow &\tilde{a}_{3\mu,x} + \beta_3\nonumber\\
 \tilde{c}_{3\mu,x} \rightarrow &\tilde{c}_{3\mu,x} - \cot( \tilde{a}_{3\mu,x}) \bar{\beta}_3 +\bar{\beta}_3 \tan \tilde{a}_{3\mu,x}+ \alpha'_2 - \alpha'_3\nonumber\\
 \alpha_1\rightarrow& \alpha_1 + \alpha'_1\nonumber\\
 \alpha_2 \rightarrow& \alpha_2 + \alpha'_2 + \bar{\beta}_3 \tan \tilde{a}_{3\mu,x}\nonumber\\
 \alpha_3 \rightarrow& \alpha_3 + \alpha'_3 - \bar{\beta}_3 \tan \tilde{a}_{3\mu,x}
\end{align}
We can then read off the transformations of the coordinates,
\begin{align}
\tilde{a}_{1\mu,x} \rightarrow &\tilde{a}_{1\mu,x} + \beta_1\nonumber\\
\tilde{c}_{1\mu,x} \rightarrow& \tilde{c}_{1\mu,x} - \bar{\beta}_1\cot( \tilde{a}_{1\mu,x}) + \bar{\beta}_1 \tan ( \tilde{a}_{1\mu,x}) +2 \gamma_3\nonumber\\
\tilde{a}_{2\mu,x}\rightarrow &\tilde{a}_{2\mu,x} + \beta_2\nonumber\\
\tilde{c}_{2\mu,x} \rightarrow& \tilde{c}_{2\mu,x} -\bar{\beta}_2\cot(\tilde{a}_{2\mu,x}) + \bar{\beta}_2 \tan(\tilde{a}_{2\mu,x}) + \sqrt{3}\gamma_8 + \gamma_3 +\bar{\beta}_1 \tan \tilde{a}_{1\mu,x}\nonumber\\
\tilde{a}_{3\mu,x} \rightarrow & \tilde{a}_{3\mu,x} + \beta_3\nonumber\\
\tilde{c}_{3\mu,x} \rightarrow & \tilde{c}_{3\mu,x} - \bar{\beta}_3 \cot( \tilde{a}_{3\mu,x}) +\bar{\beta}_3 \tan(\tilde{a}_{2\mu,x})-\gamma_3 + \sqrt{3} \gamma_8 - \bar{\beta}_1\tan \tilde{a}_{1\mu,x} + \bar{\beta}_2 \tan \tilde{a}_{2\mu,x}\nonumber\\
\tilde{d}_{3\mu,x} \rightarrow& \tilde{d}_{3\mu,x} + \gamma_3 + \frac{1}{2}(2 \bar{\beta}_1\tan \tilde{a}_{1\mu,x} + \bar{\beta}_2 \tan \tilde{a}_{2\mu,x} - \bar{\beta}_3 \tan \tilde{a}_{3\mu,x})\nonumber\\
\tilde{d}_{8\mu,x} \rightarrow & \tilde{d}_{8\mu,x} + \gamma_8 + \frac{\sqrt{3}}{2}(\bar{\beta}_3 \tan \tilde{a}_{3\mu,x} + \bar{\beta}_2 \tan \tilde{a}_{2\mu,x}).\label{eq:tranofcoord1}
\end{align}
Next we need to compare our gauge transformation with a more `neutral' one (independent of $a$ and $c$), so that we fully express the new coordinates in terms of the old ones. 

We express a transformation operator in terms of parameters $l_1,\ldots,l_8$ as
\begin{gather}
 \left( \begin{array}{ccc}1+il_3+\frac{l_8}{\sqrt{3}}&il_1 + l_2&il_4+l_5\\
 il_1-l_2&1-il_3 + \frac{l_8}{\sqrt{3}}&il_6+l_7\\
 il_4 - l_5&il_6-l_7&1-\frac{2il_8}{\sqrt{3}}\end{array}\right).
\end{gather}
By expanding the transformation matrix, we can relate an infinitesimal $\beta$ and $l$:
\begin{align}
 l_1 =&
 \cos\left( \tilde{c}_{1\mu,x}\right) \beta_1 + \sin\left( \tilde{c}_{1\mu,x}\right)\bar{\beta}_1  +\nonumber\\
 &\left(\sin\left( \tilde{a}_{2\mu,x}\right)({\sin^2\left( \tilde{a}_{1\mu,x}\right) } \sin\left( \tilde{c}_{3\mu,x}-\tilde{c}_{2\mu,x}+2\tilde{c}_{1\mu,x}\right) +
 \cos^2( \tilde{a}_{1\mu,x}) \sin\left( \tilde{c}_{3\mu,x}-\tilde{c}_{2\mu,x}\right) )\right)\beta_3 + \nonumber\\
 &\left( -\cos^2( \tilde{a}_{1\mu,x}) \sin\left( \tilde{a}_{2\mu,x}\right) \cos\left( \tilde{c}_{3\mu,x}-\tilde{c}_{2\mu,x}\right)
 -{\sin^2\left( \tilde{a}_{1\mu,x}\right) }\sin\left( \tilde{a}_{2\mu,x}\right) \cos\left( \tilde{c}_{3\mu,x}-\tilde{c}_{2\mu,x}+2\tilde{c}_{1\mu,x}\right) \right)\bar{\beta}_3\\
 l_4 = &   \beta_{2}\cos\left( \tilde{a}_{1\mu,x}\right) \cos\left( \tilde{c}_{2\mu,x}\right)  -\sin\left( \tilde{a}_{1\mu,x}\right) \cos\left( \tilde{a}_{2\mu,x}\right) \sin\left( \tilde{c}_{3\mu,x}+\tilde{c}_{1\mu,x}\right) \beta_3 +\nonumber\\
 &\bar{\beta}_2 \cos\left( \tilde{a}_{1\mu,x}\right) \sin\left( \tilde{c}_{2\mu,x}\right)  +\bar{\beta}_3 \sin\left( \tilde{a}_{1\mu,x}\right) \cos\left( \tilde{a}_{2\mu,x}\right) \cos\left( \tilde{c}_{3\mu,x}+\tilde{c}_{1\mu,x}\right)  
 \\
 l_6 = &    -\beta_2\sin\left( \tilde{a}_{1\mu,x}\right) \sin\left( \tilde{c}_{2\mu,x}-\tilde{c}_{1\mu,x}\right)  + \beta_3\cos\left( \tilde{a}_{1\mu,x}\right) \cos\left( \tilde{a}_{2\mu,x}\right) \cos\left( \tilde{c}_{3\mu,x}\right) +\nonumber\\
 &\bar{\beta}_2 \sin\left( \tilde{a}_{1\mu,x}\right) \cos\left( \tilde{c}_{2\mu,x}-\tilde{c}_{1\mu,x}\right)  +\bar{\beta}_3 \cos\left( \tilde{a}_{1\mu,x}\right) \cos\left( \tilde{a}_{2\mu,x}\right) \sin\left( \tilde{c}_{3\mu,x}\right)  
 \\
 l_2 = &  -\beta_1\sin\left( \tilde{c}_{1\mu,x}\right) +\bar{\beta}_1 \cos\left( \tilde{c}_{1\mu,x}\right)  +\nonumber\\
 &\beta_3 \left({\sin^2\left( \tilde{a}_{1\mu,x}\right) }\sin\left( \tilde{a}_{2\mu,x}\right) \cos\left( \tilde{c}_{3\mu,x}-\tilde{c}_{2\mu,x}+2\tilde{c}_{1\mu,x}\right) -\cos^2( \tilde{a}_{1\mu,x}) \sin\left( \tilde{a}_{2\mu,x}\right) \cos\left( \tilde{c}_{3\mu,x}-\tilde{c}_{2\mu,x}\right) \right) +\nonumber\\
 &\bar{\beta}_3 {\sin\left( \tilde{a}_{1\mu,x}\right) }^{2}\sin\left( \tilde{a}_{2\mu,x}\right) \sin\left( \tilde{c}_{3\mu,x}-\tilde{c}_{2\mu,x}+2\tilde{c}_{1\mu,x}\right) -\cos^2( \tilde{a}_{1\mu,x}) \sin\left( \tilde{a}_{2\mu,x}\right) \sin\left( \tilde{c}_{3\mu,x}-\tilde{c}_{2\mu,x}\right)  
 \\
 l_5 = &   -\beta_2\cos\left( \tilde{a}_{1\mu,x}\right) \sin\left( \tilde{c}_{2\mu,x}\right)   -\beta_3\sin\left( \tilde{a}_{1\mu,x}\right) \cos\left( \tilde{a}_{2\mu,x}\right) \cos\left( \tilde{c}_{3\mu,x}+\tilde{c}_{1\mu,x}\right)  +\nonumber\\
 &\bar{\beta}_2\cos\left( \tilde{a}_{1\mu,x}\right) \cos\left( \tilde{c}_{2\mu,x}\right)   -\bar{\beta}_3\sin\left( \tilde{a}_{1\mu,x}\right) \cos\left( \tilde{a}_{2\mu,x}\right) \sin\left( \tilde{c}_{3\mu,x}+\tilde{c}_{1\mu,x}\right)  
 \\
 l_7 = & -\beta_2\sin\left( \tilde{a}_{1\mu,x}\right) \cos\left( \tilde{c}_{2\mu,x}-\tilde{c}_{1\mu,x}\right)   -\beta_3\cos\left( \tilde{a}_{1\mu,x}\right) \cos\left( \tilde{a}_{2\mu,x}\right) \sin\left( \tilde{c}_{3\mu,x}\right)  -\nonumber\\
 &\bar{\beta}_2\sin\left( \tilde{a}_{1\mu,x}\right) \sin\left( \tilde{c}_{2\mu,x}-\tilde{c}_{1\mu,x}\right)  +\bar{\beta}_3 \cos\left( \tilde{a}_{1\mu,x}\right) \cos\left( \tilde{a}_{2\mu,x}\right) \cos\left( \tilde{c}_{3\mu,x}\right) 
 \\
 l_3 = &\gamma_3 + 2\beta_3 \cos\left( \tilde{a}_{1\mu,x}\right) \sin\left( \tilde{a}_{1\mu,x}\right) \sin\left( \tilde{a}_{2\mu,x}\right) \cos\left( \tilde{c}_{3\mu,x}\tilde{c}_{1\mu,x}-\tilde{c}_{2\mu,x}\right)  +\nonumber\\
 &2\bar{\beta}_3\cos\left( \tilde{a}_{1\mu,x}\right) \sin\left( \tilde{a}_{1\mu,x}\right) \sin\left( \tilde{a}_{2\mu,x}\right) \sin\left( \tilde{c}_{3\mu,x}\tilde{c}_{1\mu,x}-\tilde{c}_{2\mu,x}\right)
 \\
 l_8 = &\gamma_8
\end{align}
The inverse of this linear transformation gives us
\begin{align}
 \beta_1=&\cos\left( \tilde{c}_{1\mu,x}\right) l_1 + \frac{\sin\left( \tilde{a}_{1\mu,x}\right) \sin\left( \tilde{a}_{2\mu,x}\right) \cos\left( \tilde{c}_{2\mu,x}\right) }{\cos\left( \tilde{a}_{2\mu,x}\right) }l_4 -\nonumber\\& \frac{\cos\left( \tilde{a}_{1\mu,x}\right) \sin\left( \tilde{a}_{2\mu,x}\right) \sin\left( \tilde{c}_{1\mu,x}-\tilde{c}_{2\mu,x}\right) }{\cos\left( \tilde{a}_{2\mu,x}\right) }l_6  -\nonumber\\
 &\sin\left( \tilde{c}_{1\mu,x}\right) l_2 - \frac{\sin\left( \tilde{a}_{1\mu,x}\right) \sin\left( \tilde{a}_{2\mu,x}\right) \sin\left( \tilde{c}_{2\mu,x}\right) }{\cos\left( \tilde{a}_{2\mu,x}\right) } l_5+ \nonumber\\&\frac{\cos\left( \tilde{a}_{1\mu,x}\right) \sin\left( \tilde{a}_{2\mu,x}\right) \cos\left( \tilde{c}_{1\mu,x}-\tilde{c}_{2\mu,x}\right) }{\cos\left( \tilde{a}_{2\mu,x}\right) }l_7
 \\
   \beta_2=&\cos\left( \tilde{a}_{1\mu,x}\right) \cos\left( \tilde{c}_{2\mu,x}\right) l_4 + \sin\left( \tilde{a}_{1\mu,x}\right) \sin\left( \tilde{c}_{1\mu,x}-\tilde{c}_{2\mu,x}\right) l_6 -\nonumber\\
   &\left(\cos\left( \tilde{a}_{1\mu,x}\right) \right) \sin\left( \tilde{c}_{2\mu,x}\right)  l_5+ \left( -\sin\left( \tilde{a}_{1\mu,x}\right) \right) \cos\left( \tilde{c}_{1\mu,x}-\tilde{c}_{2\mu,x}\right) l_7
  \\
  \beta_3=& \frac{-\sin\left( \tilde{a}_{1\mu,x}\right) \sin\left( \tilde{c}_{3\mu,x}+\tilde{c}_{1\mu,x}\right) }{\cos\left( \tilde{a}_{2\mu,x}\right) }l_4 + \frac{\cos\left( \tilde{a}_{1\mu,x}\right) \cos\left( \tilde{c}_{3\mu,x}\right) }{\cos\left( \tilde{a}_{2\mu,x}\right) }l_6- \nonumber\\
  &\frac{\left( \sin\left( \tilde{a}_{1\mu,x}\right) \right) \cos\left( \tilde{c}_{3\mu,x}+\tilde{c}_{1\mu,x}\right) }{\cos\left( \tilde{a}_{2\mu,x}\right) } l_5- \frac{\cos\left( \tilde{a}_{1\mu,x}\right) \sin\left( \tilde{c}_{3\mu,x}\right) }{\cos\left( \tilde{a}_{2\mu,x}\right) }l_7
  \\
  \bar{\beta}_1=&\sin\left( \tilde{c}_{1\mu,x}\right) l_1- \frac{\sin\left( \tilde{a}_{1\mu,x}\right) \cos\left( 2 \tilde{a}_{1\mu,x}\right) \sin\left( \tilde{a}_{2\mu,x}\right) \sin\left( \tilde{c}_{2\mu,x}\right) }{\cos\left( \tilde{a}_{2\mu,x}\right) } l_4+\nonumber\\
  &\frac{\cos\left( \tilde{a}_{1\mu,x}\right) \cos\left( 2 \tilde{a}_{1\mu,x}\right) \sin\left( \tilde{a}_{2\mu,x}\right) \cos\left( \tilde{c}_{1\mu,x}-\tilde{c}_{2\mu,x}\right) }{\cos\left( \tilde{a}_{2\mu,x}\right) }l_6+ \nonumber\\
  &\cos\left( \tilde{c}_{1\mu,x}\right) l_2- \frac{\sin\left( \tilde{a}_{1\mu,x}\right) \cos\left( 2 \tilde{a}_{1\mu,x}\right) \sin\left( \tilde{a}_{2\mu,x}\right) \cos\left( \tilde{c}_{2\mu,x}\right) }{\cos\left( \tilde{a}_{2\mu,x}\right) } l_5+ \nonumber\\
  &\frac{\cos\left( \tilde{a}_{1\mu,x}\right) \cos\left( 2 \tilde{a}_{1\mu,x}\right) \sin\left( \tilde{a}_{2\mu,x}\right) \sin\left( \tilde{c}_{1\mu,x}-\tilde{c}_{2\mu,x}\right) }{\cos\left( \tilde{a}_{2\mu,x}\right) }l_7
  \\
  \bar{\beta}_2=&\cos\left( \tilde{a}_{1\mu,x}\right) \sin\left( \tilde{c}_{2\mu,x}\right) l_4 + \sin\left( \tilde{a}_{1\mu,x}\right) \cos\left( \tilde{c}_{1\mu,x}-\tilde{c}_{2\mu,x}\right) l_6 + \nonumber\\
  &\cos\left( \tilde{a}_{1\mu,x}\right) \cos\left( \tilde{c}_{2\mu,x}\right) l_5+ \sin\left( \tilde{a}_{1\mu,x}\right) \sin\left( \tilde{c}_{1\mu,x}-\tilde{c}_{2\mu,x}\right)l_7
  \\
  \bar{\beta}_3=&\frac{\left( \sin\left( \tilde{a}_{1\mu,x}\right) \right) \cos\left( \tilde{c}_{3\mu,x}+\tilde{c}_{1\mu,x}\right) }{\cos\left( \tilde{a}_{2\mu,x}\right) }l_4 + \frac{\cos\left( \tilde{a}_{1\mu,x}\right) \sin\left( \tilde{c}_{3\mu,x}\right) }{\cos\left( \tilde{a}_{2\mu,x}\right) }l_6+\nonumber\\
  &\frac{-\sin\left( \tilde{a}_{1\mu,x}\right) \sin\left( \tilde{c}_{3\mu,x}+\tilde{c}_{1\mu,x}\right) }{\cos\left( \tilde{a}_{2\mu,x}\right) }l_5+ \frac{\cos\left( \tilde{a}_{1\mu,x}\right) \cos\left( \tilde{c}_{3\mu,x}\right) }{\cos\left( \tilde{a}_{2\mu,x}\right) } l_7
  \\
  \gamma_3=&\frac{\sin\left( \tilde{a}_{1\mu,x}\right) \sin\left( 2 \tilde{a}_{1\mu,x}\right) \sin\left( \tilde{a}_{2\mu,x}\right) \sin\left( \tilde{c}_{2\mu,x}\right) }{\cos\left( \tilde{a}_{2\mu,x}\right) }l_4- \nonumber\\
  &\frac{\cos\left( \tilde{a}_{1\mu,x}\right) \sin\left( 2 \tilde{a}_{1\mu,x}\right) \sin\left( \tilde{a}_{2\mu,x}\right) \cos\left( \tilde{c}_{1\mu,x}-\tilde{c}_{2\mu,x}\right) }{\cos\left( \tilde{a}_{2\mu,x}\right) }l_6 + \nonumber\\
  &\frac{\sin\left( \tilde{a}_{1\mu,x}\right) \sin\left( 2 \tilde{a}_{1\mu,x}\right) \sin\left( \tilde{a}_{2\mu,x}\right) \cos\left( \tilde{c}_{2\mu,x}\right) }{\cos\left( \tilde{a}_{2\mu,x}\right) } l_5 - \nonumber\\
  &\frac{\cos\left( \tilde{a}_{1\mu,x}\right) \sin\left( 2 \tilde{a}_{1\mu,x}\right) \sin\left( \tilde{a}_{2\mu,x}\right) \sin\left( \tilde{c}_{1\mu,x}-\tilde{c}_{2\mu,x}\right) }{\cos\left( \tilde{a}_{2\mu,x}\right) } l_7+l_3
  \\
  \gamma_8 = &l_8
\end{align}
We write
\begin{align}
 A =& \cos (\tilde{c}_{1\mu,x}) l_1 - \sin (\tilde{c}_{1\mu,x}) l_2\nonumber\\
 B = & \cos (\tilde{c}_{2\mu,x}) l_4 - \sin (\tilde{c}_{2\mu,x}) l_5\nonumber\\
 C =& \cos (\tilde{c}_{1\mu,x}-\tilde{c}_{2\mu,x}) l_7-\sin(\tilde{c}_{1\mu,x}-\tilde{c}_{2\mu,x}) l_6\nonumber\\
 D = &\cos (\tilde{c}_{3\mu,x} + \tilde{c}_{1\mu,x}) l_5 + \sin(\tilde{c}_{3\mu,x}+\tilde{c}_{1\mu,x}) l_4\nonumber\\
 E = &\cos (\tilde{c}_{3\mu,x}) l_6 - \sin (\tilde{c}_{3\mu,x}) l_7
\end{align}
which gives
\begin{align}
 \delta \tilde{a}_{1\mu,x} = & A + \sin \tilde{a}_{1\mu,x} \tan \tilde{a}_{2\mu,x} B + \cos \tilde{a}_{1\mu,x} \tan \tilde{a}_{2\mu,x} C\nonumber\\
 \delta \tilde{a}_{2\mu,x} =& \cos \tilde{a}_{1\mu,x} B - \sin \tilde{a}_{1\mu,x} C\nonumber\\
 \delta \tilde{a}_{3\mu,x} = & -\frac{\sin \tilde{a}_{1\mu,x}}{\cos \tilde{a}_{2\mu,x}} D + \frac{\cos \tilde{a}_{1\mu,x}}{\cos \tilde{a}_{2\mu,x}} E
\end{align}
The relevant terms which contribute to the Jacobian are
\begin{align}
 \frac{\partial \beta_1}{\partial \tilde{a}_{1\mu,x}} =& B \cos \tilde{a}_{1\mu,x} \tan \tilde{a}_{2\mu,x} - C \tan \tilde{a}_{2\mu,x} \sin \tilde{a}_{1\mu,x}\nonumber\\
 \frac{\partial \bar{\beta}_1}{\partial \tilde{c}_{1\mu,x}} = & A + C \cos \tilde{a}_{1\mu,x} \cos 2 \tilde{a}_{1\mu,x} \tan \tilde{a}_{2\mu,x}\nonumber\\
 \frac{\partial \gamma_3}{\partial \tilde{c}_{1\mu,x}} =& - C \cos \tilde{a}_{1\mu,x} \sin 2 \tilde{a}_{1\mu,x} \tan \tilde{a}_{2\mu,x}\nonumber\\
 \frac{\partial \gamma_3}{\partial \tilde{c}_{2\mu,x}} =& C \cos \tilde{a}_{1\mu,x} \sin 2 \tilde{a}_{1\mu,x} \tan \tilde{a}_{2\mu,x} + B \sin \tilde{a}_{1\mu,x} \sin 2 \tilde{a}_{1\mu,x} \tan \tilde{a}_{2\mu,x} \nonumber\\
 \frac{\partial\bar{\beta}_2}{\partial \tilde{c}_{2\mu,x}} =& B \cos \tilde{a}_{1\mu,x} - C \sin \tilde{a}_{1\mu,x} = \delta \tilde{a}_{2\mu,x}\nonumber\\
 \frac{\partial\bar{\beta}_3}{\partial \tilde{c}_{3\mu,x}} = & -\frac{\sin \tilde{a}_{1\mu,x}}{\cos \tilde{a}_{2\mu,x}} D + \frac{\cos \tilde{a}_{1\mu,x}}{\cos \tilde{a}_{2\mu,x}} E = \delta \tilde{a}_{3\mu,x}.
\end{align}
 Multiplying each of these terms by the coefficients implied by equation (\ref{eq:tranofcoord1}) and summing them up gives the Jacobian to first order in $l_i$ (at $\tilde{a}_i = 0$ or $\tilde{a}_i = \pi/2$ this calculation breaks down since the expansion is in in $l \ll \cos \tilde{a}$ and $l \ll \sin \tilde{a}$; however these points will be forbidden by the measure so we can neglect them):
 \begin{align}
  J =& 1 - 2\cot 2 \tilde{a}_{3\mu,x} \delta \tilde{a}_{3\mu,x} - 2 \cot 2 \tilde{a}_{2\mu,x} \delta \tilde{a}_{2\mu,x} - 2\cot 2 \tilde{a}_{1\mu,x} A + \nonumber\\
  &B\tan \tilde{a}_{2\mu,x} (\cos \tilde{a}_{1\mu,x} - \frac{\sin^2 \tilde{a}_{1\mu,x}}{\cos \tilde{a}_{1\mu,x}} \cos 2 \tilde{a}_{1\mu,x} + 2 \sin^2 \tilde{a}_{1\mu,x} \cos \tilde{a}_{1\mu,x}) + \nonumber\\
  &C\tan \tilde{a}_{2\mu,x} (-\sin \tilde{a}_{1\mu,x} - 2 \cot 2 \tilde{a}_{1\mu,x} \cos \tilde{a}_{1\mu,x} \cos a \tilde{a}_{1\mu,x} -\nonumber\\&
  \phantom{spacespacespacespace}\sin \tilde{a}_{1\mu,x} \cos 2 \tilde{a}_{1\mu,x} - \cos \tilde{a}_{1\mu,x} \sin 2 \tilde{a}_{1\mu,x})\\
  =& 1 - 2\cot 2 \tilde{a}_{3\mu,x} \delta \tilde{a}_{3\mu,x} - 2 \cot 2 \tilde{a}_{2\mu,x} \delta \tilde{a}_{2\mu,x} - 2\cot 2 \tilde{a}_{1\mu,x} A +\nonumber\\
  &B\frac{\tan \tilde{a}_{2\mu,x}}{\cos \tilde{a}_{1\mu,x}}(\cos^2 \tilde{a}_{1\mu,x} - \sin^2 \tilde{a}_{1\mu,x} (\cos^2 \tilde{a}_{1\mu,x} - \sin^2 \tilde{a}_{1\mu,x}) + 2 \sin^2 \tilde{a}_{1\mu,x} \cos^2 \tilde{a}_{1\mu,x}) + \nonumber\\
  &C\frac{\tan \tilde{a}_{2\mu,x}}{\sin \tilde{a}_{1\mu,x}} (-\sin^2 \tilde{a}_{1\mu,x} -\nonumber\\
  & \phantom{spacespa}(\cos^2 \tilde{a}_{1\mu,x} - \sin^2 \tilde{a}_{1\mu,x}) (\cos^2 \tilde{a}_{1\mu,x} - \sin ^2 \tilde{a}_{1\mu,x} + \sin^2 \tilde{a}_{1\mu,x} - 2\sin^2 \tilde{a}_{1\mu,x} \cos^2 \tilde{a}_{1\mu,x}))\\
  =&1 - 2\cot 2 \tilde{a}_{3\mu,x} \delta \tilde{a}_{3\mu,x} - 2 \cot 2 \tilde{a}_{2\mu,x} \delta \tilde{a}_{2\mu,x} - 2\cot 2 \tilde{a}_{1\mu,x} A +\nonumber\\
  &B\tan \tilde{a}_{2\mu,x} \left(2 \cos \tilde{a}_{1\mu,x}  -  \frac{\cos 2 \tilde{a}_{1\mu,x}}{\cos \tilde{a}_{1\mu,x}}\right) - C\tan \tilde{a}_{2\mu,x} \left(2 \sin \tilde{a}_{1\mu,x} + \frac{\cos 2 \tilde{a}_{1\mu,x} }{\sin \tilde{a}_{1\mu,x}}\right)\\
  =& 1 - 2\cot 2 \tilde{a}_{3\mu,x} \delta \tilde{a}_{3\mu,x} - 2 \cot 2 \tilde{a}_{2\mu,x} \delta \tilde{a}_{2\mu,x} -2 \cot 2 \tilde{a}_{1\mu,x} \delta \tilde{a}_{1\mu,x} + 2 \tan \tilde{a}_{2\mu,x} \delta \tilde{a}_{2\mu,x}\nonumber\\
  =& 1 - \delta \log (\sin 2 \tilde{a}_{3\mu,x}) - \delta \log (\sin 2 \tilde{a}_{1\mu,x}) - \delta (\log (\sin 2\tilde{a}_{2\mu,x}) + 2\delta \log (\cos \tilde{a}_{2\mu,x})).
 \end{align} 
Therefore, since the measure satisfies
\begin{gather}
 (1+\frac{\partial \log \mu}{\partial \tilde{a}_{1\mu,x}} \delta \tilde{a}_{1\mu,x} +\frac{\partial \log \mu}{\partial \tilde{a}_{2\mu,x}} \delta \tilde{a}_{2\mu,x} +\frac{\partial \log \mu}{\partial \tilde{a}_{3\mu,x}} \delta \tilde{a}_{3\mu,x}) J = 1, 
\end{gather}
we find
\begin{gather}
 \mu = \sin 2 \tilde{a}_{1\mu,x} \sin 2\tilde{a}_{2\mu,x} \sin 2 \tilde{a}_{3\mu,x} \cos^2 \tilde{a}_{2\mu,x}.
\end{gather}

\bibliographystyle{elsart-num.bst}
\bibliography{weyl}

\begin{thebibliography}{10}
\expandafter\ifx\csname url\endcsname\relax
  \def\url#1{\texttt{#1}}\fi
\expandafter\ifx\csname urlprefix\endcsname\relax\def\urlprefix{URL }\fi

\bibitem{Vortices}
G.~{'t Hooft}, Nucl. Phys. B190 (1981) 455.

\bibitem{Monopoles1}
G.~{'t Hooft}, Nucl. Phys. B79 (1974) 276.

\bibitem{Monopoles2}
A.~M. Polyakov, JETP Lett. 20 (1974) 194.

\bibitem{Mandelstam:1976}
S.~Mandelstam, Phys. Reports 23C (1976) 245.

\bibitem{thooft:1976}
G.~{'t Hooft}, in: A.~Zichichi (Ed.), {High Energy Physics}, Editrice
  Comprostrini, Bologna, 1976.

\bibitem{Cundy:2015caa}
N.~Cundy, Y.~Cho, W.~Lee, J.~Leem, {The Static Quark Potential from the Gauge
  Independent Abelian Decomposition}, Nucl.Phys. B895 (2015) 64--131.

\bibitem{Cundy:2013xsa}
N.~Cundy, {Y. M. Cho}, W.~Lee, J.~Leem, {The static quark potential from the
  gauge invariant Abelian decomposition}, Phys.Lett. B729 (2014) 192--198.

\bibitem{Cho:1980}
Y.~M. Cho, Phys. Rev. D 21 (1980) 1080.

\bibitem{Cho:1981}
Y.~M. Cho, Phys. Rev. D 23 (1981) 2415.

\bibitem{F-N:98}
L.~Faddeev, A.~Niemi, Phys. Rev. Lett. 82 (1999) 1624.

\bibitem{Shabanov:1999}
S.~Shabanov, Phys. Lett. B 458 (1999) 322.

\bibitem{Duan:1979}
Y.~Duan, M.~Ge, Sci. Sinica 11 (1979) 1072.

\bibitem{Kondo:2008su}
K.-I. Kondo, {Wilson loop and magnetic monopole through a non-Abelian Stokes
  theorem}, Phys. Rev. D77 (2008) 085029.

\bibitem{Shibata:2009af}
A.~Shibata, K.-I. Kondo, T.~Shinohara, {The Exact decomposition of gauge
  variables in lattice Yang-Mills theory}, Phys. Lett. B691 (2010) 91--98.

\bibitem{Kondo:2010pt}
K.-I. Kondo, A.~Shibata, T.~Shinohara, S.~Kato, {Non-Abelian Dual
  Superconductor Picture for Quark Confinement}, Phys. Rev. D83 (2011) 114016.

\bibitem{Kondo:2005eq}
K.-I. Kondo, T.~Murakami, T.~Shinohara, {Yang-Mills theory constructed from
  Cho-Faddeev-Niemi decomposition}, Prog.Theor.Phys. 115 (2006) 201--216.

\bibitem{Shibata:2007pi}
A.~Shibata, et~al., {Toward gauge independent study of confinement in SU(3)
  Yang-Mills theory}, POS LATTICE-2007 (2007) 331.

\bibitem{Wilson:1974}
K.~G. Wilson, Phys. Rev. D10 (1974) 2445.

\bibitem{Bali:2005fu}
G.~S. Bali, H.~Neff, T.~Duessel, T.~Lippert, K.~Schilling, {Observation of
  string breaking in QCD}, Phys.Rev. D71 (2005) 114513.

\bibitem{HMC}
S.~Duane, A.~Kennedy, B.~Pendleton, D.~Roweth, Phys. Lett. B195 (1987) 216.

\bibitem{TILW}
M.~L{\"u}scher, P.~Weisz, {}, Commun Math Phys 97 (1985) 59.

\bibitem{Morningstar:2003gk}
C.~Morningstar, M.~J. Peardon, Phys. Rev. D69 (2004) 054501.

\bibitem{Moran:2008ra}
P.~J. Moran, D.~B. Leinweber, Phys. Rev. D77 (2008) 094501.

\bibitem{Greensite:2014gra}
J.~Greensite, R.~Höllwieser, {Double-winding Wilson loops and monopole
  confinement mechanisms}, Phys. Rev. D91~(5) (2015) 054509.

\end{thebibliography}

\end{document}